\documentclass[12pt, preprint]{aastex}
\usepackage{amsmath}
\usepackage{amssymb}
\usepackage{graphicx}
\usepackage{color}
\usepackage{xcolor}
\usepackage[caption=false]{subfig}

\newcommand{\be}{\begin{equation}}
\newcommand{\ee}{\end{Fuation}}
\newcommand{\bea}{\begin{eqnarray}}
\newcommand{\eea}{\end{eqnarray}}

\begin{document}

\slugcomment{FERMILAB-PUB-17-187-AE, UCI-TR-2017-05}

\title{Milky Way Tomography with K and M Dwarf Stars: \\ 
the Vertical Structure of the Galactic Disk}

\author{
Deborah Ferguson$^{1,\ast}$, Susan Gardner$^{1,2}$, and Brian Yanny$^{3}$ 
}
\affil{
${}^{1}$Department of Physics and Astronomy, 
University of Kentucky, Lexington, KY 40506-0055, USA\\
${}^{2}$Department of Physics and Astronomy, University of California, Irvine, CA 92697, USA\\
${}^{3}$Fermi National Accelerator Laboratory, Batavia, IL 60510, USA
}

\begin{abstract}
We use the number density distributions of K and M dwarf stars with vertical height 
from the Galactic disk, determined using observations from the Sloan Digital Sky Survey, 
to probe the structure of the Milky Way disk across the survey's footprint. 
Using photometric parallax as a distance estimator we analyze a sample
of several million disk stars in matching footprints above and
below the Galactic plane, and we determine the location and extent
of vertical asymmetries in the number counts in 
a variety of thin- and thick- disk subsamples 
in regions of some 
200 square degrees 
within 2 kpc in vertical distance from the Galactic disk. 
These disk asymmetries present 
wave-like features as previously observed 
on other scales
and at other distances from the Sun. 
We additionally explore 
%variations in 
the scale height of the disk and
the implied offset of the Sun from the Galactic plane at different locations, 
noting that the scale height of the disk can differ significantly when measured
using stars only above or only below the plane. 
Moreover, we compare the shape of the number density 
distribution in the north for different latitude ranges with a fixed 
range in longitude 
and find the shape to be sensitive to the selected latitude window.  
We explain why this may be indicative of 
a change in stellar populations in the latitude regions compared, 
possibly 
allowing access to the 
systematic metallicity difference between thin- and 
thick- disk populations through photometry.
\end{abstract}

% note from http://journals.aas.org/authors/keywords2013.html#Galaxies
% fundamental parameters?

%\date{\today} 

\vfill
\eject 

\section{Introduction}

The study of the structure of the Milky Way galaxy emerges 
as a key tool in the analysis of the nature of its dark halo 
and the most massive dwarf galaxies within it, through the manner 
in which they interact with the visible gas, dust, and stars of the Galactic disk. 
Astrometric observations at large scales support a cosmology 
of cold dark matter (CDM) and dark energy, the so-called $\Lambda$CDM model~\citep{planck16}, and 
an ``inside-out'' 
formation history of the cosmos~\citep{white78,frenk83} and
indeed  of the Milky Way. 
At small scales this model predicts 
many more satellite galaxies in the Milky Way than have been 
observed~\citep{kauffmann93,klypin99}. 
Discoveries of faint Milky Way satellites continue 
to be made (see, e.g., \citet{belokurov07}), 
however, and the technical limitations of current observations 
explain at least 
some of the mismatch~\citep{tollerud08,walsh09,bullock10}
in the numbers of observed and expected satellite galaxies, 
or subhalos. 
Nevertheless, questions persist concerning the number, evolution, and mass distribution 
of Milky Way subhalos (see \citet{kravtsov10} for a review), 
as well as their stellar content~\citep{boylan-kolchin_big11}.
Yet more dwarf galaxies could potentially be detected through a search 
for their tidal imprint on outer gaseous disks~\citep{chakrabarti11}, and low-mass subhalos
can be destroyed through interactions with the disk~\citep{donghia10}: 
these disjoint ideas argue for the importance of the observational study we effect here, namely, 
of the vertical structure of the Galactic disk within 2 kpc of the Sun in height and in-plane distance.

The hierarchical nature of structure formation, i.e., the 
ongoing merging of clumps with time, suggests that dark matter could have surviving
phase-space structure, which could also be imprinted on the stars. This could occur, e.g., 
through the formation of stellar streams in the halo. In this paper we are interested
in identifying possible traces of halo-disk interactions and thus wish to search for 
the appearance of stellar number 
count distributions that break the spatial symmetries expected from the integral
of motions embedded in virialized probability distribution functions. 
To do this we focus on an observational study of that 
%the latter 
possibility through use of a sample of some 3.6 million K and M
dwarfs from the Sloan Digital Sky Survey (SDSS). 

The appearance of asymmetric structures in the Milky Way has been 
noted through HI gas, dust, and stellar tracers, particularly in 
association with the warping of the disk~\citep{binney92}. 
Various scenarios have been proposed for the appearance of warps, though analytic~\citep{nelson95}
and numerical~\citep{shen06} studies suggest a dynamical origin: warps can appear and disappear 
through interactions of the disk with the halo and/or the satellites it contains 
over time scales short with respect to the 
age of the universe. 
The stellar disk is more complicated and thus exhibits additional features. 
Rings~\citep{newberg03,morganson16}, as well as ripples~\citep{price-whelan15,xu15}, 
have been noted, with the latter 
at distances in excess of 10 kpc from the Galactic center, 
where the disk is thinning out. In the solar neighborhood 
vertical asymmetries have been noted in the number counts~\citep{widrow12,yanny13}, 
as well as in the radial velocities~\citep{widrow12}, 
which have been confirmed by other studies~\citep{carlin13,williams13}. 
These asymmetries could be temporally recent as well and 
be generated by non-axisymmetric features in the disk, 
such as the spiral arms or galactic bar~\citep{debattista14,faure14,monari15,monari16}, 
or by interactions of the disk with the halo and its 
embedded satellite galaxies~\citep{gomez13,widrow14,gomez15b,gomez16,laporte16}. 
It is thought that the 
asymmetries in the number
counts are more suggestive of a dynamical origin~\citep{laporte16}, 
although, on the other hand, it appears 
that the spiral arms possess out-of-plane structure as well~\citep{camargo15}. 

Only two of the Milky Way's satellites are known to be possibly 
massive enough and close enough to 
be able to perturb 
the structure of the dark halo and Galactic disk: the Sagittarius dwarf spheroidal (dSph) 
galaxy at 20 kpc and the Large Magellanic Cloud at 50 kpc from the 
Sun~\citep{weinberg98,jiang99,bailin03,purcell11,gomez13}. Such massive objects are 
believed to have formed at late cosmic times~\citep{boylan-kolchin11}. 
Numerical simulations of the tidal interactions of the satellites reveal significant vertical 
perturbations of the disk~\citep{gomez13,widrow14,gomez15b,gomez16,laporte16}. 
However, simulations of such effects 
do not yield the precise vertical asymmetries observed in the 
existing data~\citep{laporte16}. 

In this paper we revisit and expand the earlier studies of \citet{widrow12} and \citet{yanny13}
to scrutinize the vertical structure of the Galactic disk
and its variation across the Galactic plane. We 
use a larger stellar photometric sample to 
subdivide the disk of the Milky Way within $|z| < 2$ kpc in vertical distance from the
Galactic plane into bins of longitude $l$ and latitude $b$ 
to explore the extent to which the large-scale
        asymmetries persist to smaller scales.  We show that these asymmetries are more
        pronounced toward the Galactic anticenter, 
   and we find similarities to the ripples 
seen at further distances from the Sun by \citet{xu15} and \citet{price-whelan15}. 
        We provide quantitative locations and amplitudes of these asymmetries so that
        those who model Milky Way structure can compare with them; associated with these
features are changes in the disk scale heights, north and south --- we report these as well. 
Interestingly, numerical simulations of the 
tidal interactions of the 
disk-satellite-halo system reveal an induced spiral arm and
barred structure in the Galactic disk~\citep{purcell11,laporte16}. 
This also motivates our observational study, and 
we compare the variations we do observe with the known spiral arm structure of 
the Galactic disk~\citep{camargo15}. 

Our studies are also pertinent to the long-standing problem of 
the determination of the matter density in the vicinity of the Sun~\citep{kapteyn22}, as
inferred from the measured kinematics of the local stars~\citep{oort32}. 
In this so-called Oort problem, an assumed gravitationally relaxed population of stars
is used to trace the local gravitational potential and to infer the local matter
and dark matter densities, where we note \citet{binney08} and \citet{read14} for reviews. 
%{\svg{add these ``Oort'' refs? 
%bahcall bienayme kujiken gilmore holmbergy bovy \& treimaine *gabari* bovy \& rix}}
The appearance of 
vertical oscillations in the stellar number counts and 
velocity distributions of the Milky Way~\citep{widrow12,yanny13} suggests
that the local stars may not be sufficiently gravitationally relaxed, incurring 
additional uncertainty in the assessment of the local dark matter density~\citep{banik16}, 
a parameter key to 
current efforts to detect dark matter directly~\citep{peter10}. 
We believe that further observational study and analysis, as we help realize in this paper, 
will be able to resolve the origin of the 
vertical asymmetries and ultimately that of their impact 
on the assessment of the local dark matter density. 
Further studies with {\it Gaia}~\citep{gaia16a,gaia16b} will also be key~\citep{banik16}. 

The SDSS~\citep{york2000} is a mature observational platform that gives us 
access to a very large photometric sample of stars, observed with the same telescope under
carefully calibrated conditions. The uniformity of the photometric calibration is ensured
through multiple observations of the same stars, rendering it nearly free of 
astrophysical assumptions~\citep{padmanabhan08}. 
The SDSS is the first survey to provide significant  
coverage in the south, allowing us to study and compare regions in 
both the north and the south. We study red, main-sequence stars, particularly K and M dwarfs,  
because main-sequence stars vary little in their intrinsic luminosity 
for a given color and metallicity, and the population of faint red stars is less likely 
to be infiltrated by non-main-sequence stars, such as giants.
Our largest analysis sample contains 3.6 million of such stars. By breaking the sky into regions, 
we study the vertical structure of the Galaxy and search for changes across the footprint. 
We use Galactic longitude and latitude with the Sun at the center 
such that the longitude is measured counterclockwise 
from the line toward the Galactic center 
and latitude is measured from the plane of the Galactic equator 
such that positive latitude is north, noting the standard transformation 
from equatorial to Galactic coordinates~\citep{binneytext}. 
As in \citet{widrow12} and \citet{yanny13}, 
we study the vertical symmetry of the Galactic disk 
by comparing the number of stars observed north and south. 
For continuity and simplicity we follow 
\citet{widrow12} and \citet{yanny13} 
in our modelling of 
the vertical structure of the disk. 
That is, we modify the solution of \citet{spitzer42} for 
the vertical structure of an axially symmetric thin disk 
to give a parameterization of the stellar density that has both a 
thin disk and a thick disk, as inferred from observations~\citep{gilmore83}. 
Apparently, the two disks have distinct 
metallicities and scale lengths~\citep{bensby14,bensby13}, though we
refer the reader to \citet{bovy13} for further discussion; 
we do not include these refinements, however, in our analysis. 
We probe the local structure of the stellar disk 
by determining 
the parameters of our model across the footprint, or, specifically, as we
change the chosen range of longitude for fixed latitude. 

In this paper we not only compare the structure, north and south, 
with longitude, across the footprint, 
but we also exploit the complete northern coverage of the SDSS to determine 
(i) the rate at which the mass in the stars changes with 
distance from the Galactic center, to compare with galactic models, 
and (ii) how changing the selected latitude interval 
for a fixed longitude interval impacts the shape of the vertical stellar number 
count distribution. If there were no change in vertical structure across the Galactic plane
and if the vertical 
metallicity 
distribution were uniform, then the shape of the vertical distribution in number counts
would not change when the latitude window is changed, once corrections are made for the
geometric acceptance. We do, however, observe definite changes in shape; consequently, we have 
a photometric proxy for changes in in-plane structure and/or metallicity. 
We believe that such studies can be refined and sharpened with further observations from 
the {\it Gaia} mission~\citep{gaia16a,gaia16b}, 
and not only through more photometric observations, because 
many more spectra will also become available. Spectral information 
makes it possible to measure the vertical gradient 
in metallicity~\citep{hayden14}, and further studies with better
resolution  across the Galactic plane should help reveal 
the specific origin of the shape differences.

We begin by discussing our data selection and consider the various 
systematic effects that 
must be understood to determine the vertical distribution 
of stellar counts. We then turn to a discussion of our fitting procedure 
before presenting the results of our 
north-south combined analysis and north-only analysis. 

\section{Data Selection and Photometric Distance Assessment}
\label{photod}
This study uses data from the SDSS DR9~\citep{ahn12}, 
the Ninth Data Release of the SDSS, which was
also used in \citet{widrow12} and \citet{yanny13}. As in our earlier papers, the
errors are predominantly systematic and derive from the use of 
a photometric ``parallax'' relation to relate the color of a
main-sequence star to its intrinsic luminosity and thus to determine its distance.  
Particular sources of error include those associated with 
stellar identification, as well as possible inadequacies in the application 
of corrections for reddening and absorption due to dust. The first source of error
includes not only the possible confusion of dwarfs with giants, or the appearance
of unresolved binaries, but also the possibility of admixtures of stellar populations of
varying age and metallicity. We refer to 
\citet{yanny13} for an extended discussion and for tests of our corrections for dust effects. 
We follow the procedures described in \citet{yanny13} 
and now summarize them briefly. 

To minimize dust effects, we analyze data only above $30^\circ$ in absolute latitude, and 
we correct the data for the reddening and absorption due to dust 
by using the maps of \citet{schlegel98}. 
In selecting the stars for our analysis, 
we require $15 < r_0 < 21.5$, where $r_0$ is the apparent brightness of the star, 
and $1.8 < (g-i)_{0} < 2.4$. Note that $g-i$ is the color defined as the ratio of the 
intensities of the $g$-band and the $i$-band, which are parts of a five-band survey 
where each band is associated with a range of frequencies, whereas 
$(g-i)_0$ is the color once the effect of reddening from dust has been removed.
For our selection of $|b| > 30^\circ$,  the color excess $E(B-V)$ is typically less
        than 0.03 mag.  This translates to a color correction in $g-r$ of 
        about 0.03 mag, and an error on that correction of
        much less than that amount.  Recall that a color error of 0.01 mag corresponds to
        a distance error of $< 5\%$ and that the typical SDSS photometric color errors are
        about 0.02 mag for fainter stars.

\begin{figure}[hp!]
\centering
\includegraphics[scale=0.65]{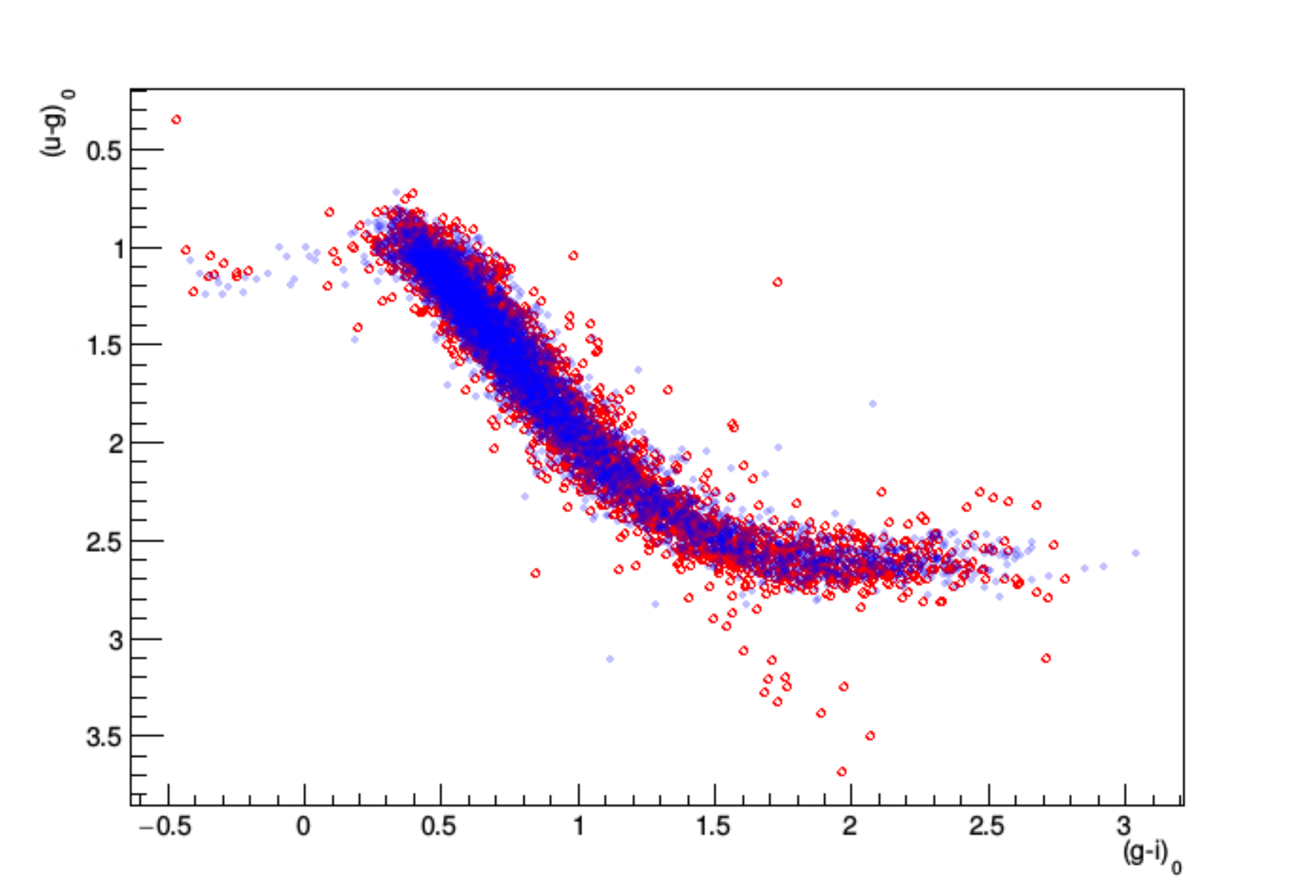}
\caption[Plot of $(u-g)_{0}$ vs. $(g-i)_{0}$ to discern non-main-sequence stars]{
Plot of $(u-g)_{0}$ versus $(g-i)_{0}$ to discern non-main-sequence stars. 
Stars in the north are denoted by small, filled, blue circles, whereas stars in the south 
are denoted by large, open, red circles. 
The plotted stars are all those in 
the region $150^\circ<l<170^\circ$ and $51^\circ<\lvert b\rvert<65^\circ$ 
with $u$-band errors less than 0.05 mag and with $14.9<r_{0}<15.4$ --- 
because this selection favors giants. The 
trail of stars within the color-color region bounded by 
$((u-g)_0, (g-i)_0)=(3,1.4)$  and $(4,2.2)$ as in \citet{yanny2009} and \citet{yanny13}
are the
identified giants; our selection admits the appearance of giants from the Sagittarius stream, 
which is in the south. 
For our analysis, we use $15<r_{0}<21.5$, which will contain an even smaller proportion of giants. 
}
\label{fig:giants}
\end{figure}

The majority of stars in our data selection are K- and M-type main-sequence stars, 
which are the reddest and coldest. They 
make up the majority of the main sequence, with M-type stars being the most prevalent. 
By using 
the reddest stars, we reduce the possibility of pollution 
from non-main-sequence stars.
However, it is still possible for giants to infiltrate the sample. 
Figure \ref{fig:giants} shows the results of a photometric test for the presence of giants. 
This plot analyzes stars with $14.9 < r_0 < 15.4$ in a region of the sky where
giants are expected to occur; we thus expect our test sample to 
have a much greater proportion 
of giants than our analysis sample. 
We have found 
the number of giants in our photometric test sample to be a very small fraction 
of the total number of stars and thus conclude that the giant admixture 
in our data set is trivially small.

%Figure 2

In order to determine the distance to each of our selected stars, we use the 
photometric parallax relations devised by \citet{juric08} for $(r-i)_{0}$ color 
and by \citet{ivezic08} for $(g-i)_{0}$ color, along with the refinements 
of \citet{yanny13} from fits to globular clusters 
for the red stars we consider here --- we refer the reader to \citet{yanny13}
for a detailed discussion and comparison with \citet{widrow12}.
The photometric parallax method determines the distance to a star from its apparent brightness and 
its intrinsic brightness --- the latter is inferred from its color and metallicity. 
The relation we have used in $(r-i)_{0}$ color for the absolute magnitude 
$M_{r}$ is that of \citet{juric08} but also includes a metallicity correction
$\Delta M_r$: 
\begin{eqnarray}
M_r &=& \Delta M_r ([\hbox{Fe/H}]) + 3.2 + 13.30(r-i)_0 \nonumber \\
&& -11.50(r-i)_0^2+5.40(r-i)_0^3-0.70(r-i)_0^4 \,, 
\label{pprmi0}
\end{eqnarray}
with 
%a metallicity correction 
$\Delta M_r ([\hbox{Fe/H}])= -1.11 [\hbox{Fe/H}] - 0.18 [\hbox{Fe/H}]^2$ \citep{ivezic08}.  
We have taken $[\hbox{Fe/H}]$ to be $-$0.3 universally 
 because we are unable to use the photometric metallicity assessment of \citet{ivezic08} 
due to the absence of sufficiently precise 
$u$-band information for the faint, red stars we consider in this paper. 
Moreover, for calibration purposes, only the stellar clusters 
M67 and NGC 2420 are sufficiently red and
out of the Galactic plane~\citep{yanny13}, and their metallicities are much larger
than expected for thick-disk stars. (In Section \ref{Nonlyscale} we report evidence for 
decreasing metallicity in the K/M dwarfs as we sample them well above the Galactic plane.)
A similar color-magnitude equation, adapted from \citet{ivezic08}, 
has been used in $(g-i)_{0}$ color: 
\begin{eqnarray}
M_r &=& \Delta M_r ([\hbox{Fe/H}])  -0.50 + 14.32(g-i)_0 \nonumber \\
&& -12.97(g-i)_0^2+6.127(g-i)_0^3-1.267(g-i)_0^4+0.0967(g-i)_0^5 \,,
\label{ppgmi0}
\end{eqnarray}
with the same metallicity correction. 
Using the intrinsic brightness $M_r$ and the apparent brightness $r_0$ in magnitudes, 
we calculate the distance $d[r_0,M_r]$ in kiloparsecs 
using the relation $r_0 - M_r = -5 + 5 \log_{10} d[r_0,M_r]$. 
We refine these distances using the analysis of \citet{yanny13}, so that,  
for $(r-i)_0$ color, the distance $d$ is finally determined to be 
\begin{equation}
d
 = d[r_0, M_r( (r-i)_0,[{\rm Fe/H}])]
+ 0.1415 (r-i)_0 + 0.0436 \,, 
\label{drmi0tune}
\end{equation}
whereas in $(g-i)_0$ color, it is 
\begin{equation}
d 
 = d[r_0, M_r( (g-i)_0,[{\rm Fe/H}])]
+ 0.08982 (g-i)_0 - 0.0726   \,. 
\label{dgmi0tune}
\end{equation}
For the faint, red stars we analyze in this paper 
the color-color diagram, as shown in Figure 14 of \citet{yanny13}, in $(r-i)_0$ versus 
$(g-i)_0$ 
%color 
shows
a tight correlation. We thus expect the two photometric parallax relations to work comparably well, and
this is borne out by both our findings and those of previous studies \citep{widrow12,yanny13}. 

  The color precision of the SDSS photometry ($\pm 0.02$ mag) 
leads to typically $\pm 0.2$ mag precision in $M_r$ 
and thus $\pm 10\%$ distance errors~\citep{juric08}.  
This assumes that the assigned metallicity is correct. 
If an incorrect metallicity were used, then the distance errors would be larger. 
Supposing, as usual, that the stars close to the plane (thin disk) have an [Fe/H] metallicity of 
$-0.3$, but the stars far from the plane (thick disk) have an [Fe/H] 
metallicity of $-0.8$, then one has an 
additional error of up to $+0.4$ mag (-20\% in distance) 
for a full mismatch of $\Delta {\rm [Fe/H]} =-0.5$~\citep{ivezic08}. 
We thus expect the reported distances to become 
        gradually too large 
%by about 20\% 
as the population shifts from mostly thin-disk stars at  $|z| < 0.5$ kpc 
to that of mostly thick-disk stars at 
$|z| > 1.5$ kpc because the distances have been overestimated
        by 10\%-20\% as a consequence of our  metallicity assumption. 

In this paper we consider a sample of stars several times larger than those analyzed previously, 
%\citep{widrow12,yanny13}, 
to the end of discerning and interpreting variations in the 
stellar number distributions across the footprint. As a result, in the current analysis, 
systematic errors associated with 
the possible inadequacy of the dust corrections for reddening and extinction across the sky, 
as well as with any nonuniformity in the photometric calibration itself,  
become potentially 
pertinent and need to be considered. We now address each of these effects in turn. 

We have used the dust maps of \citet{schlegel98} to assess the effects of dust; they 
are constructed from 
observations of dust emission in the far infrared after correcting for its temperature. 
Photometric~\citep{schlafly10} and 
spectroscopic~\citep{schlafly11} tests have shown these maps to be 
accurate for the higher-latitude ($|b| > 30^\circ$) 
sightlines we employ in our study, 
though this work has also revealed, working in $(g-r)_0$ color and averaging over large
angular scales, that the stars in the south are redder than those in the north. 
Quantitatively the two assessments are not the same
(note Table 5 in \citet{schlafly11}), with the photometric assessment using blue-tip stars
being larger, yielding a result of 
21.8 mmag redder in the south rather than $8.8\pm1.5$ mmag. 
% Table 4 Schlagel 10
We believe that the common reddening difference is better attributed to an inadequacy in the 
calibration of the $g_0$-band magnitude in the south, rather than to one 
in the dust reddening 
correction per se~\citep{betoule13,yanny13}, though metallicity variations in the photometric
sample can appear as well. Indeed \citet{schlafly11} suggests that the blue-tip population
could be fundamentally different in the north, 
due to age and metallicity differences.  
Subsequent work on different fronts supports this view. 

Photometric studies with Pan-STARRS1 of stellar reddening 
under assumptions of variation in 
stellar metallicity and population across the Galactic disk 
have recently been employed to construct an independent dust map, based on 
dust absorption~\citep{schlafly14,green15}; these maps agree closely with the older 
emission-based maps if compared out of the Galactic plane~\citep{schlafly14}. 
using Pan-STARRS1 data, and its claimed stability of $(20, 10, 10, 10, 20)$ mmag in 
$ugriz$ bands has been confirmed in the northern hemisphere by \citet{finkbeiner16}. 
Additionally, they note small differences in the median magnitudes 
of Pan-STARRS1 and SDSS data, north and south, of less than $\pm10\, {\rm mmag}$
in the rms for $|b|>20^\circ$ and $griz$ colors. 
Were these shifts uniform over smaller
angular scales, then they would be of little relevance. However, their maps of the 
differences between Pan-STARRS1 and SDSS data reveal that they can differ up to 
$\pm 40$ mmag over $15'$ pixels. Nevertheless, the shifts in $(r-i)$ color
are smaller than those in $(g-i)$ color, even in localized regions~\citep{finkbeiner16}. 
Thus we think that 
if our conclusions are insensitive to the choice 
of the photometric parallax scheme then they 
should also be robust with respect to a refinement of the 
photometric calibration. 
Nevertheless, in later sections we consider how the errors 
we discuss compare to the size of 
the effects we discover. 

\section{Functional Fits}
\begin{figure}[h!]
\begin{center}
\includegraphics[scale=0.5]{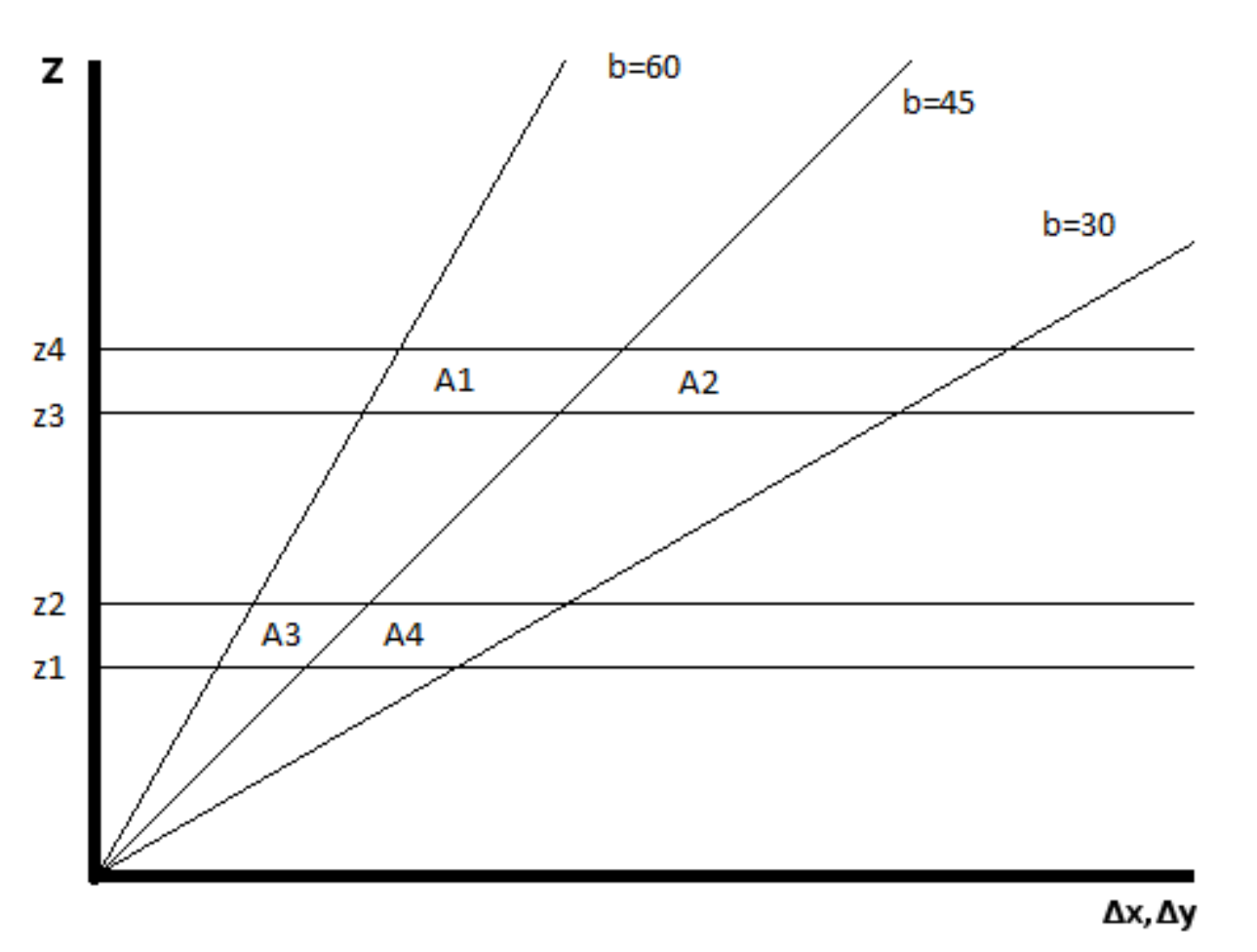}
\caption[Selection function]{
Plot of $z$ vs. $\Delta x$ or $\Delta y$ to show 
the different projected areas associated with various slices in latitude and in $z$. 
}
\label{fig:selectionfunction}
\end{center}
\end{figure}

A geometric selection function is used in  
order to correct for geometric effects 
and to calculate a stellar density. The selection function determines 
the effective volume for each bin of observed stars in height $z$ from the Galactic plane. 
Each bin is divided by the associated selection function to determine the number density 
at that $z$. 
% (based on the middle z of the bin). 
The selection function is 
\begin{equation}
{\cal V}(z) =
\frac{1}{2} 
\delta (l_2 - l_1) z^2 \left(\frac{1}{\sin^2 b_1} - 
\frac{1}{\sin^2 b_2} \right) \,,
\label{selection}
\end{equation} 
where $\delta$ is the width of the bin in kpc and $z$ is in 
the middle of the bin and is given in kpc. 
Note that $l_1 < l <l_2$, $b_1 < |b| <b_2$, and both quantities are in radians. 
The coordinates 
\begin{equation} 
x = -8 + d \cos b \cos l \quad;\quad 
y = d \cos b \sin l \quad;\quad 
z = d \sin b 
\end{equation}
are all measured from the Galactic center with the convention 
that the Sun is located at (-8, 0, 0) kpc in this coordinate system. 
We define distances from the Sun in the Galactic plane as $\Delta x$ and $\Delta y$. 
Figure~\ref{fig:selectionfunction} shows a plot of $z$ versus $\Delta x$ or $\Delta y$, 
demonstrating how different slices in latitude and $z$ yield a different projected area 
on the $z-x$ or $z-y$ plane and impact the volume ${\cal V}(z)$, 
Slices of different width in longitude 
also affect  ${\cal V}(z)$, 
but the specific longitudes themselves do not. 

%figure 3

The stellar densities $n(z)\equiv n_{\rm raw}(z)/{\cal V}(z)$ are then plotted 
and fit to the thin- and thick-disk model of \citet{widrow12}, 
a modification of the model of \citet{spitzer42} that adds a second disk:
\begin{equation}
n(z)=n_0 \left({\rm sech}^2\left(\frac{(z+z_{\odot})}{2H_1}\right) 
+ f {\rm sech}^2\left(\frac{(z+z_{\odot})}{2H_2}\right) 
\right) \,,
\label{fit}
\end{equation}
where $n_0$ is a normalization factor, 
$H_1$ is the scale height of the thin disk, $H_2$ is the scale height of the thick disk, 
and $f$ is the fraction of the stellar population that belongs to the 
thick disk. The parameter $z_{\odot}$ is the height of the Sun 
above the Galactic plane, though we emphasize that 
$z_\odot$ does not correspond to a point-to-point distance because the Galactic plane
is a locally determined quantity: the Galactic disk need not be perfectly flat. 
Our model assumes a two-disk structure 
that is north-south-symmetric with respect to the Galactic plane. 
Due to brightness saturation effects, 
we analyze only those stars that possess a vertical height 
$|z| > 0.35\,{\rm kpc}$ 
from the Galactic plane. Thus we believe that we are considering a mix of thin- and thick-disk 
stars
across the SDSS footprint. It is pertinent to compare our choice with other recent models 
of the structure
of the Galactic disk. The recent work of \citet{robin14} and \citet{bovy17}, for example, also 
supports the use of a ${\rm sech}^2$ vertical distribution for thick-disk stars 
in the Milky Way, though 
the thin-disk model of \citet{czekaj14} employs Einasto ellipsoids (generalized exponentials). 
In addition, the study of \citet{schwarzkopf2000} suggests, rather, that the 
vertical distribution of stars in galaxies with a merger history 
might be better described by a ${\rm sech}^1$ distribution. 
Since our particular purpose is to 
study the appearance of north-south (N/S) symmetry breaking across the 
Galactic plane, we believe that our
conclusions do not rely on the particular N/S symmetric parameterization of the vertical 
distributions that we use. If the stellar disk is flat over small variations in $x$ and $y$, 
%in $R$ and $\phi$, 
rather than merely flat on average, then one can expect $z_\odot$ to be universal for all lines
of sight. However, if the stellar disk has local warps or ripples, then the determined $z_\odot$ need not
be universal, and can serve as a proxy for such local variability. 
Thus variations in the determination of 
$z_{\odot}$ in the context of a fit to a distribution that is N/S symmetric 
probe how well one 
can meaningfully define the Galactic ``plane.'' 

To fit Eq.~(\ref{fit}) to the vertical distribution of star counts, 
we optimize $\chi^2$, namely, 
\begin{equation} 
\chi^2 = \sum_{i=1}^{N_{\rm bin}} \left( \frac{\left( N_i - n(z_i) \right)}{\sigma_i} \right)^2\,, 
\end{equation}
over $N_{\rm bin}$ bins, where $N_i$ stars are in bin $i$ centered on a height $z_i$ 
and $n(z_i)$ is the theoretical model of Eq.~(\ref{fit}). 
The weights $\sigma_i$ are given by $\sqrt{N_i}$ of $N_i$ stars. We 
use a combination of C and Python and  
the fitting routines of 
ROOT~\footnote{https://root.cern.ch}, a powerful data analysis framework.
At low $z$, the selection function becomes 
ineffective if the stars cannot be seen because they are too bright. 
A minimum $z$ cut is employed to remove this effect. 
This cut was determined for each data selection 
by fitting the model with trial cuts and minimizing the $\chi^2/{\rm dof}$ of the fit; 
we have determined that 
a minimum $z$ cut of $\lvert z \rvert>0.35\,{\rm kpc}$ can be used for all our fits. 

Thus far we have addressed the vertical structure of the disk exclusively. However, 
in certain stellar data sets, the variation in in-plane radial distance $R$, 
noting $R\equiv\sqrt{x^2 + y^2}$, 
can exceed 
1.5 kpc. To investigate the consequences of such variations, we also employ fits in 
which we bin our data set in $R$ as well and scale out the $R$ variation, noting
a similar procedure in \citet{banik16}, by modifying the selection 
function of Eq.~(\ref{selection}) to include the factor 
\begin{equation} 
\exp((R - R_0)/(2.7 {\rm kpc})) \,,
\label{selectionR}
\end{equation} 
where $R_0 = 8\,{\rm kpc}$ and $2.7\, {\rm kpc}$ is the central value of the radial scale
length (with an uncertainty of 4\%) 
found by \citet{bovy13}. That is, after binning the data in $R$ and $z$, where 
$R_i$ is the coordinate at the center of a bin in $R$, we construct 
\begin{equation}
n_{\rm eff}(z) = \sum_{i} n_{\rm raw}(z,R_i) \exp((R_i-R_0)/(2.7\,{\rm kpc}))\,,
\end{equation}
after summing over all over bins in $R$ for fixed $z$, and then compute 
$n(z)=n_{\rm eff}(z)/{\cal V}(z)$. As the result of this procedure we end up
with larger relative errors in $n(z)$, though we find no indications that such an
analysis is warranted. We find that using $n_{\rm eff}(z)$ in place of $n(z)$ 
does not affect our extracted parameters in a significant way and thus does 
not affect our conclusions. Thus in what follows, 
we do not employ a simultaneous analysis in $R$ and
$\phi$ outright, but rather bin our data in the vertical coordinate using the
selection function of Eq.~(\ref{selection}) to determine $n(z)$. 

%Figure 4
\begin{figure}[h!]
\begin{center}
\subfloat[]{\includegraphics[scale=0.33]{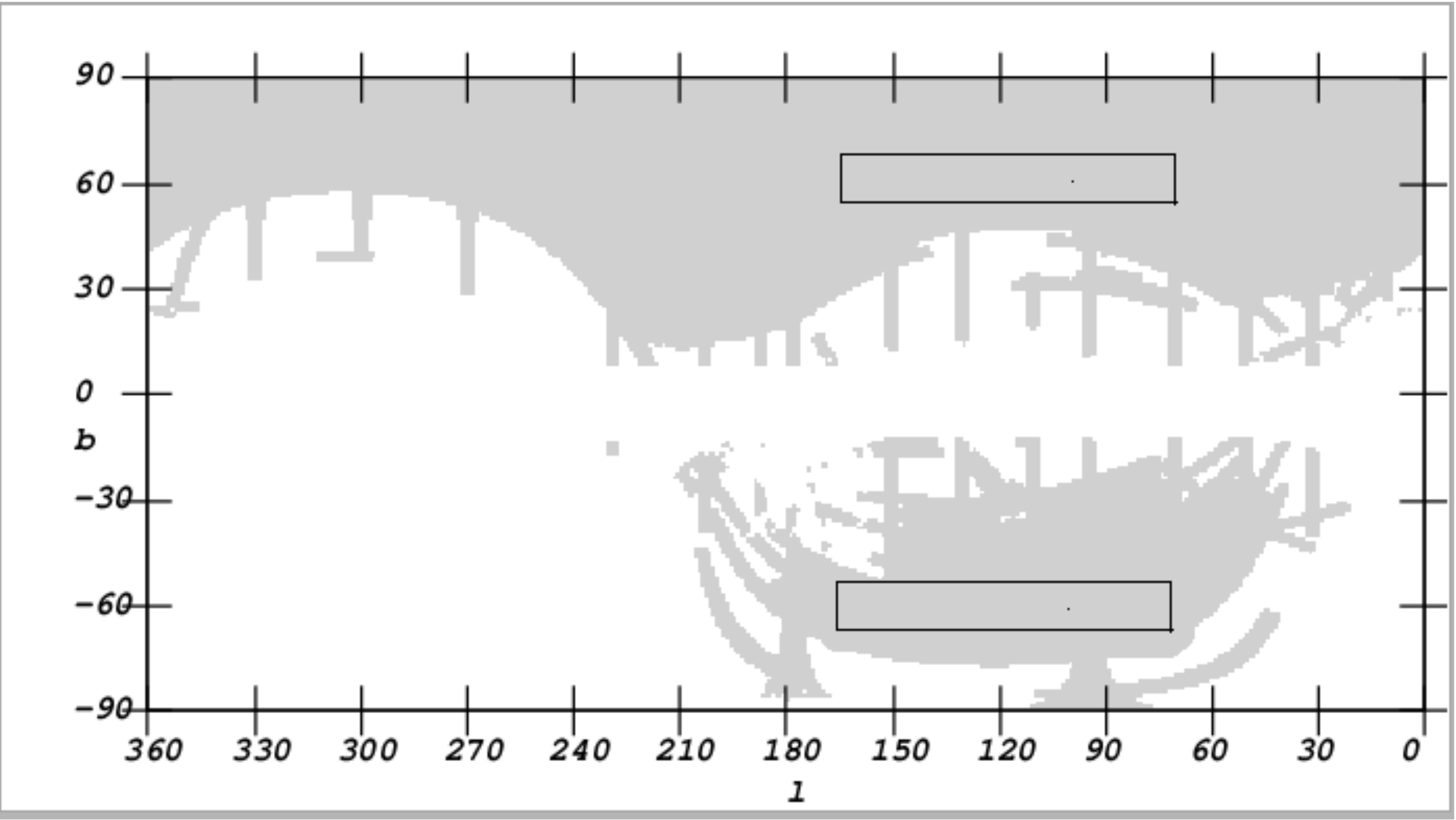}} 
\subfloat[]{\includegraphics[scale=0.33]{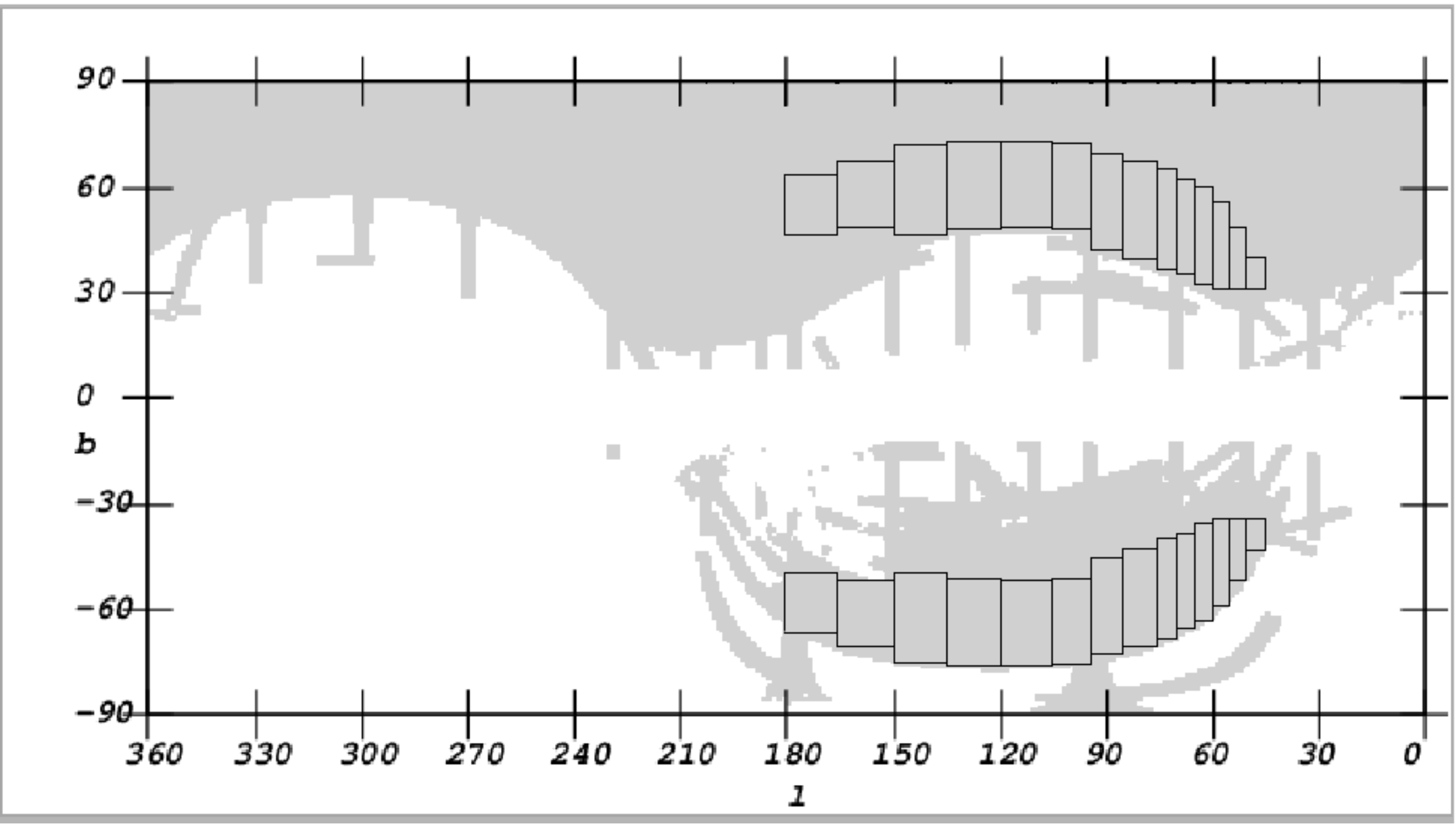}}
\caption[Sloan Digital Sky Survey footprint for the north and south matched analysis]{
The SDSS footprint~\citep{aihara11} 
as a map of $b$ vs. $l$. The blocked-in 
regions represent the selections analyzed in the matched north and south study. 
%Figure~\ref{fig:footprintmatched}a shows a 
(a) A
region with $70^\circ<l<165^\circ$ and $51^\circ<\lvert b\rvert <65^\circ$. 
%Figure~\ref{fig:footprintmatched}b shows 
(b) Selections over $45^\circ<l<180^\circ$ 
for which each wedge region in $l$ has a maximum range in $b$.
}
\label{fig:footprintmatched}
\end{center}
\end{figure}

\begin{figure}[h!]
\begin{center}
\includegraphics[scale=0.45]{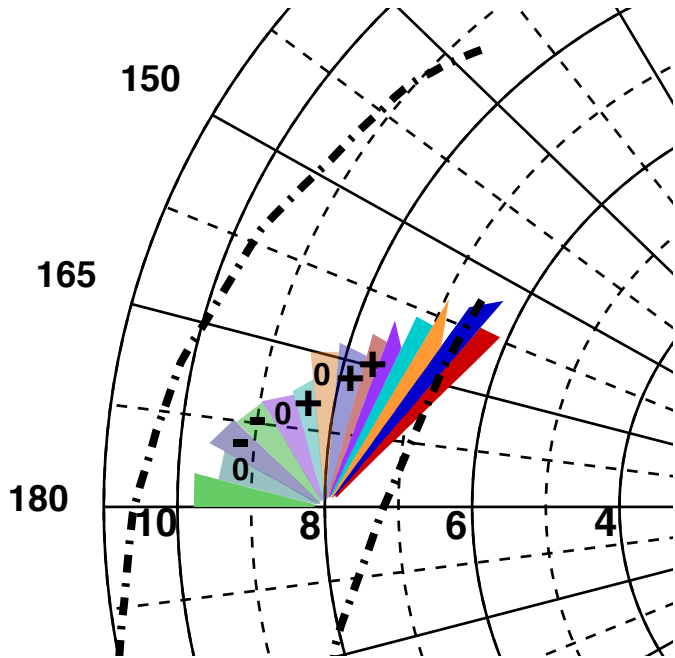} 
\caption[$R$ and $\phi$ distribution of SDSS footprint]{
The sample in Figure \ref{fig:footprintmatched}(b), broken into wedges in 
$l$ of roughly equal population, with those stars that have 
$1.8 < (g-i)_0 < 2.4$ and $|z|< 2\,{\rm kpc}$ 
plotted in terms of their Galactocentric coordinates $R$ and $\phi$. The indicated 
radial distances $R$ are in kpc, and ``180'' marks the azimuth 
$\phi=180^\circ$ that extends through the location of the Sun to the Galactic center. 
The different $l$ wedges, of which there are 14 in all, span different distances in the 
$(R,\phi)$ plane because each 
has a different latitude wedge. 
The spiral arm structures inferred from distances to embedded clusters
determined by \citet{camargo15} are also indicated as dashed-dotted lines, with the inner arm 
being that of Sagittarius-Carina and the outer arm being Perseus. 
The Orion spur, in which the Sun is located, is not apparent, however. 
The wedges correspond to those analyzed in Figure \ref{fig:histogramsmatchedset}. 
For subsequent reference, we have also superimposed 
the $z_\odot$ results of Figure \ref{fig:zsun}, with 
 ``$\pm$'' denoting a $z_{\odot}$ greater (less) than the global average at 1$\sigma$ and 
 ``$0$'' denoting no significant change.}
\label{fig:polarplot}
\end{center}
\end{figure}

\begin{figure}
\begin{center}
\includegraphics[scale=0.27]{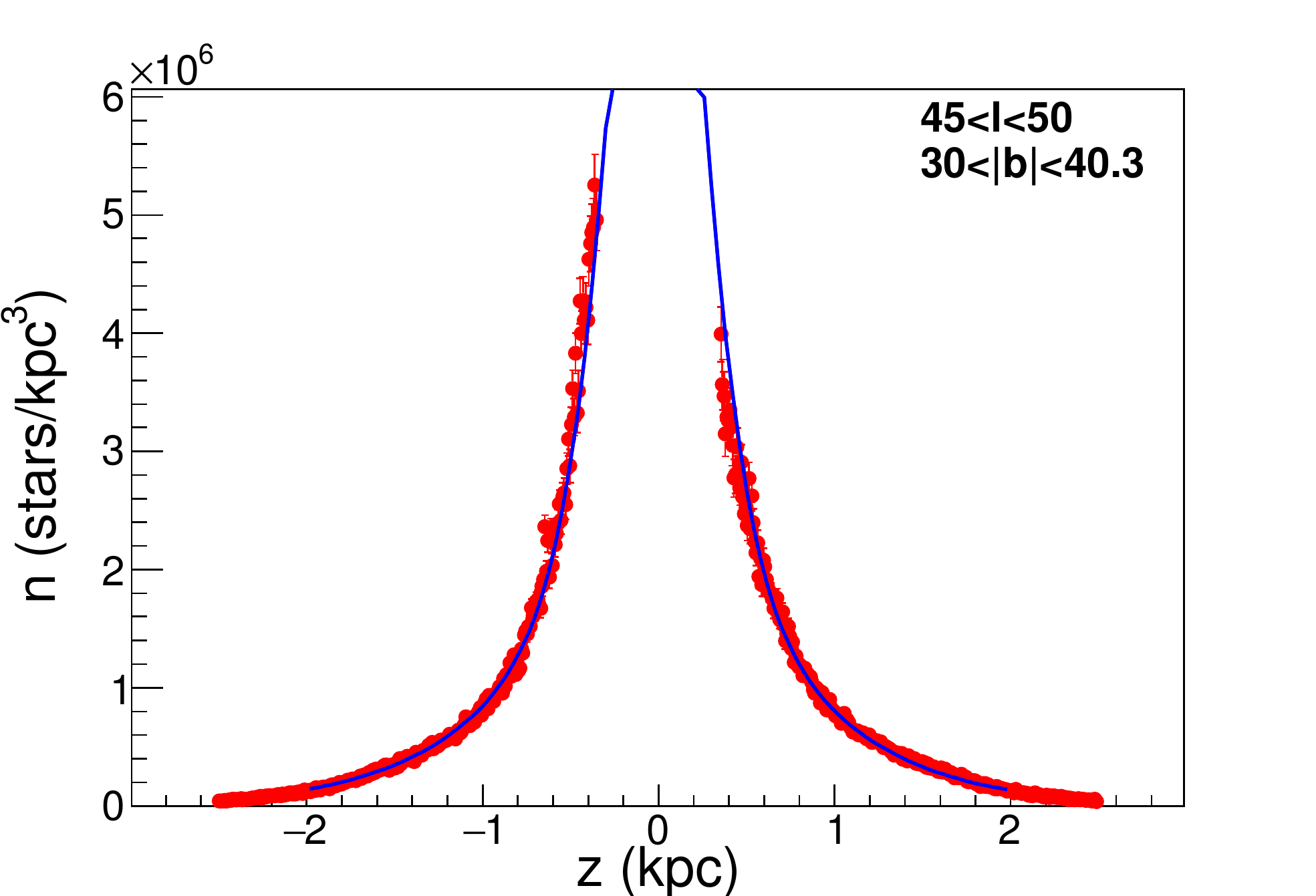}
\includegraphics[scale=0.27]{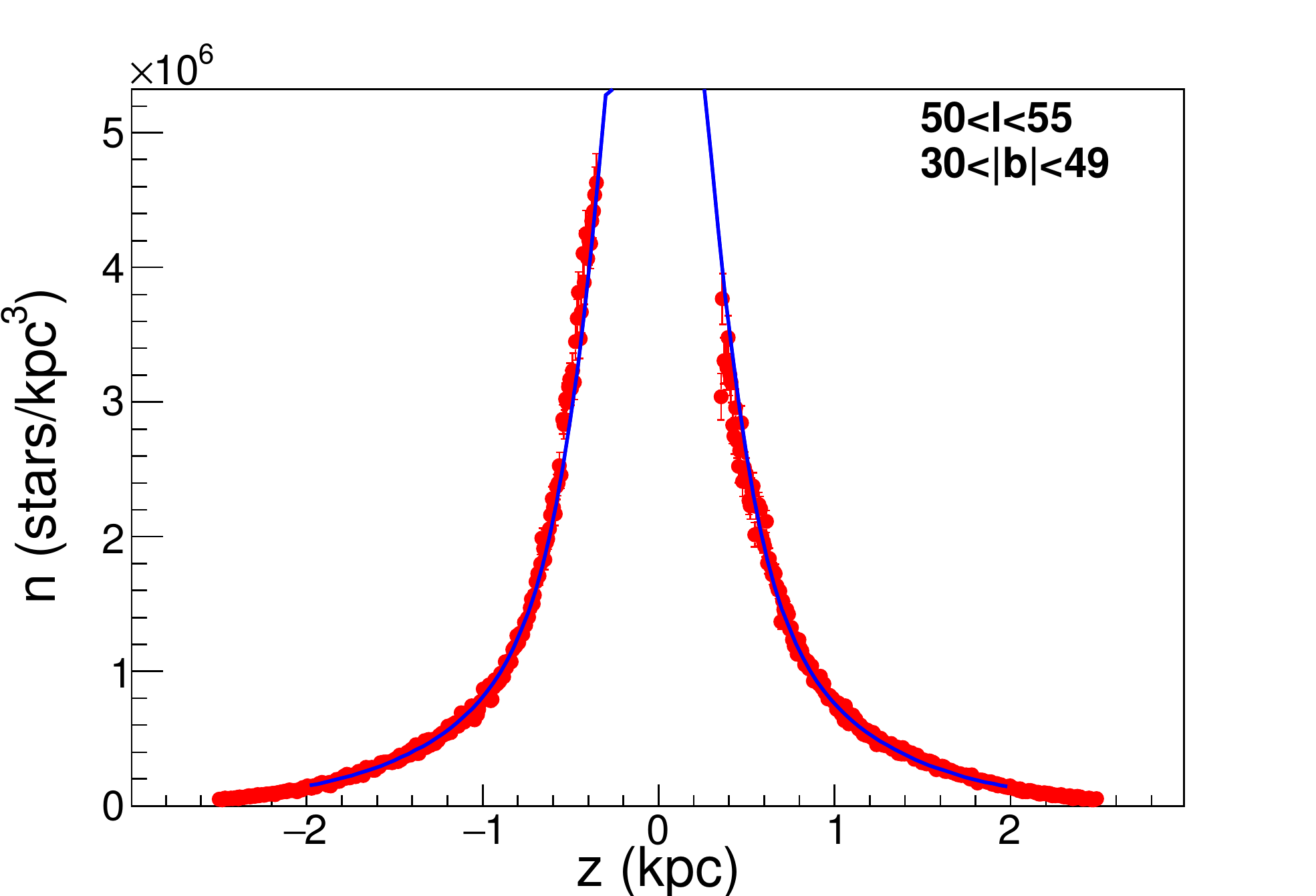} 
\includegraphics[scale=0.27]{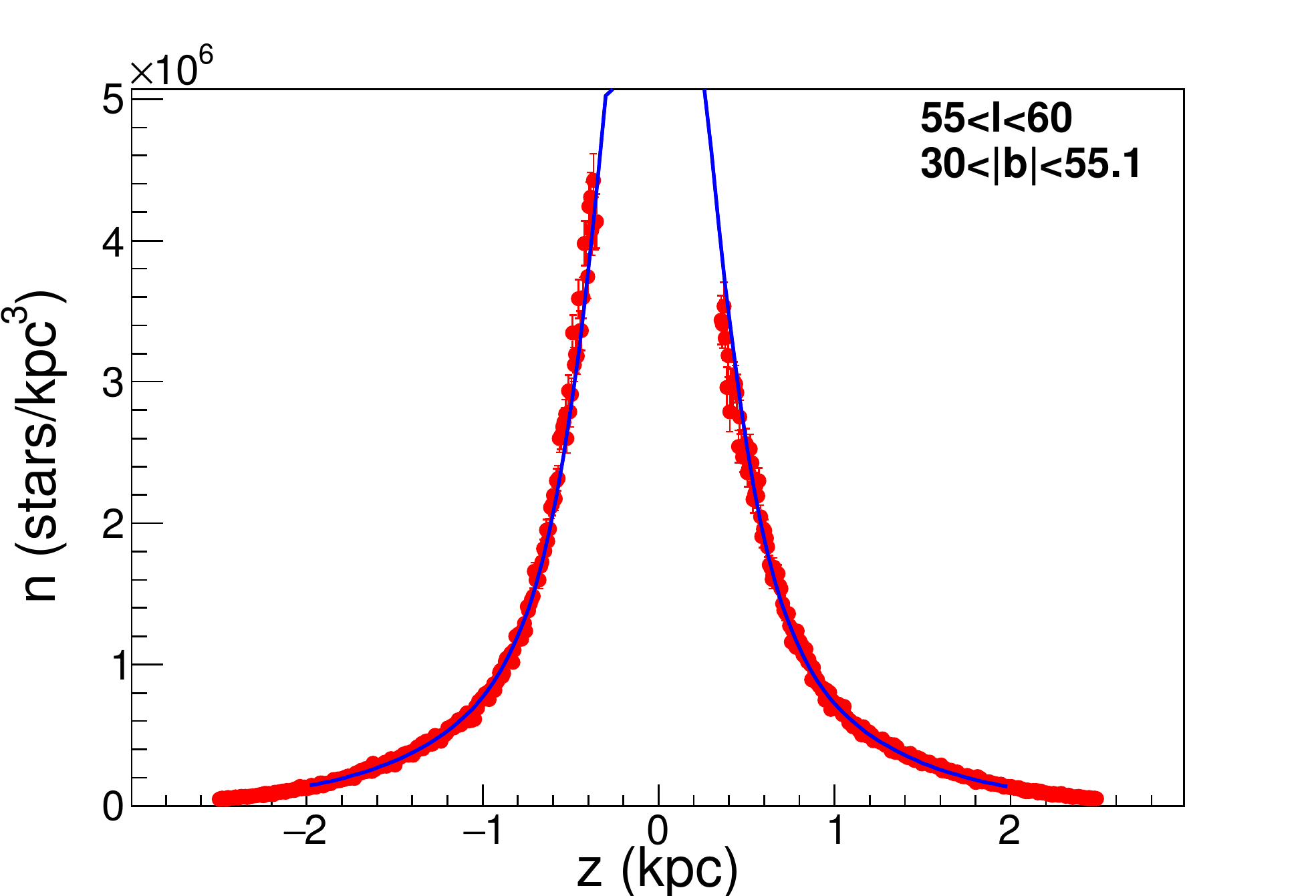}
\includegraphics[scale=0.27]{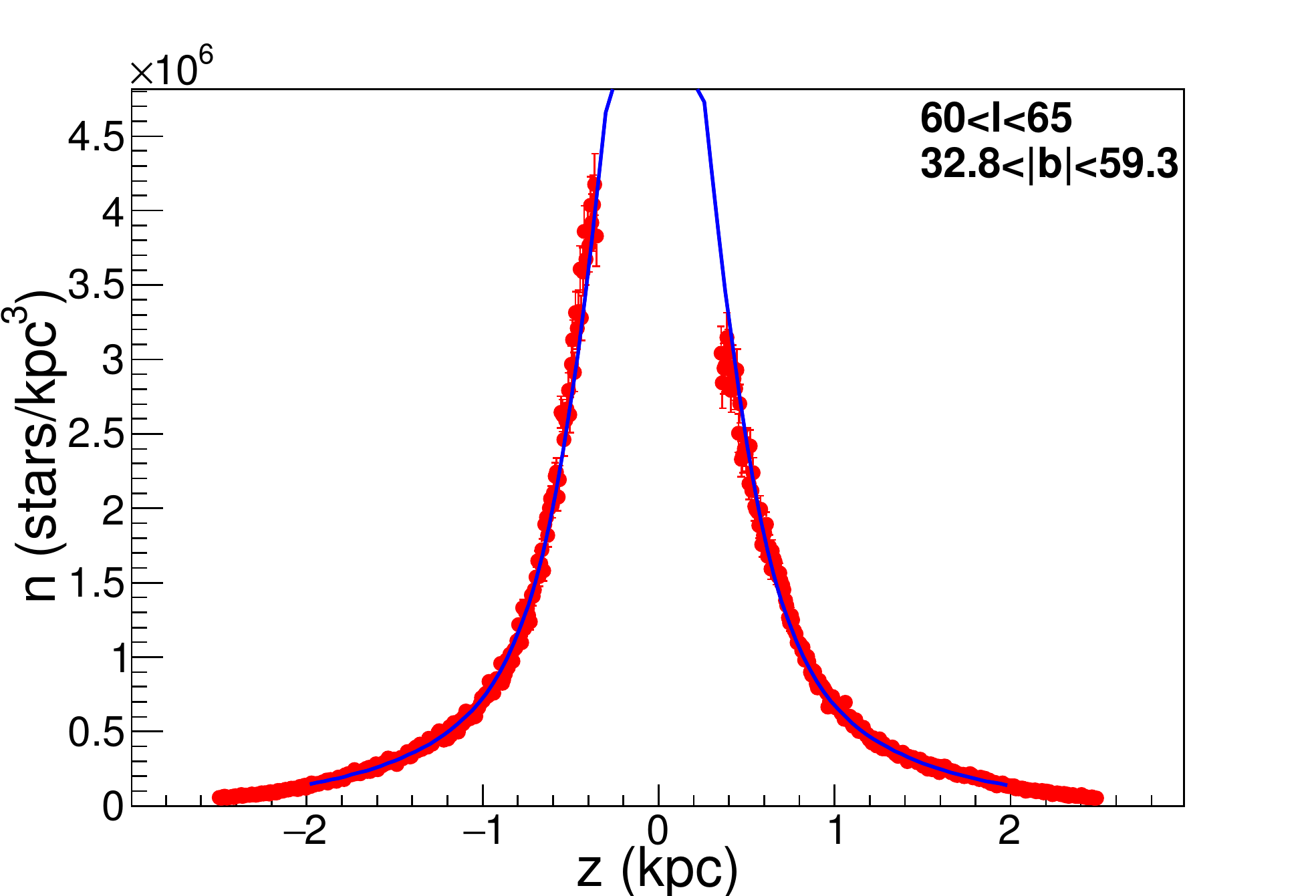} 
\includegraphics[scale=0.27]{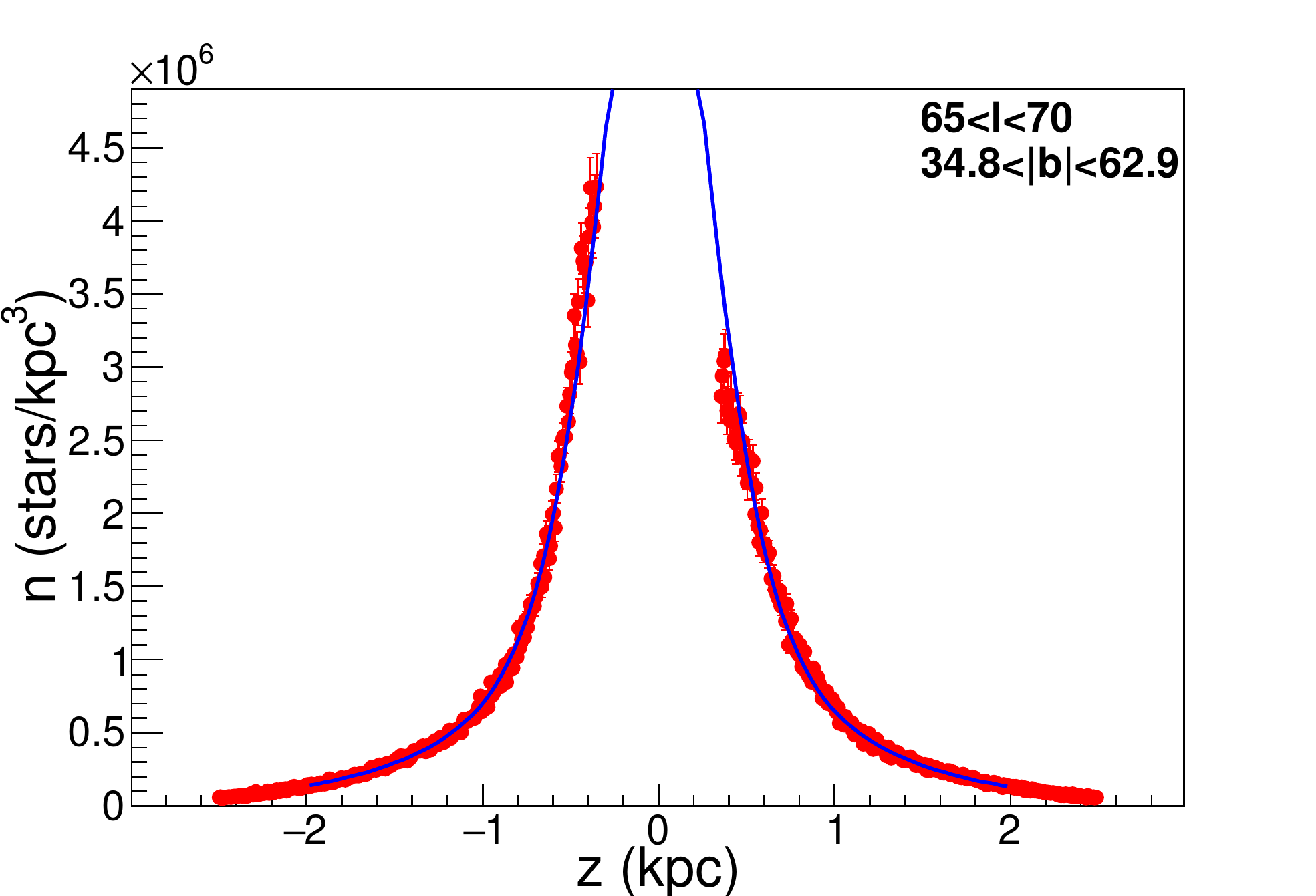}
\includegraphics[scale=0.27]{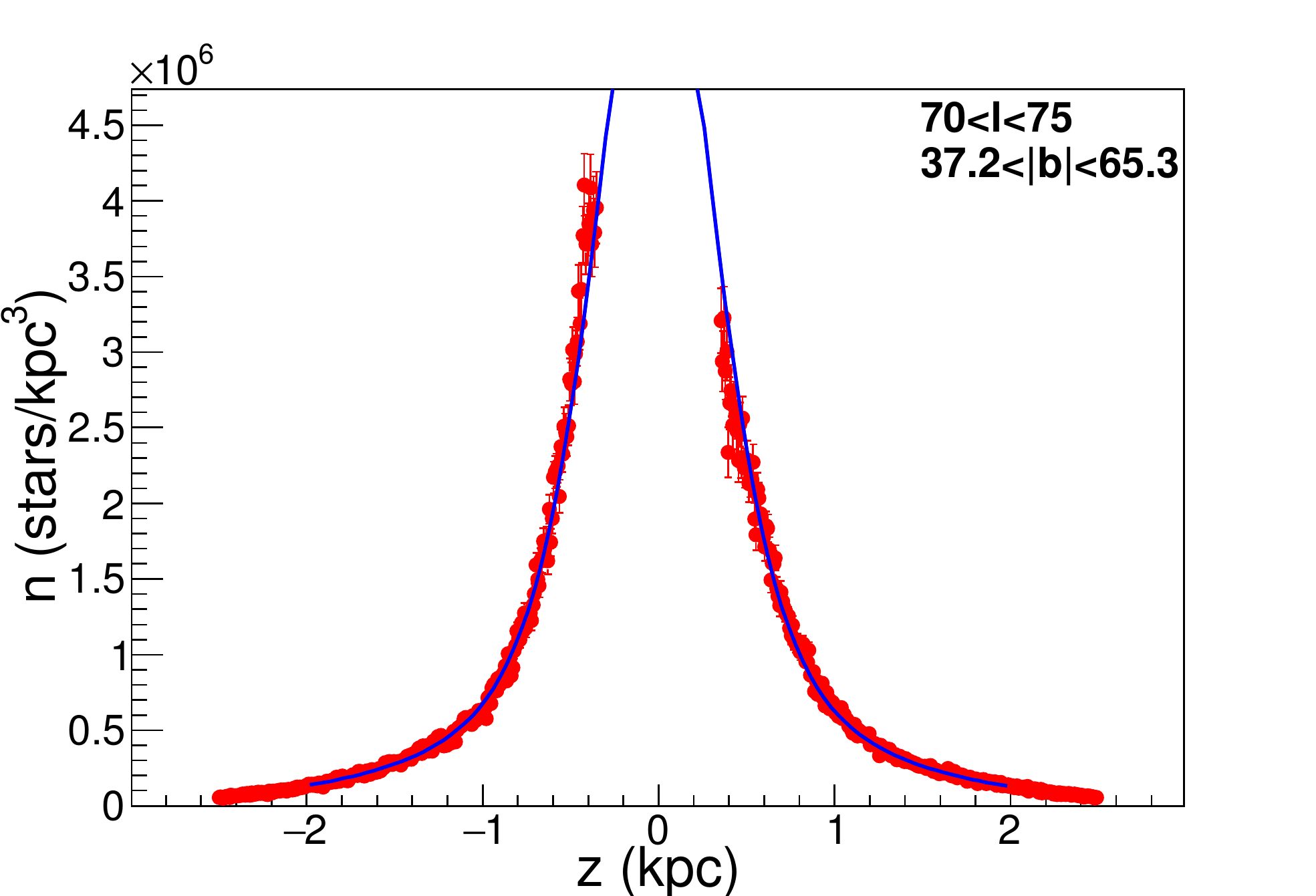} 
\includegraphics[scale=0.27]{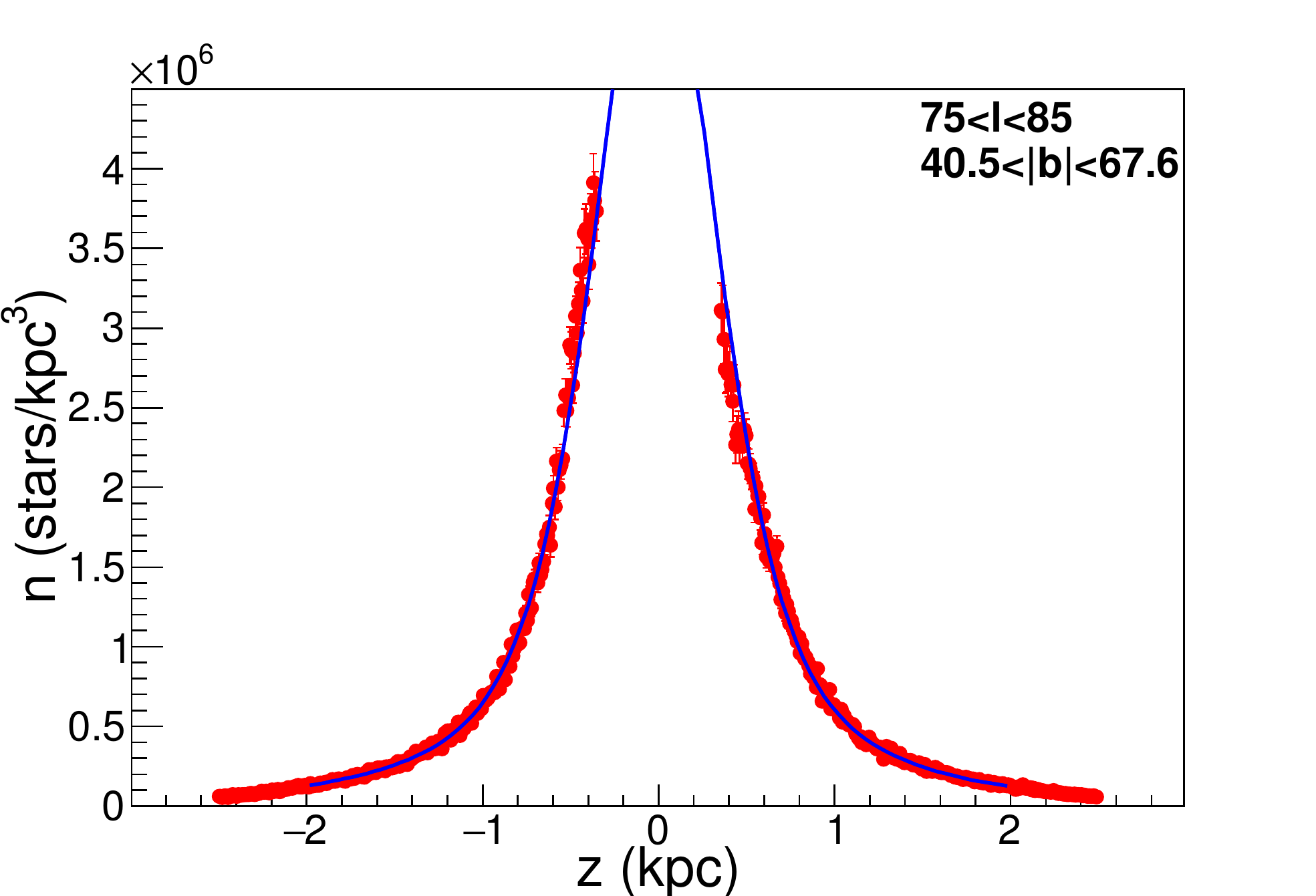}
\includegraphics[scale=0.27]{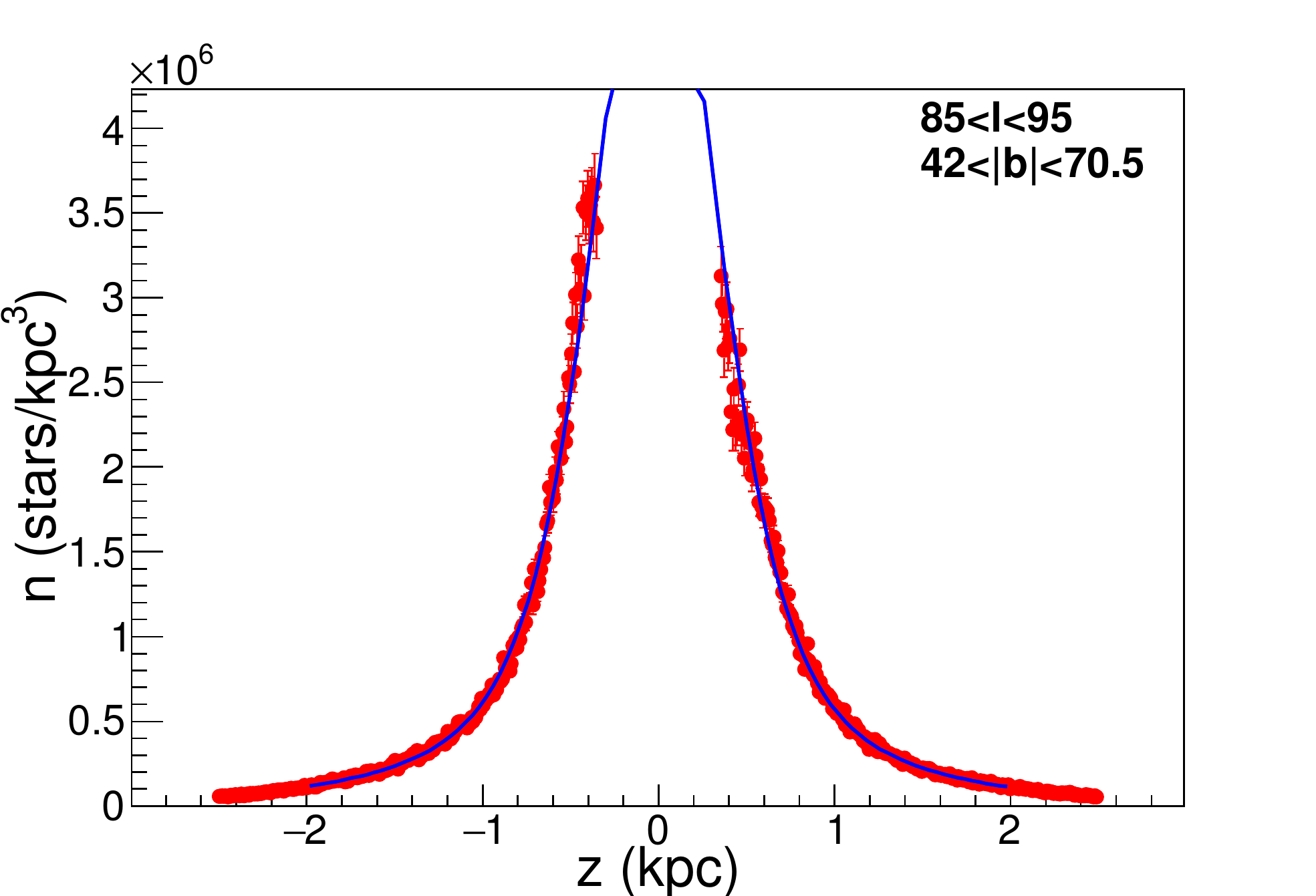} 
\includegraphics[scale=0.27]{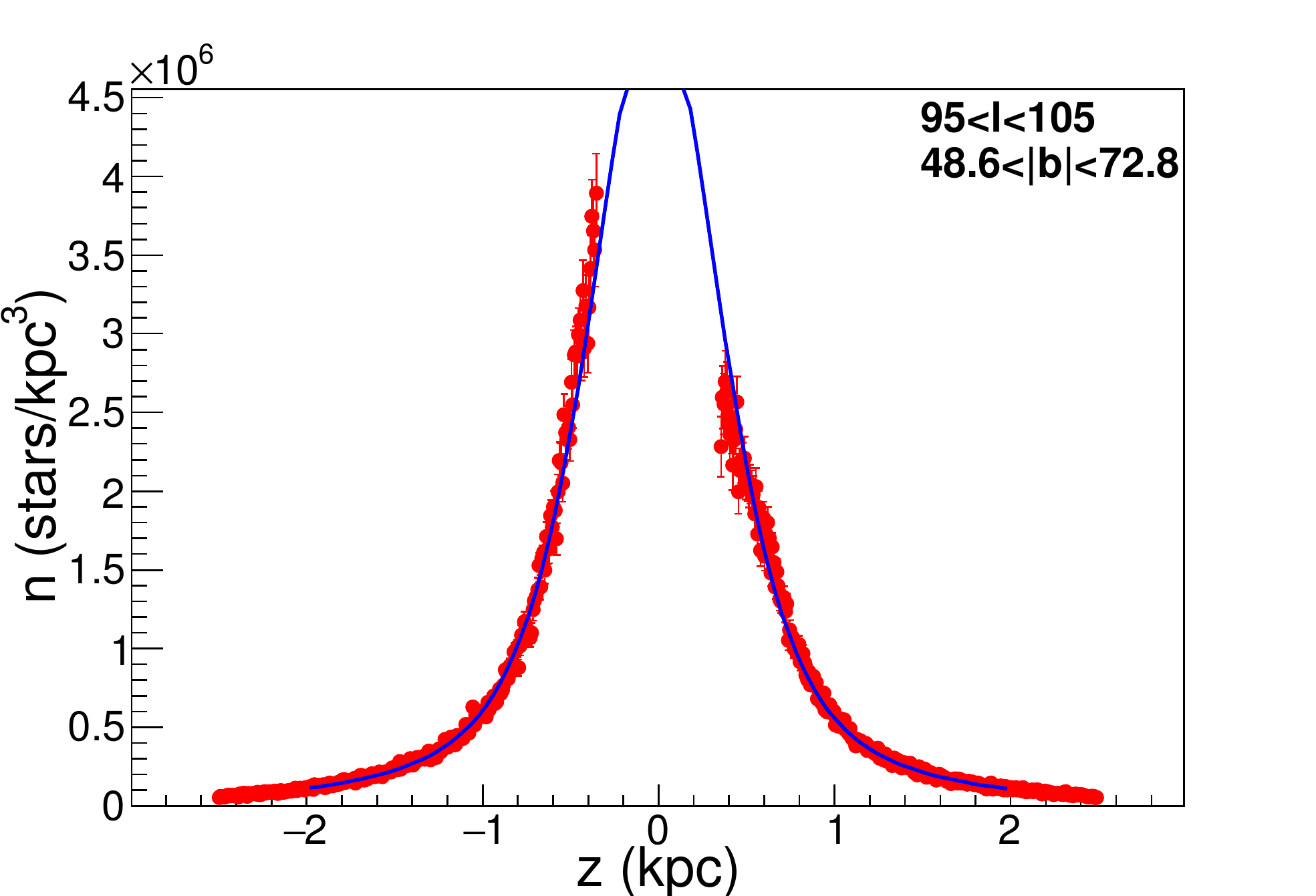}
\includegraphics[scale=0.27]{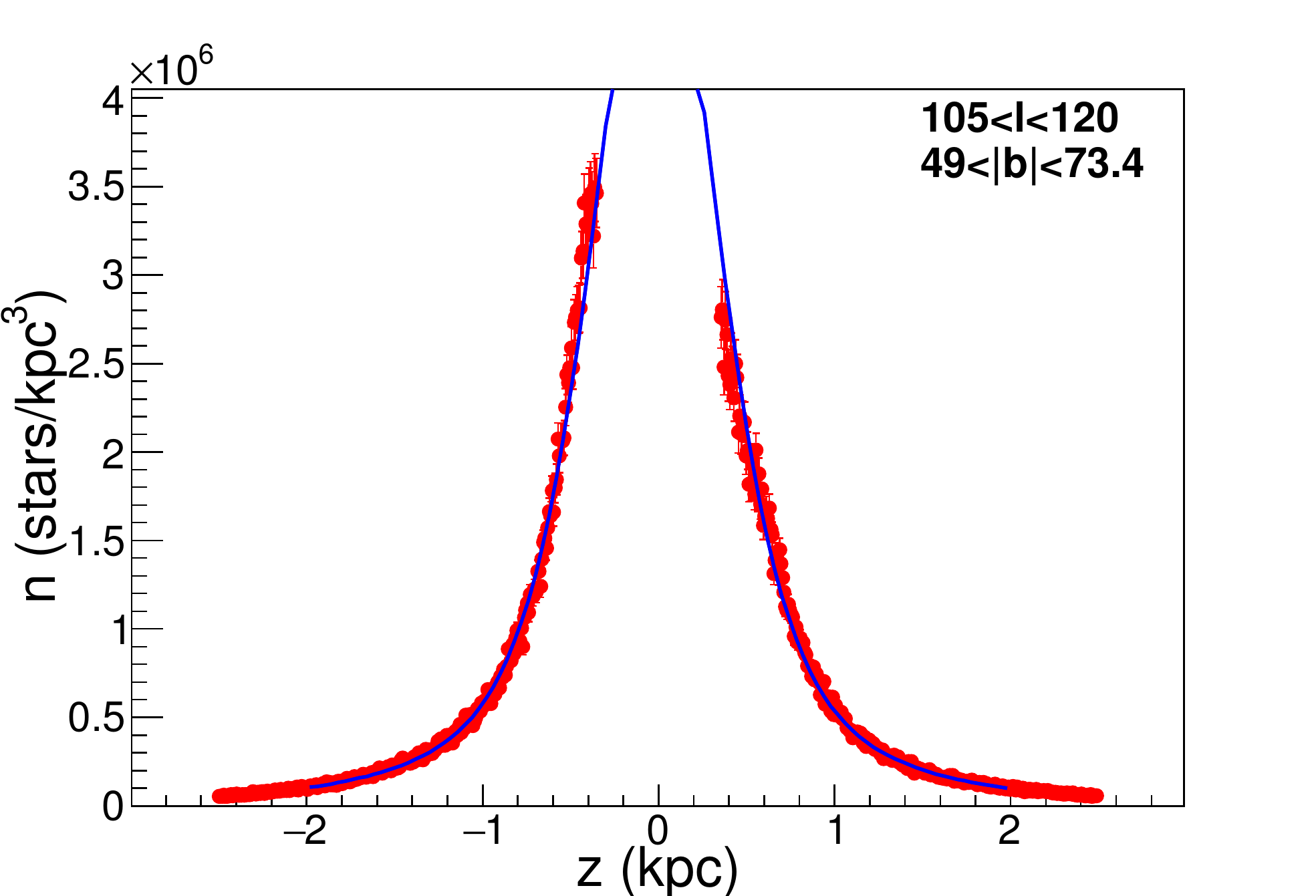} 
\includegraphics[scale=0.27]{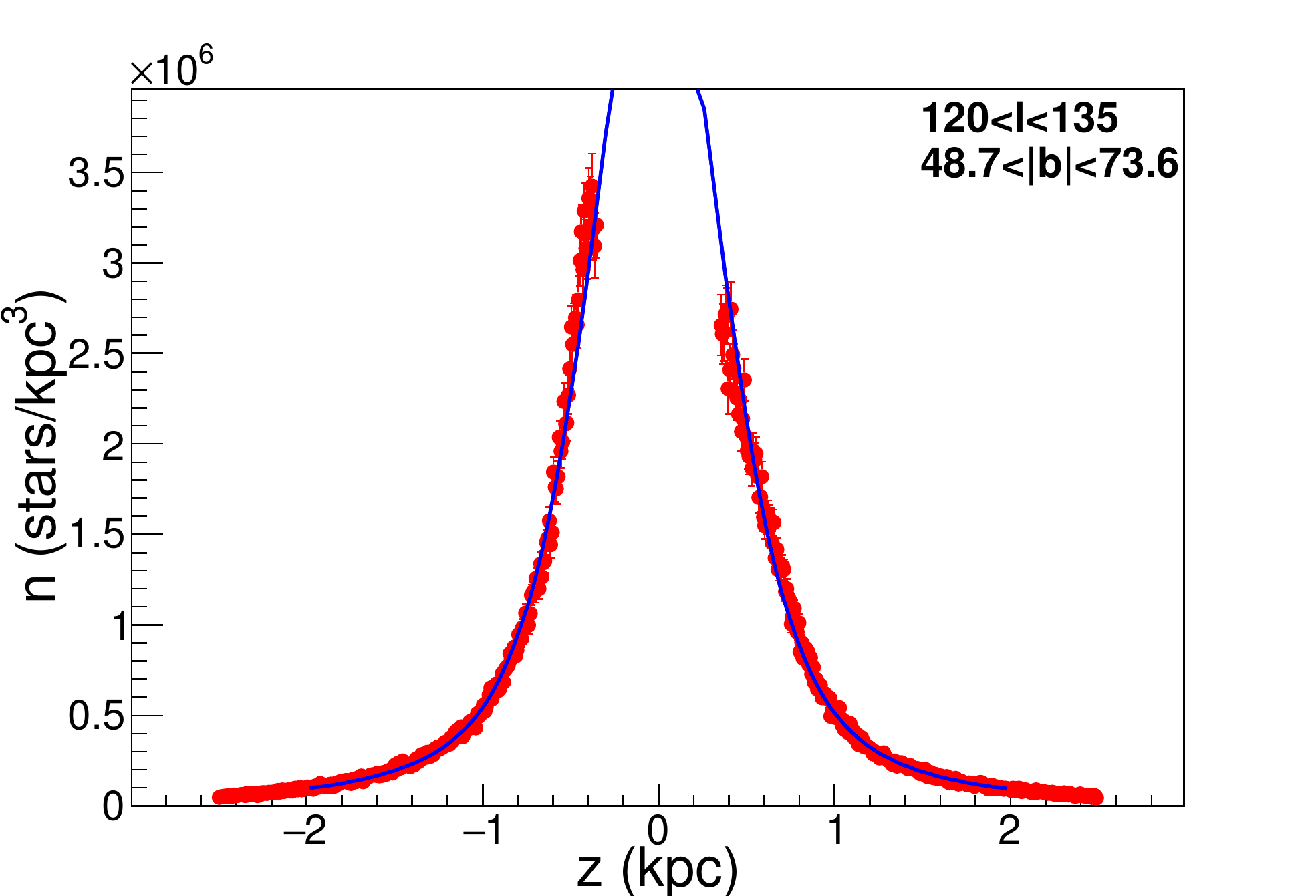}
\includegraphics[scale=0.27]{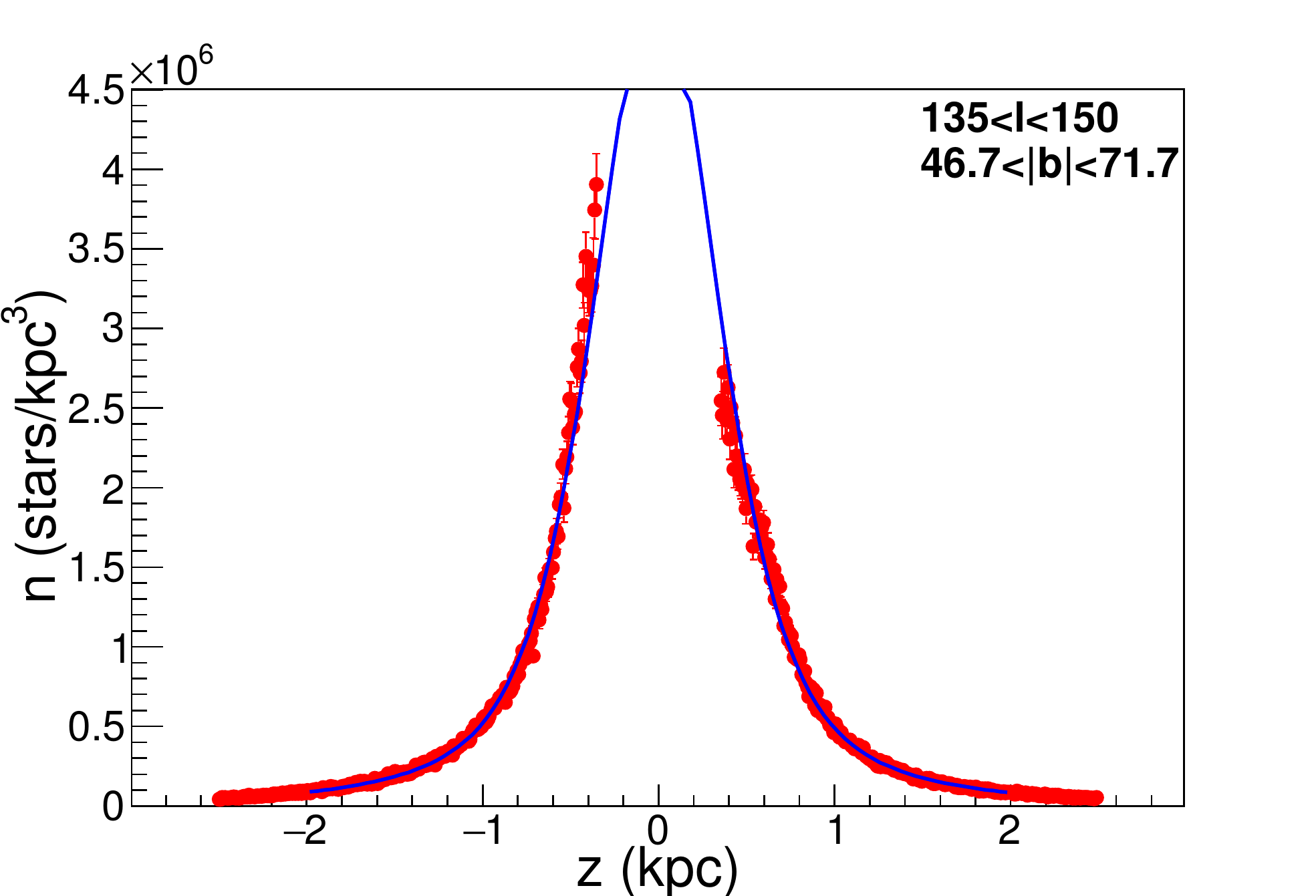} 
\includegraphics[scale=0.27]{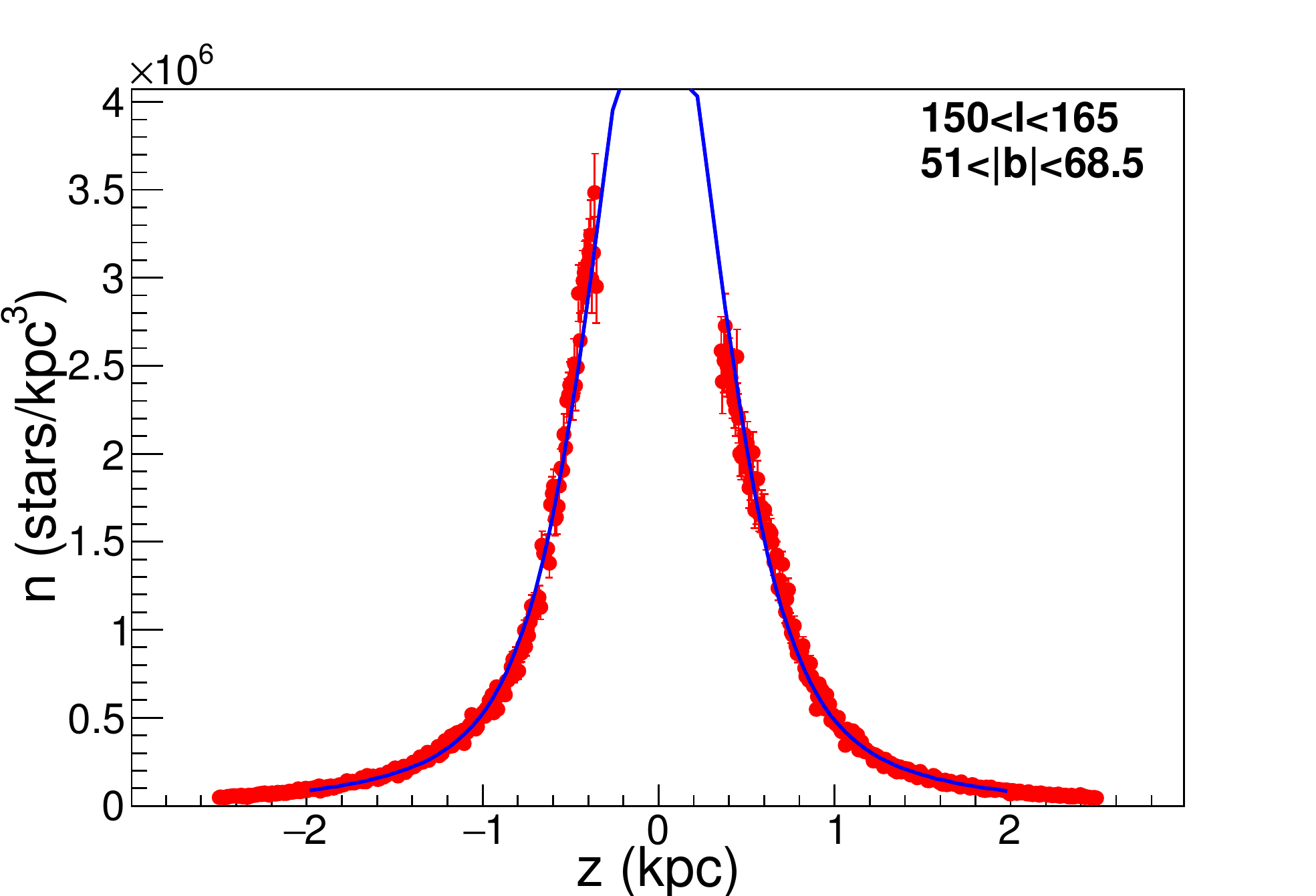}
\includegraphics[scale=0.27]{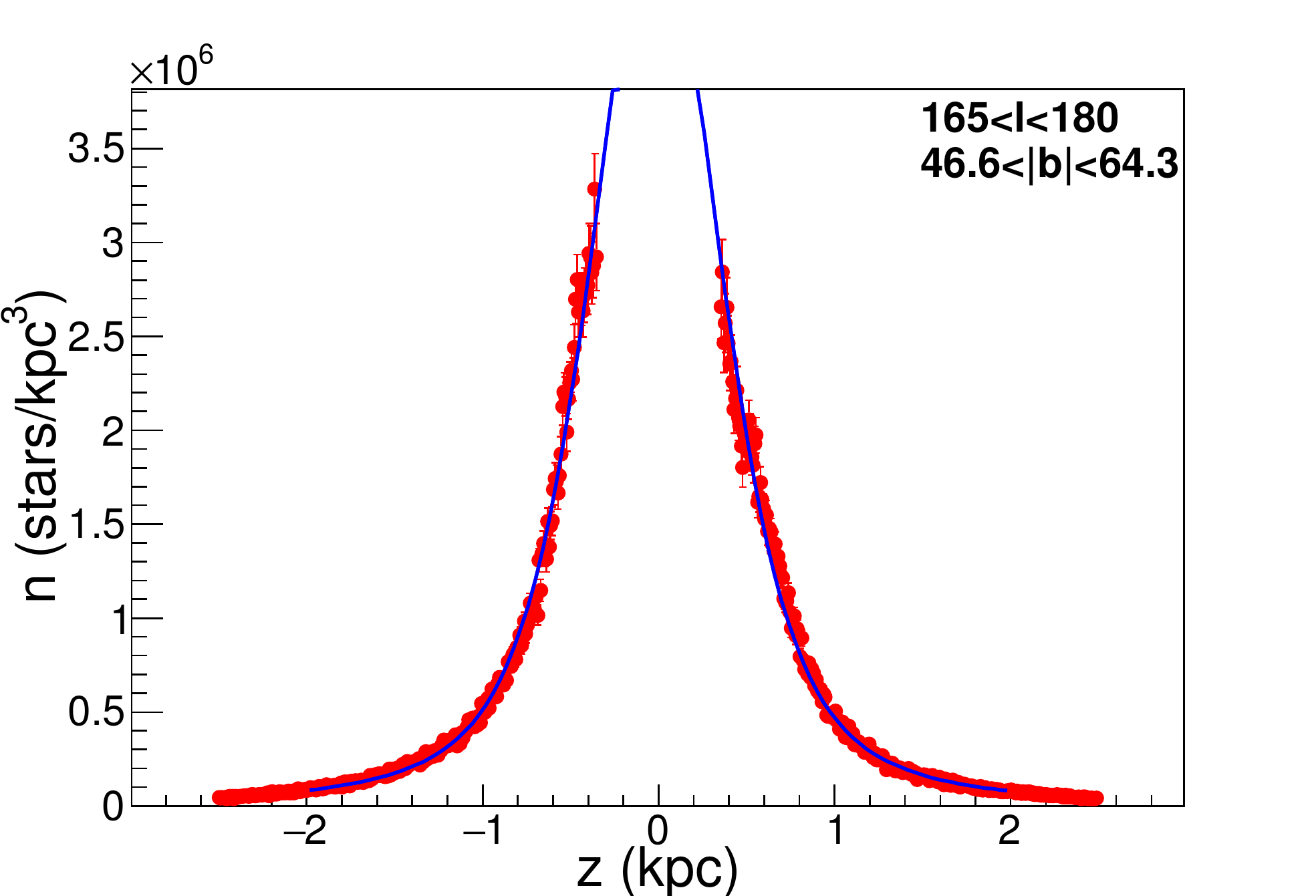} 
\caption[Stellar density as a function of vertical displacement z from the Sun]{
Stellar density as a function of vertical displacement $z$ from the Sun. 
These plots employ the expanded ranges in $|b|$ for the north and south matched set (Figure \ref{fig:footprintmatched}(b)). Each histogram represents a specific region in $l$ and $|b|$ (as shown 
in degrees in the upper-right corner of each image). 
}
\label{fig:histogramsmatchedset}
\end{center}
\end{figure}

%Figure 6
\begin{figure}[hp!]
\centering
\includegraphics[scale=0.50]{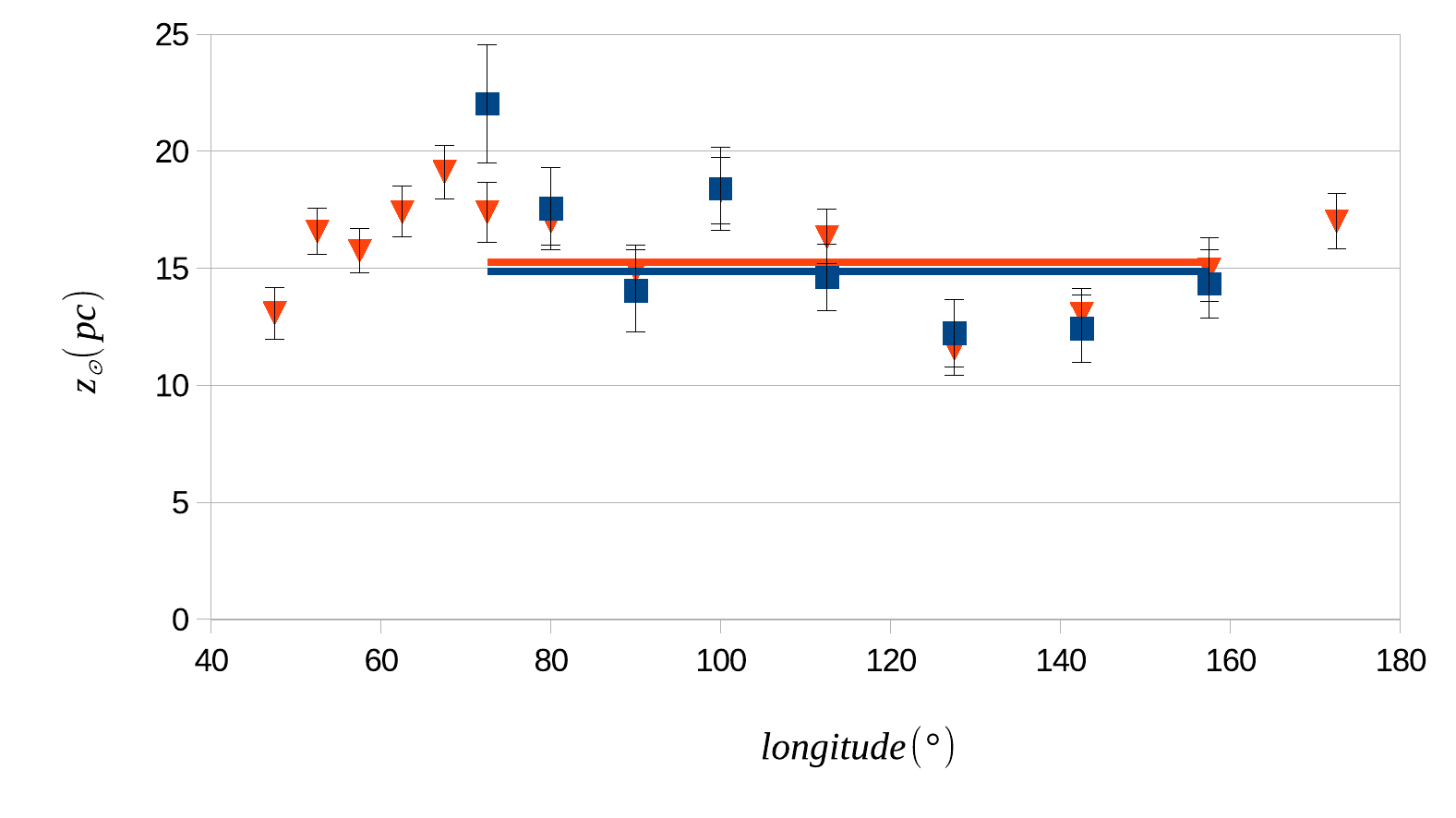}
\caption[$z_{\odot}$ as a function of longitude for north and south matched analysis]{
The Sun's height above the Galactic plane, $z_{\odot}$, in pc as a function of longitude 
obtained from the best-fit models of the stellar density histograms. 
The graph overlays the values obtained for the combined north and south analysis 
for each region in latitude: with the uniform selection 
as the blue squares (region shown in Figure \ref{fig:footprintmatched}(a)) and 
the expanded selection as the red triangles (region shown in Figure \ref{fig:footprintmatched}(b)). 
The blue and red lines, respectively, 
show the average $z_{\odot}$ 
obtained for the uniform and expanded latitudes within the region $70^\circ \le l \le 165^\circ$. 
}
\label{fig:zsun}
\end{figure}

%Figure 7
\begin{figure}[hp!]
\begin{center}
\subfloat[]{\includegraphics[scale=0.5]{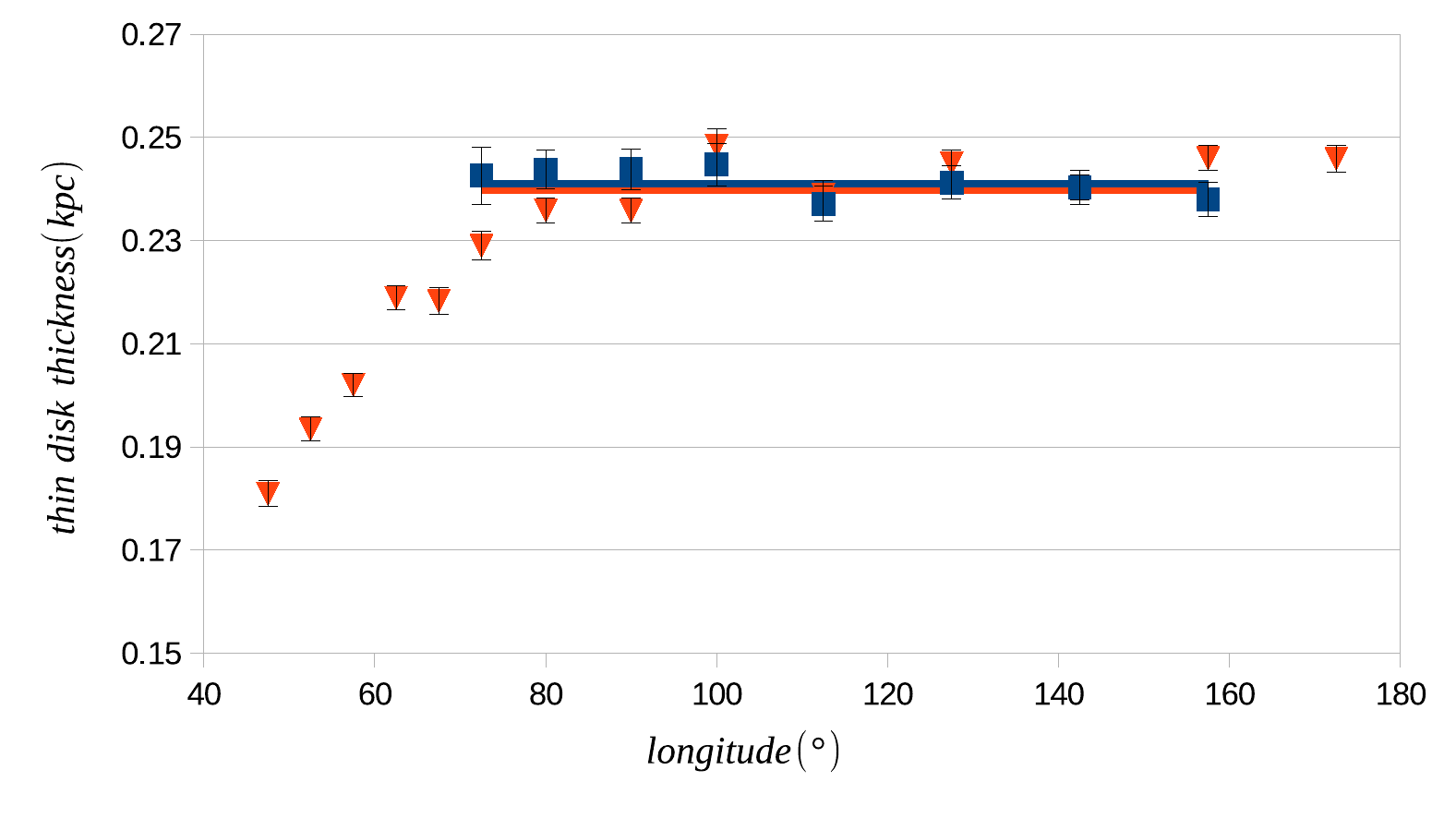}}
\subfloat[]{\includegraphics[scale=0.5]{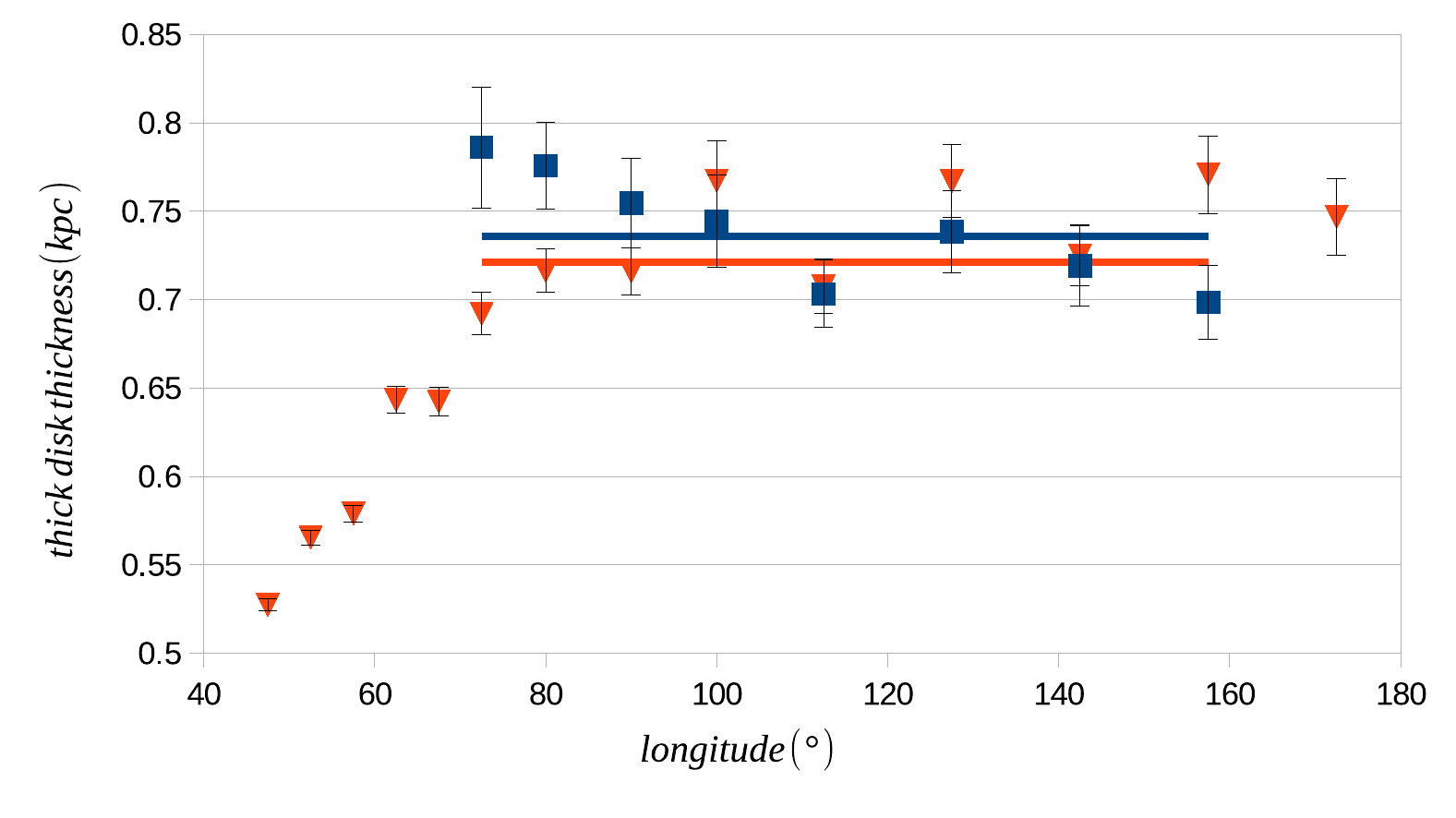}}
\caption[Thin and thick disk thicknesses as functions of longitude for north and south matched analysis]{
Thickness $H_1$ and $H_2$, respectively,  of (a) the thin dick and (b) the thick disk in kpc, 
%Thin (a) and thick (b) disk thicknesses in kpc, 
as functions of longitude as 
obtained from the best-fit models of 
the stellar density histograms for the uniform latitude selections (shown in 
Figure \ref{fig:footprintmatched}(a)) and the expanded latitude selections 
(shown in Figure \ref{fig:footprintmatched}(b)). 
The notation and conventions are identical to those in Figure \ref{fig:zsun}; 
the blue and red lines show the average thicknesses. 
The thicknesses are correlated with the thick-disk 
fraction $f$, so that $f$ tends to be smaller when the thicknesses are larger. 
}
\label{fig:matchedthinthick}
\end{center}
\end{figure}

\subsection{North and South Combined Analysis}
\label{NScomb}
We begin by studying 
regions selected such that the north and south have completely matched coverage. 
The first selection has been designed for uniformity 
by selecting stars with Galactic coordinates in $70^\circ < l < 165^\circ$ 
and $51^\circ < |b| <65^\circ$, which are then analyzed 
in longitude increments of $5^\circ$, $10^\circ$, or $15^\circ$ so that each selection has
roughly the same number of stars (Figure \ref{fig:footprintmatched}(a)). 
This uniform latitude cut creates a sample that can be directly compared as a function of 
longitude without the added complication of changing latitude. 
The second selection has been made to maximize the number of stars included in the analysis, 
by dividing the sample into regions 
that cover $5^\circ$, $10^\circ$, or $15^\circ$ of longitude, and choosing 
the maximum range in latitude possible in each wedge 
while maintaining complete coverage in the north and south (Figure \ref{fig:footprintmatched}(b)). 
The $R$ and $\phi$ distributions of the $(l,b)$ wedges in Figure \ref{fig:footprintmatched}(b), 
for the particular range of colors $1.8 < (g-i)_0 < 2.4$ that we analyze, are shown in Figure \ref{fig:polarplot}. 

Using the previously discussed cuts and selections, 
we apply the selection function in Eq.(\ref{selection}) and employ 
the fitting function of Eq.~(\ref{fit}) with the photometric parallax method 
in $(g-i)_0$ color as described in Section \ref{photod} to determine the distance
to each star. We require $0.35\,{\rm kpc} \le z \le 2.0\,{\rm kpc}$, where the 
maximum value of $z$ is determined by color completeness~\citep{yanny13}. 
We first analyzed 
the region shown in Figure \ref{fig:footprintmatched}(a)
to determine what variations occur when
changing the selected longitude range and then repeated our analysis with the 
expanded latitude sample of Figure \ref{fig:footprintmatched}(b). There are 1.05 and 3.59 
million stars in total in the selections of Figures~\ref{fig:footprintmatched}(a) 
and ~\ref{fig:footprintmatched}(b), respectively. 
The changes in Galactic parameters for the two sets 
are sufficiently small that 
we have shown the stellar density profiles for the expanded latitude selections in 
Figure \ref{fig:footprintmatched}(b). 
Figure~\ref{fig:histogramsmatchedset} 
shows the results, displaying the stellar number density as a function of vertical distance. 
We find
the quality of fit ($\chi^2/{\rm dof}$) to range from 1.4 to 2 for the fits shown. 
Although 
the vertical stellar distribution is grossly smooth and symmetric, the 
plots do show a definite north-south asymmetry of ${\cal O}(10\%)$ in the stellar 
number counts, with the effect
becoming more marked at larger $l$, as noticed in \citet{yanny13}, though the 
expanded latitude data set is four times larger. 
The wave-like nature of the departures from the north-south-symmetric fit argues against
either metallicity variations or 
a failure of the north-south photometric calibration in explaining this result; indeed 
the possible sizes of such effects are too small to have an impact. 
We discuss the residuals, 
north and south, in Section \ref{cfNS}. We note that the quality of fit is poorest 
in regions where the observed wave-like nature of the N/S asymmetry is most prominent. 
In the remainder of this section we consider the 
changes in the Galactic parameters across the regions described in 
Figures~\ref{fig:footprintmatched}(a) and (b), to provide a context
for our analysis of the N/S variations.

Figures~\ref{fig:zsun}, \ref{fig:matchedthinthick}(a), and \ref{fig:matchedthinthick}(b) 
show the fit parameters $z_{\odot}$, thin-disk thickness, and thick-disk thickness, 
respectively, as functions of longitude. 
Our analysis shows only modest variation in $z_{\odot}$ across 
the footprint, both for the selections with a 
uniform latitude-cut (which we henceforth refer to as ``uniform'') 
and for the expanded latitude selections (which we henceforth refer to as ``expanded''). 
The average $z_\odot$ values obtained for the uniform and expanded analyses, 
respectively, are $z_{\odot} = 14.9 \pm 0.5$ pc and $z_{\odot} = 15.3 \pm 0.4$ pc. 
The average for the uniform analysis is consistent 
with the value of $z_{\odot} = 14.3 \pm 0.6$ pc found by \citet{yanny13}. 
Fitting the expanded results with a straight line 
reveals a $\chi ^2/{\rm dof}$ of 3.53, 
so that there appears to be a genuine variation with longitude. However, there is little
difference in the average if the uniform or expanded latitude selection is 
employed.    Our values for $z_\odot$ are about 2$\sigma$ lower than the
recent literature using similar stars, noting $z_\odot \sim 25$ pc~\citep{juric08} and 
        $z_\odot = 27.5 \pm 6 \rm pc$~\citep{chen99,chen01}.  
Other determinations of $z_\odot$, such as the value of 
$26 \pm 3$ pc found in a study of Cepheid variables~\citep{majaess09}, or that 
of        \citet{joshi07}, using younger population tracers, 
with values ranging
        from 6 to 20 pc, 
differ
from these results and support the existence of the 
 ``environmental'' sensitivity we have found (see also \citet{bovy17}). 
Figure~\ref{fig:matchedthinthick} shows the analogous results 
for the disk thicknesses. 
The differences between the averages for uniform and expanded thickness are 
small, particularly for the thin disk, which is most pertinent to the north-south
asymmetry seen in Figure \ref{fig:histogramsmatchedset}, as will become clear in 
Section \ref{cfNS}. Recall, too, that the 
errors in photometric distances are at least $\pm 10\%$. 
The evidence for variations with $l$ is also weaker than in the case of $z_\odot$,  
though a change in the vertical scale height farther from the 
Galactic center --- and hence at larger $l$ --- is expected~\citep{kent91,narayan02}. 
Sensitivity to such variations
is evidently limited by our statistics, 
so that we will return to this point in our 
north-only analysis, for which we have a vastly larger number 
of stars. 

Although other analyses have revealed a range in the value of $z_\odot$, determined
from stars of differing spectral class in differing regions of the sky, 
our analysis is the first to reveal variations in its value %$z_\odot$ 
across the Galactic plane with stars of the same spectral class. 
Barring the existence of significant 
in-plane metallicity gradients, which would be at odds with 
the results of existing observational studies~\citep{hayden14}, 
we believe the variations we have found speak to a Galactic disk that possesses ripples
and thus is not globally flat. 
We have confirmed that this feature, and indeed all the 
features we have found thus far, also appear in our analyses 
using photometric parallax with $(r-i)_0$ color. 
We note, moreover, that a positive warp in the north, 
toward $l\sim 90^\circ$, has also been observed~\citep{russeil03},
though this does not explain the variations in $z_\odot$ that we have observed. 
The ripples could potentially arise from tidal effects on the disk, which appear 
in numerical simulations of disk-satellite 
encounters~\citep{gomez13,widrow14,gomez15b,gomez16,laporte16}. 
However, it is also important to compare our results with known in-plane features that 
vary across the SDSS footprint, and thus we consider 
the Milky Way's spiral arm structure. 

The dust obscuring the Milky Way's disk has limited our ability to study its spiral arms. 
Questions persist as to their nature and origin~\citep{baba09,sellwood11,sellwood14}, 
as well as to their precise structure~\citep{georgelin76,levine06,hou09,lepine11,francis12,camargo13,camargo15,bobylev14,griv14,pettitt14,vallee14}. 
The absence of bright, massive $^{13}$CO clouds in the interarm 
space~\citep{roman-duval09} promotes the association of these apparently short-lived 
features with the spiral arms themselves. As a result, young, red embedded clusters 
of stars should 
act as tracers of the spiral arms~\citep{camargo13,camargo15a,camargo15b,camargo15}. 
\citet{camargo15a} 
have challenged the notion of spiral arms as in-plane structures 
with north-south-symmetric vertical 
extent~\citep{cox02,monari16}, 
revealing embedded clusters with locations extending both above and 
below the plane. These authors also note that cluster formation
can be associated with the halo as well~\citep{camargo15b,camargo16}. 
Nevertheless, it is possible that the variations in $z_{\odot}$ are 
related to an out-of-plane structure of the spiral arms. 
Figure~\ref{fig:polarplot} 
shows the spiral arm structure detailed in \citet{camargo15} overlaid on the 
analysis sample of Figure \ref{fig:footprintmatched}(b) plotted in Galactocentric coordinates $R,\phi$, and 
the analyzed wedges in longitude 
have been marked 
with a $+$ in regions where the value of $z_{\odot}$ is greater than the global average 
and with a $-$ in regions where it is less than the global average. 
While the work of \citet{camargo15a,camargo15b,camargo15} 
shows evidence for some parts of the
        Milky Way's stellar spiral arms being 
located  a few hundred parsecs above or
        below the Galactic plane, 
we do not see any 
correlation between these out-of-plane
        stellar spiral arms and variations in $z_\odot$ with $l$ or 
any correlation with 
regions of the Galactic plane where significant north-south 
        asymmetries are apparent at $|z| > 0.35$ kpc.

\subsection{North and South Comparative Analysis} 
\label{cfNS}

\begin{figure}[hp!]
\begin{center}
\includegraphics[scale=0.267]{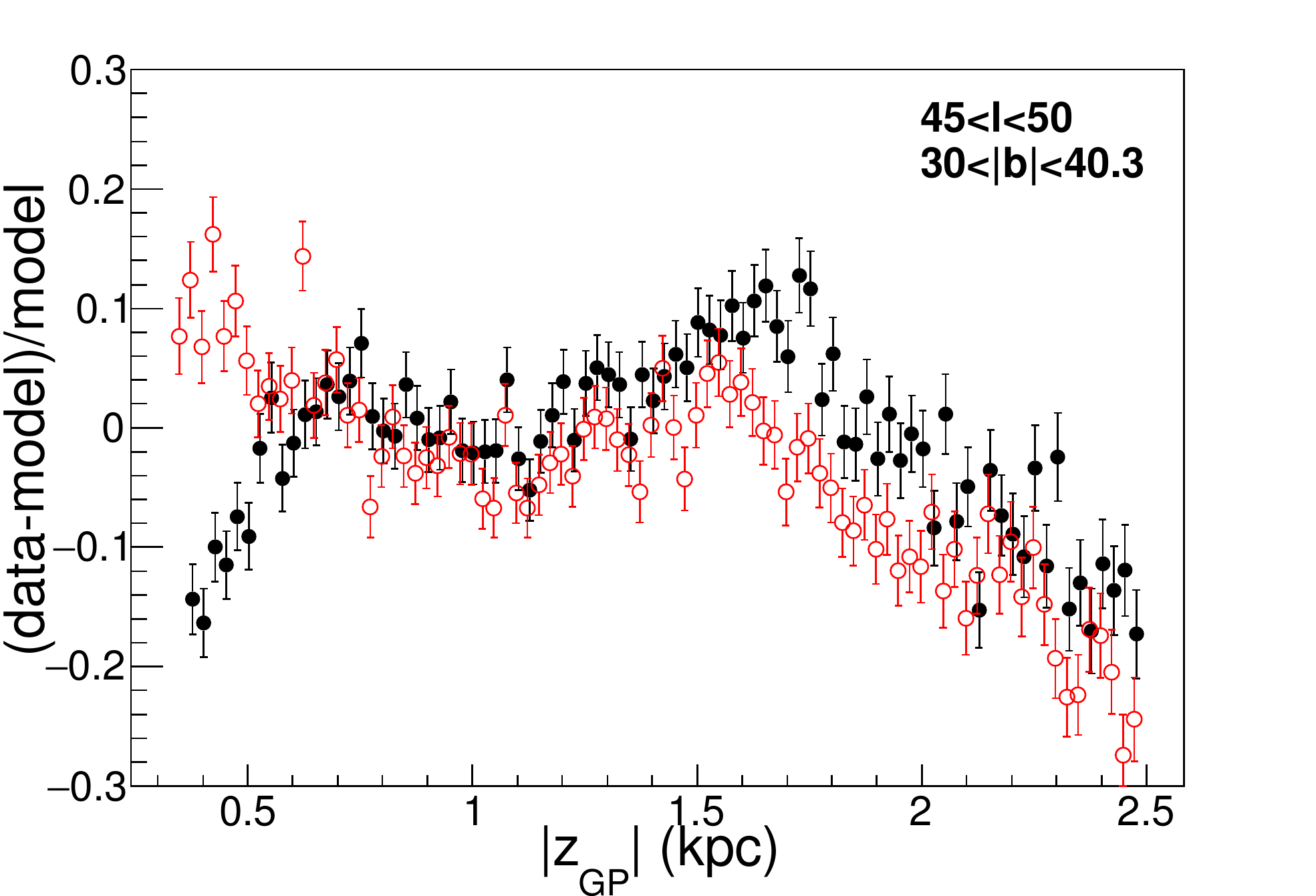}
\includegraphics[scale=0.267]{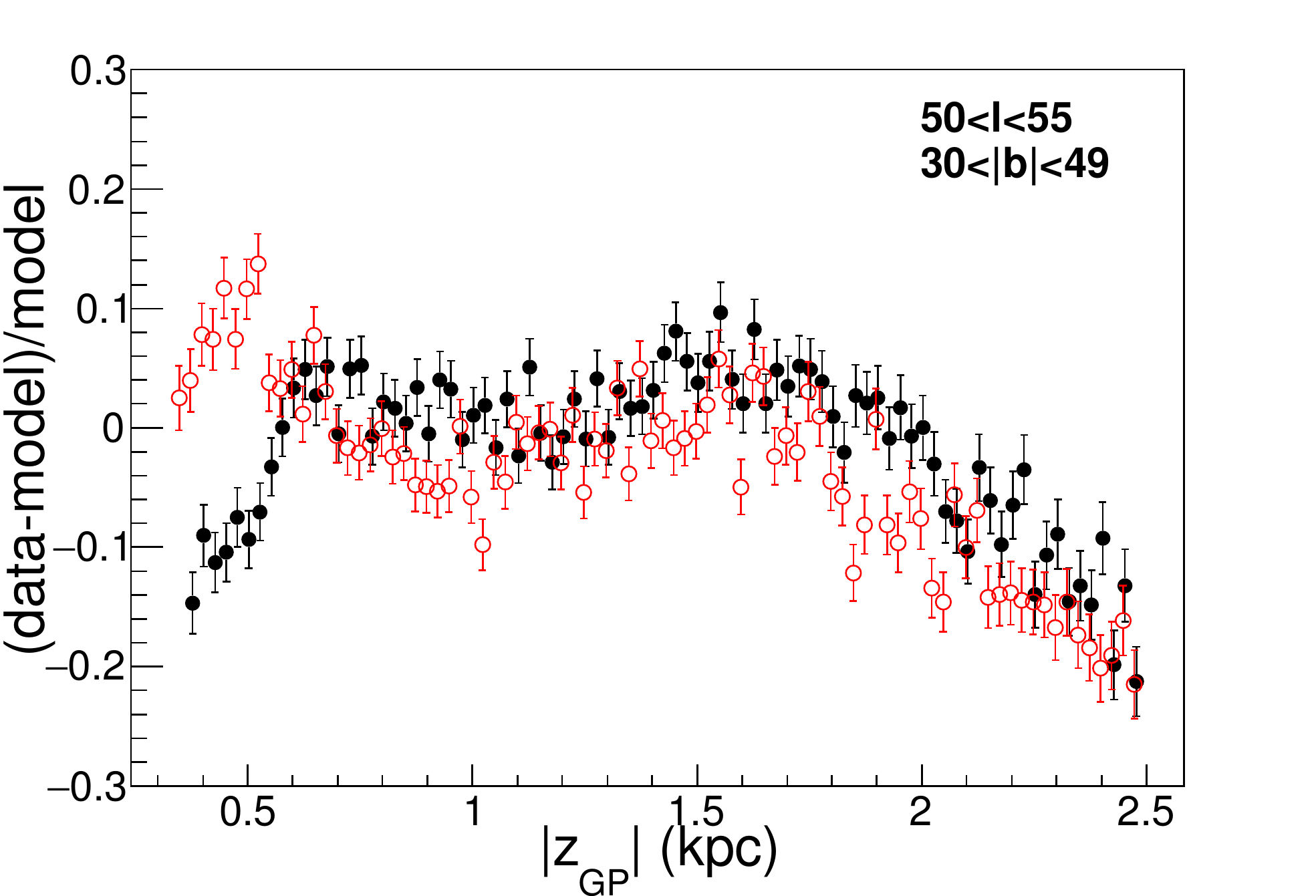} 
\includegraphics[scale=0.267]{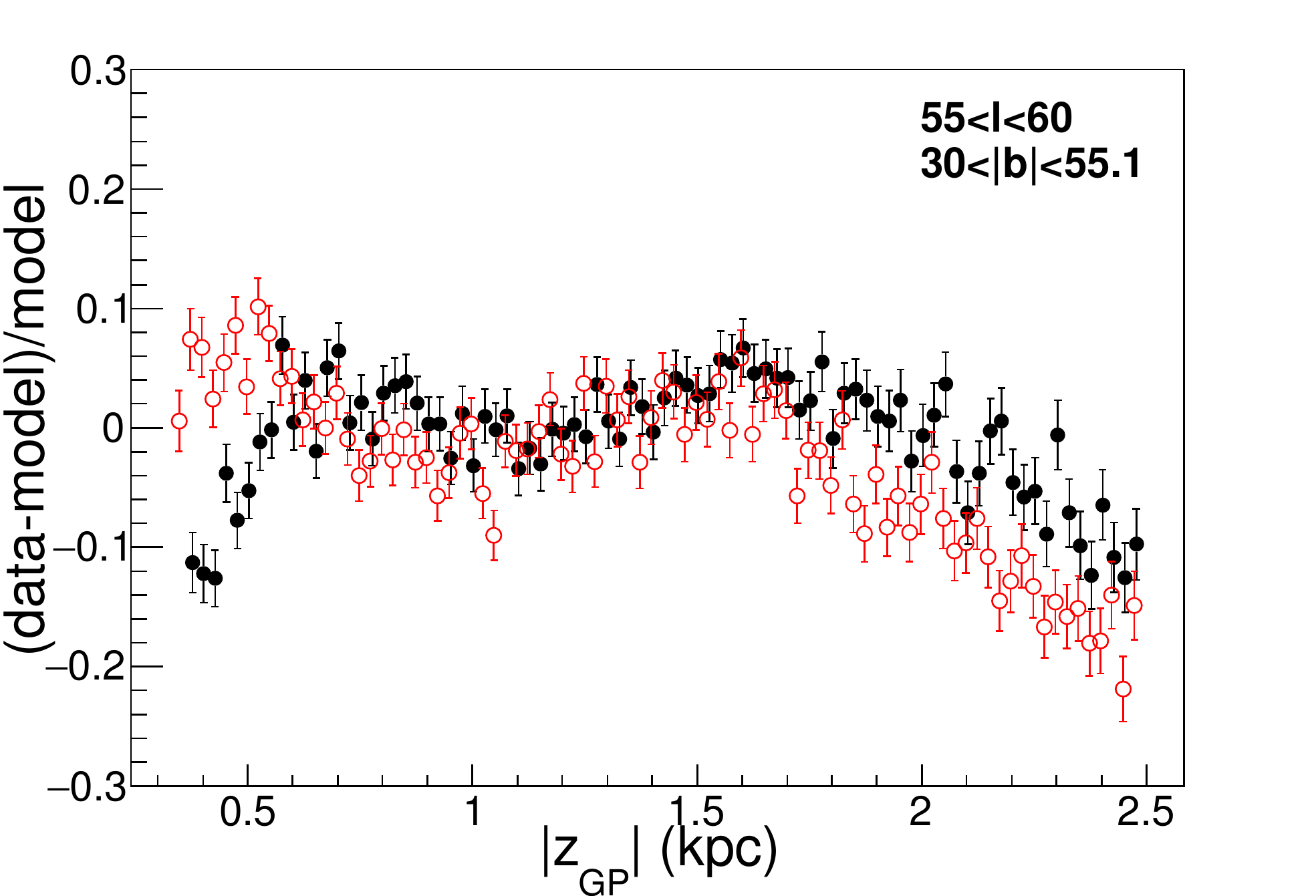}
\includegraphics[scale=0.267]{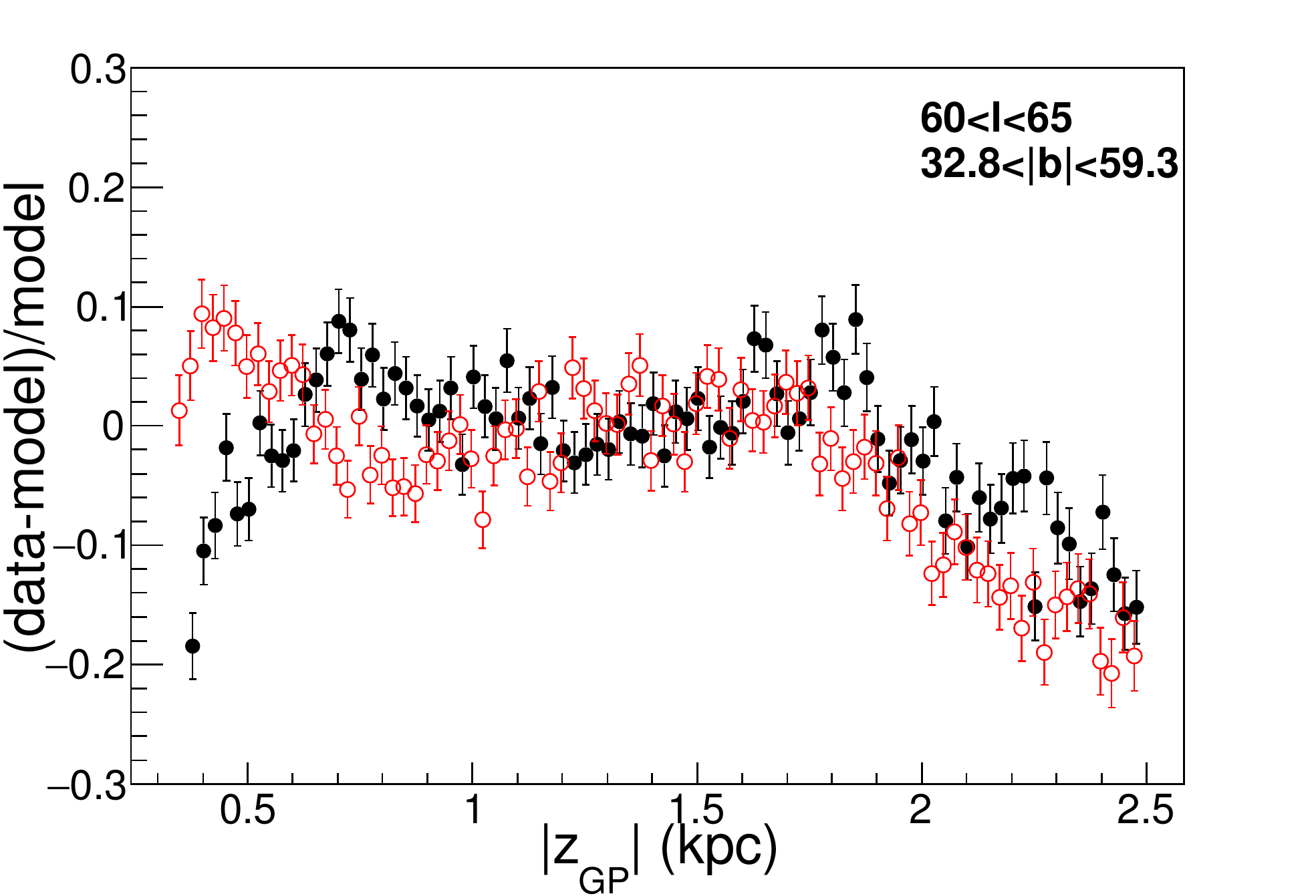} 
\includegraphics[scale=0.267]{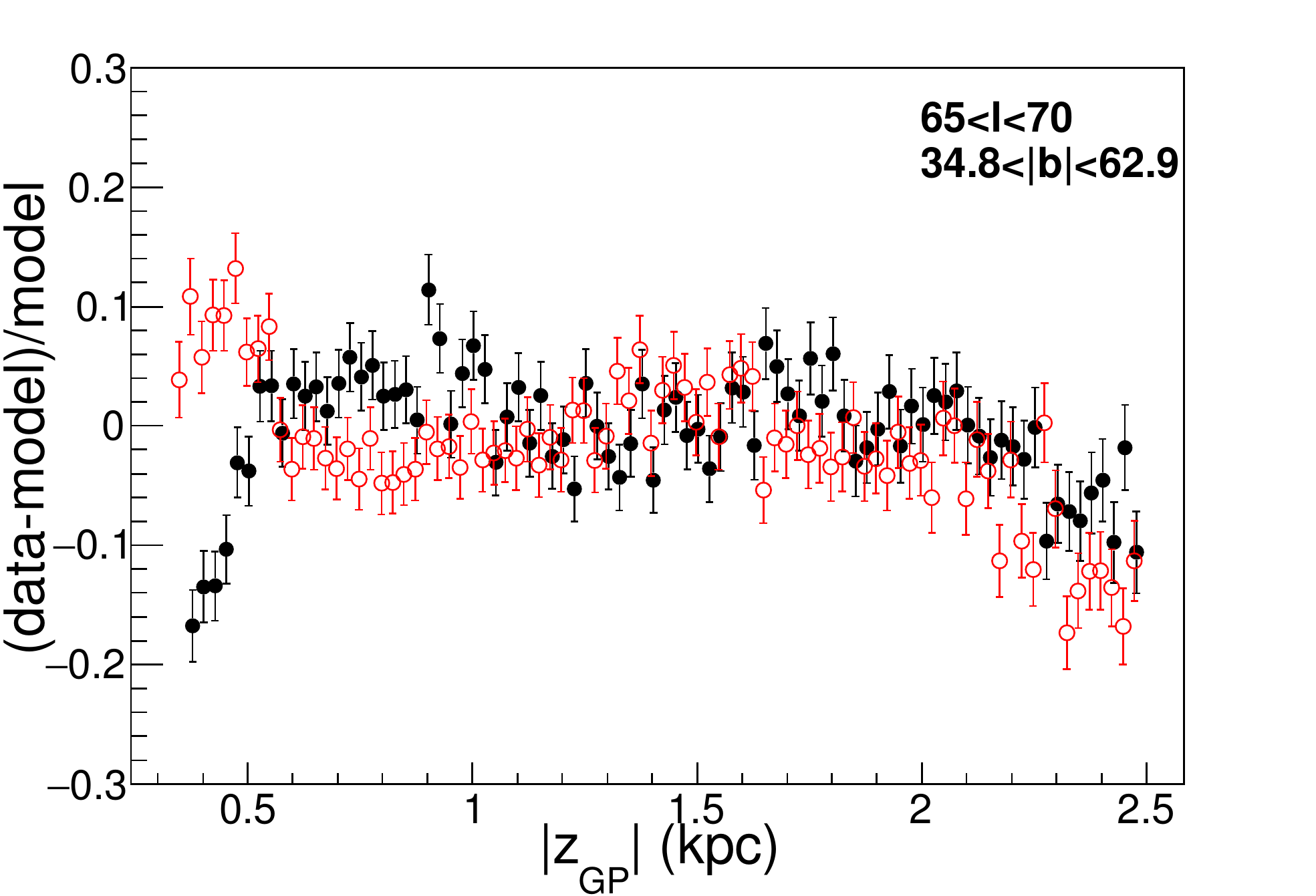}
\includegraphics[scale=0.267]{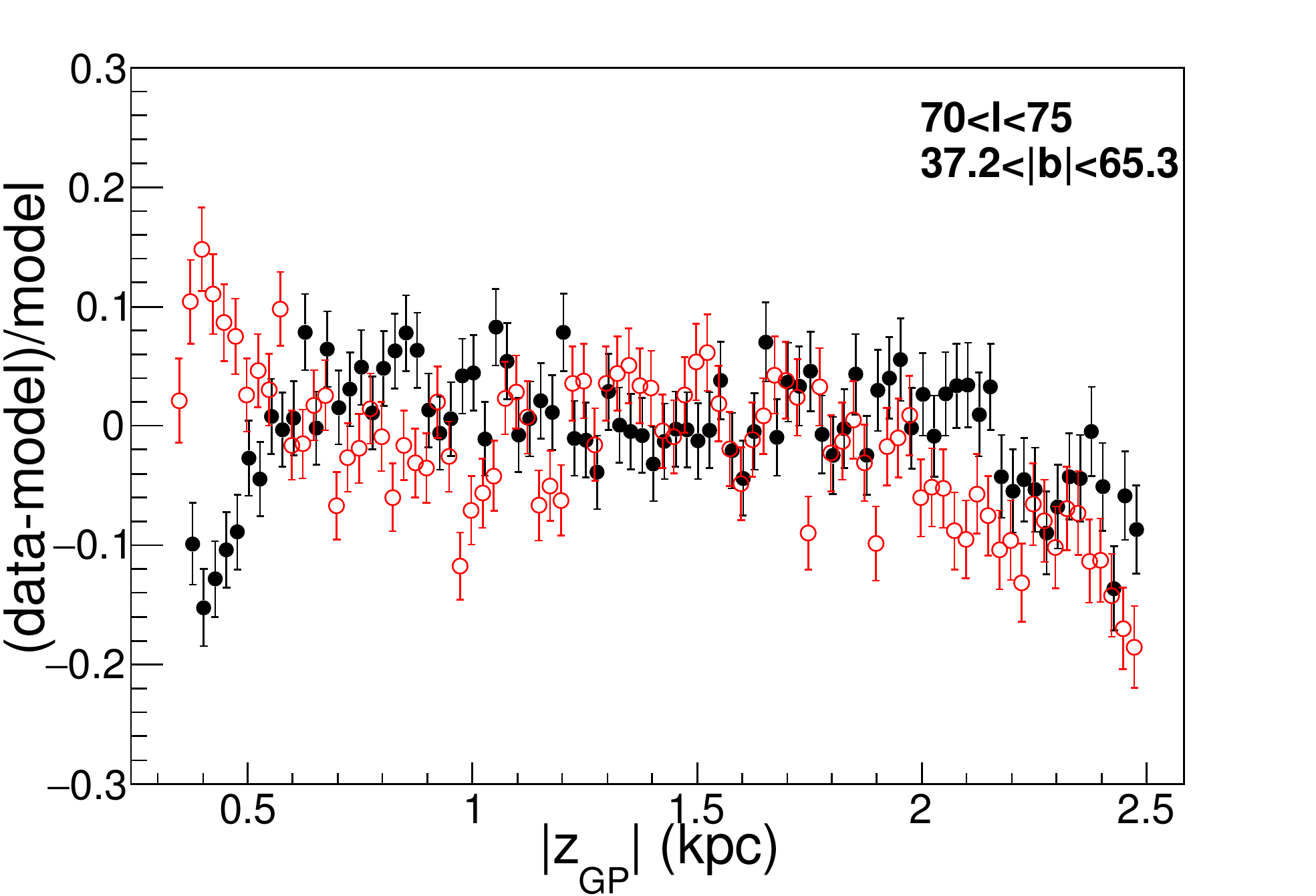} 
\includegraphics[scale=0.267]{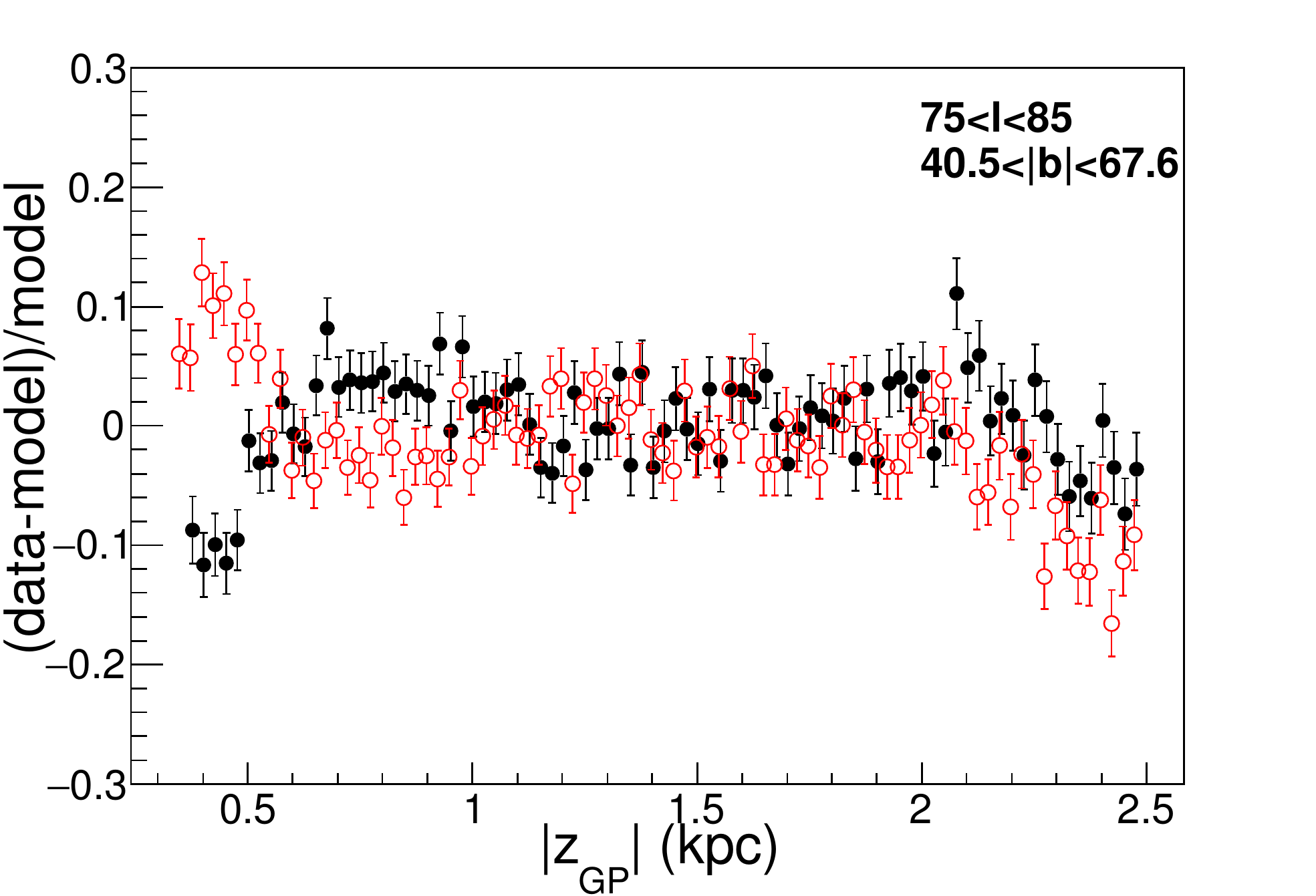}
\includegraphics[scale=0.267]{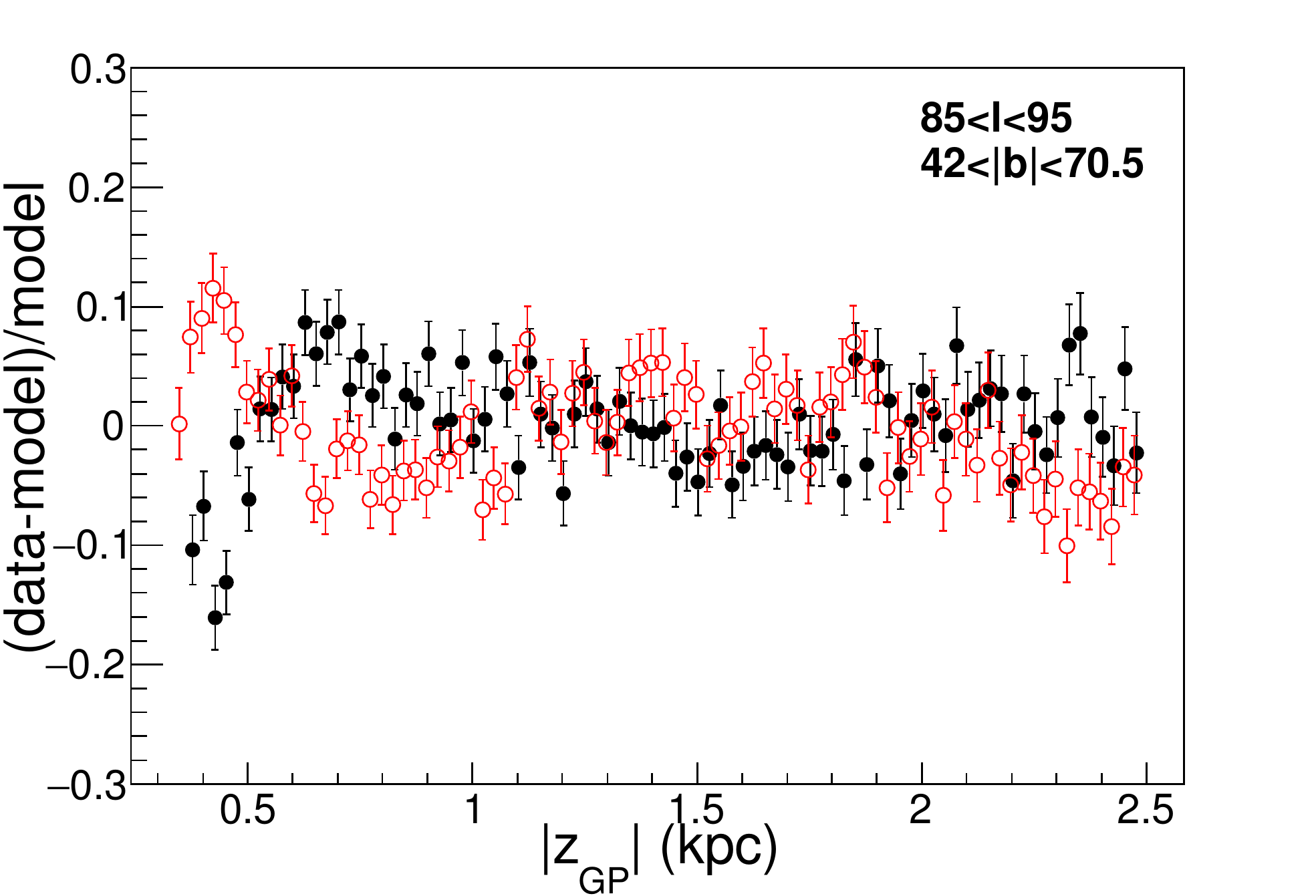} 
\includegraphics[scale=0.267]{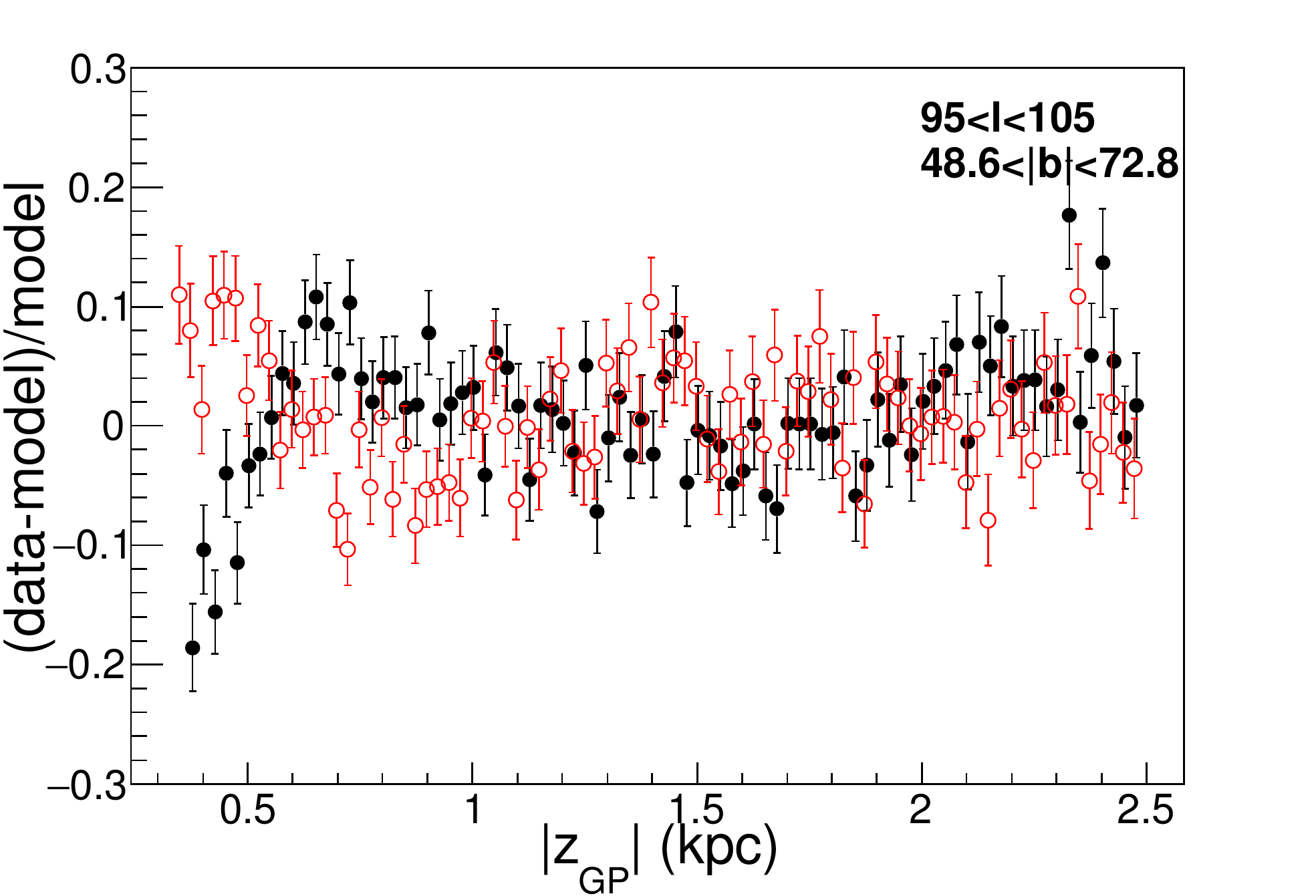}
\includegraphics[scale=0.267]{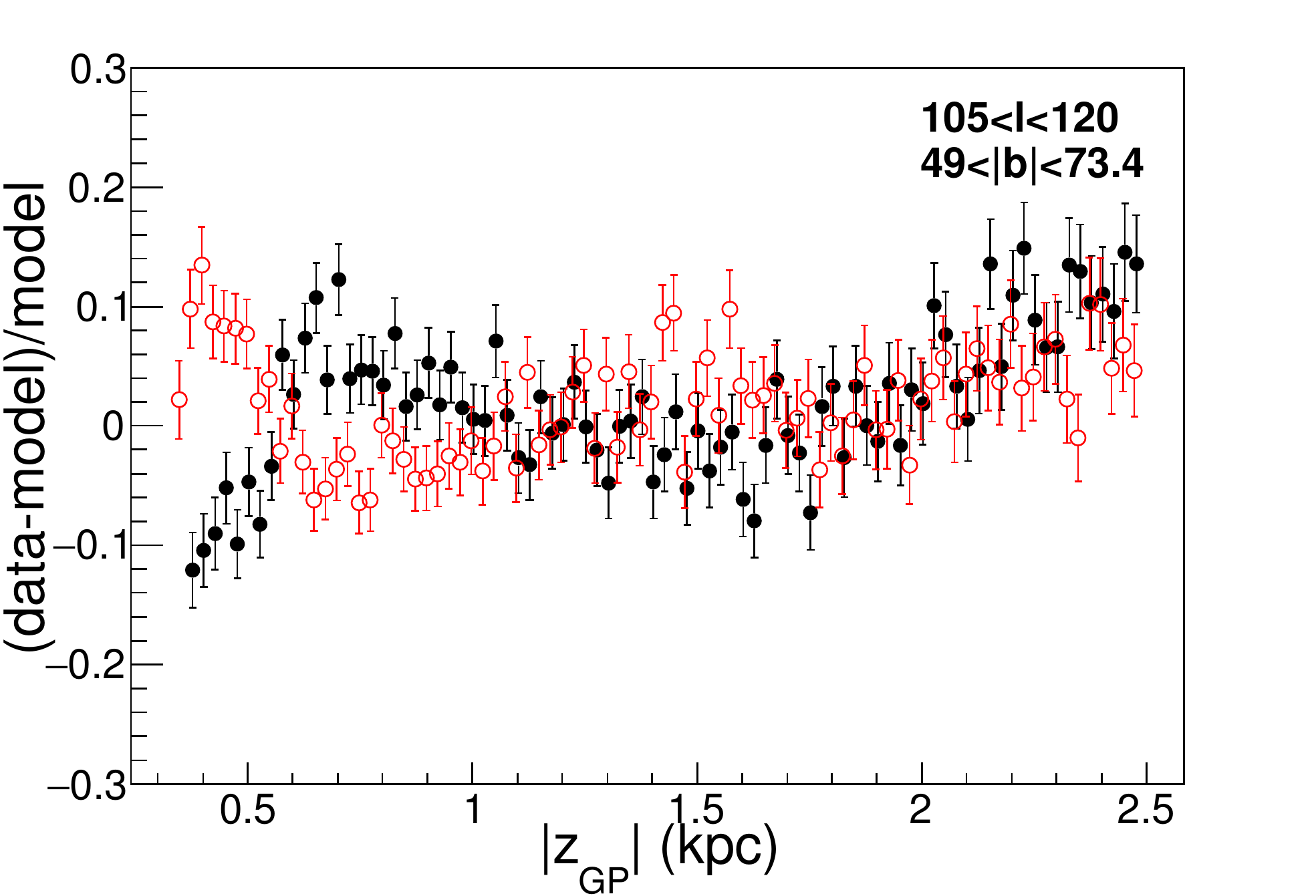} 
\includegraphics[scale=0.267]{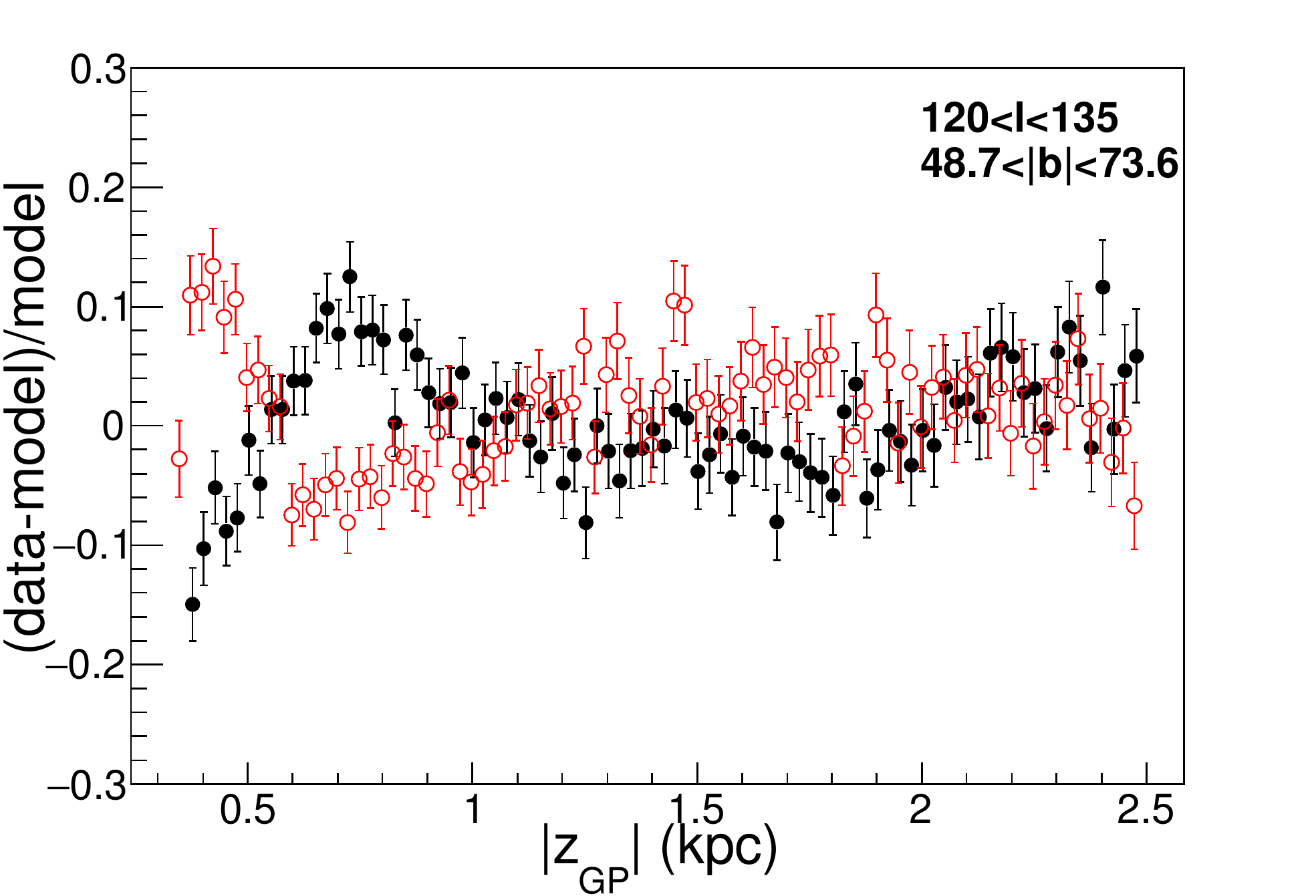}
\includegraphics[scale=0.267]{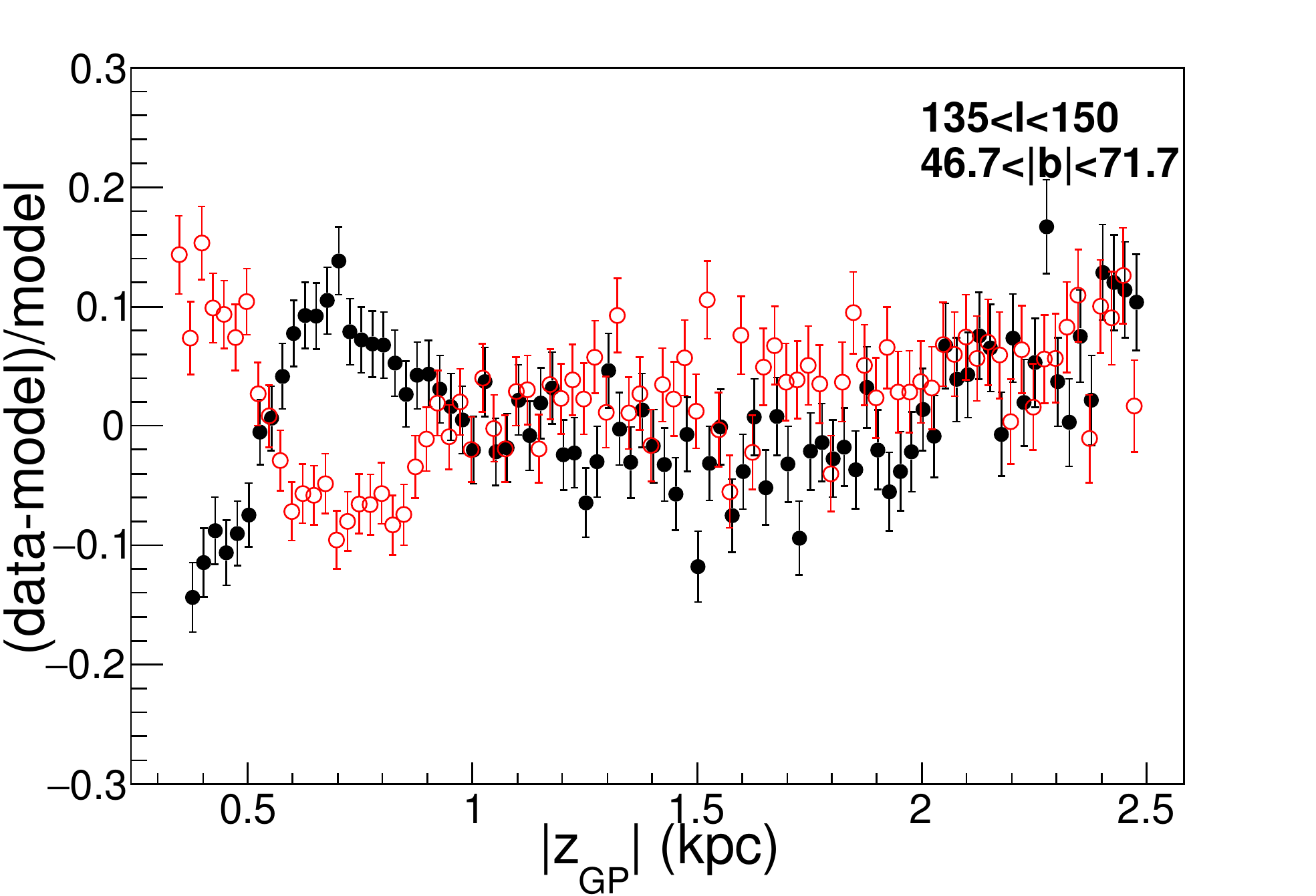} 
\includegraphics[scale=0.267]{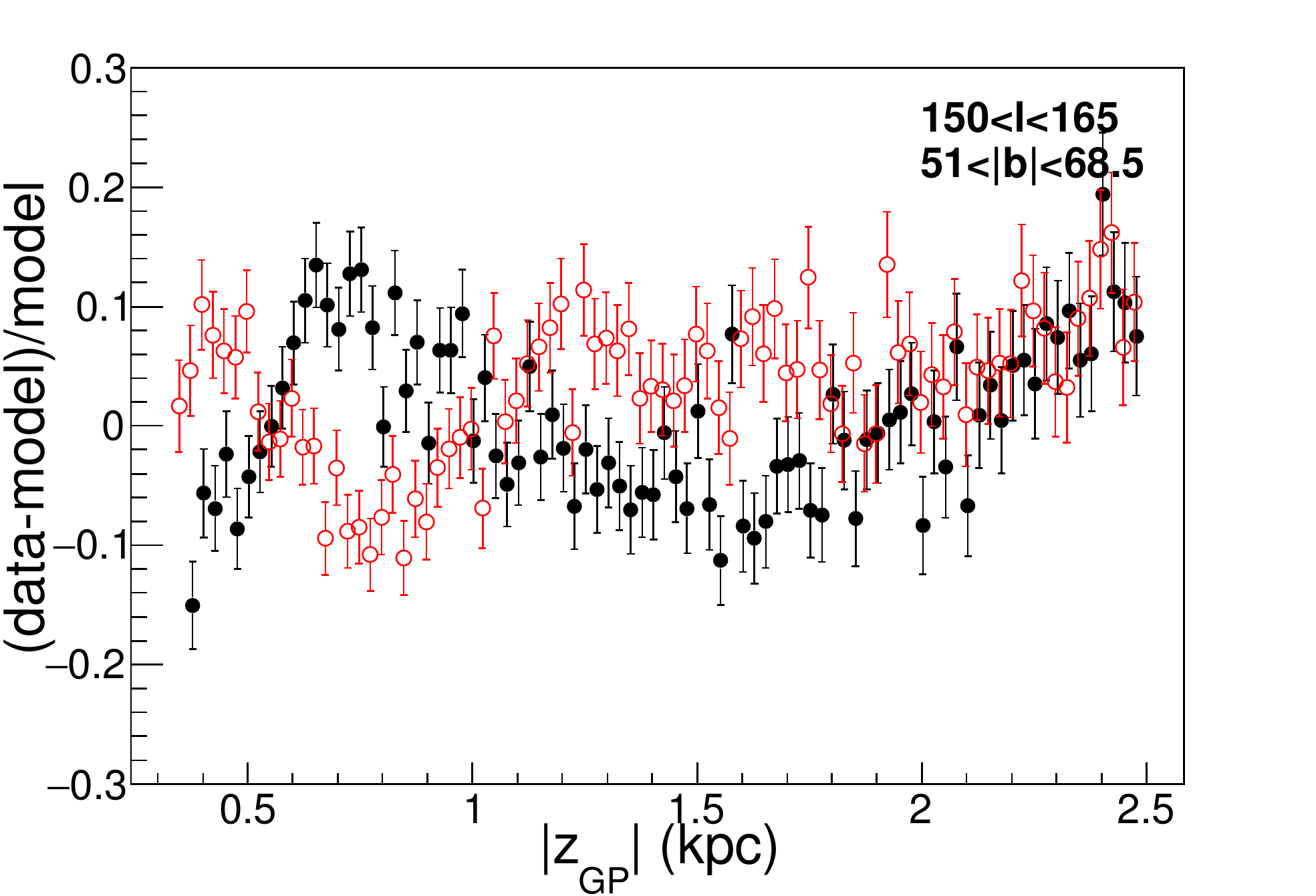}
\includegraphics[scale=0.267]{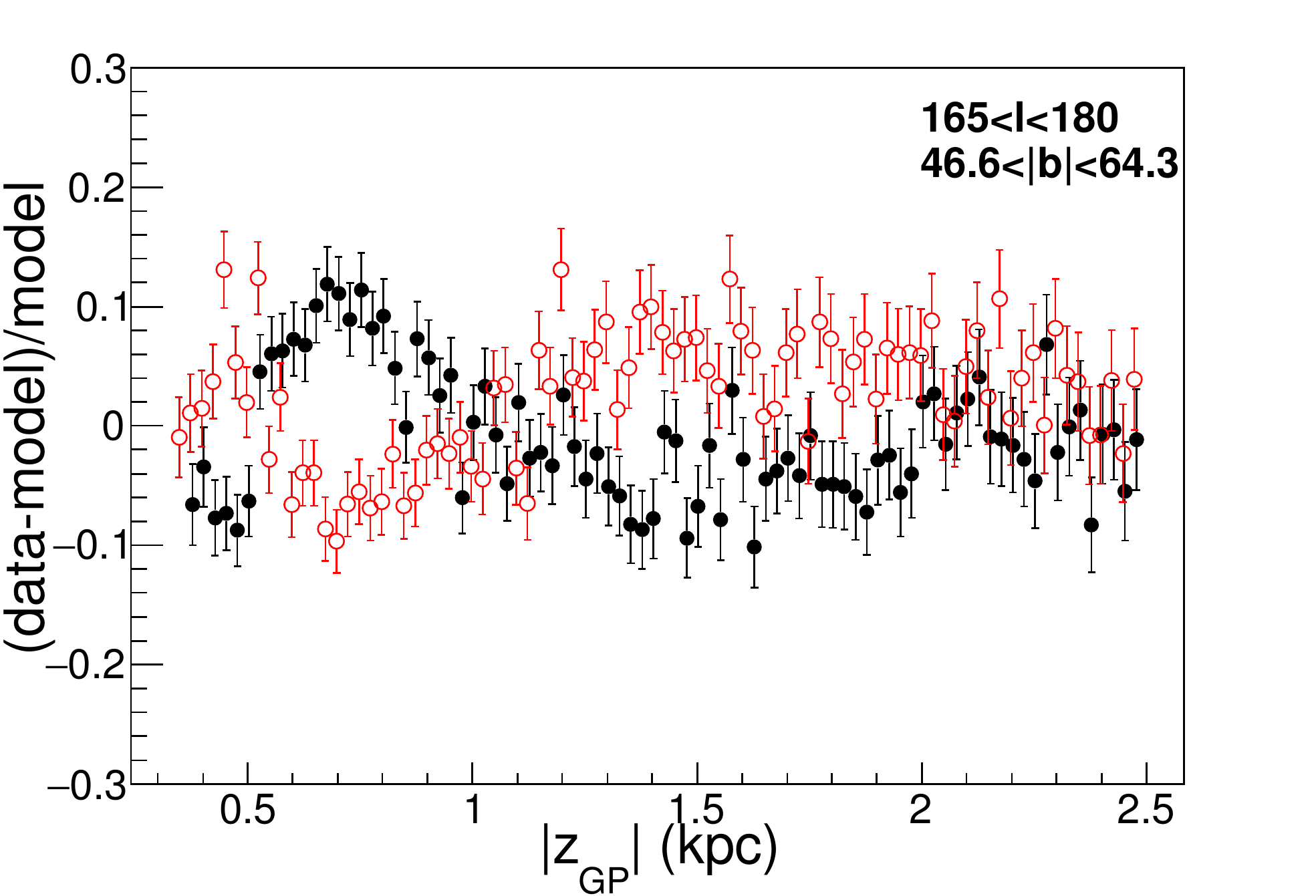} 
\caption[Residuals of the stellar densities overlaying north and south]{
Residuals of the stellar densities, namely, 
the difference between the observed 
stellar density and the best-fit model divided by the best-fit model, are 
shown with respect to 
the magnitude of 
the vertical distance 
to the Galactic plane, $|z_{\rm GP}| = |z + z_\odot|$ with $z_\odot$ = 14.9 pc.
The plots overlay the residuals in the north (filled, black) 
with those in the south (open, red). 
Each panel corresponds to the histogram covering the same region 
in $l$ and $b$ in Figure \ref{fig:histogramsmatchedset}, which is the result of 
using the matched north and south regions with expanded latitude shown in 
Figure \ref{fig:footprintmatched}(b). 
}
\label{fig:residualoverlays}
\end{center}
\end{figure}

We now turn to a detailed analysis of the north-south differences shown in 
Figure \ref{fig:histogramsmatchedset}. Working in a coordinate system 
with the Galactic plane as its origin (by shifting $z$ so that $z\rightarrow z +z_\odot$), 
we show residuals calculated as (data-model)/model as a function of vertical distance 
in Figure \ref{fig:residualoverlays}. 
We overlay the residual in the north and that 
in the south in  each region of Figure \ref{fig:histogramsmatchedset} using the parameters
of its fit. This visualizes the north-south asymmetry of Figure \ref{fig:histogramsmatchedset} and
confirms the earlier results of \citet{widrow12} and \citet{yanny13}. 
The residuals have wave-like features that 
grow in amplitude and vertical extent 
with increasing longitude, refining and extending what was observed by \citet{yanny13}. 
Interestingly, at low $z$ the south always has a positive residual, 
whereas the north always has a negative residual. 
Note that although the figures go out to 2.5 kpc, 
the fits themselves only extend to 2 kpc. 
Repeating this analysis on regions with uniform latitude 
reveals the same pattern; the changes due to latitude are small 
compared to the effects seen in the figure. The data set with the 
expanded latitude selection is used here because it allows for a larger range in longitude and better 
statistics. We find that similar results emerge when we repeat our analysis using distances computed
using $(r-i)_0$ color. 

%Figure 9

%Figure 10
\begin{figure}[hp!]
\centering
\begin{center}
\subfloat[]{\includegraphics[scale=0.5]{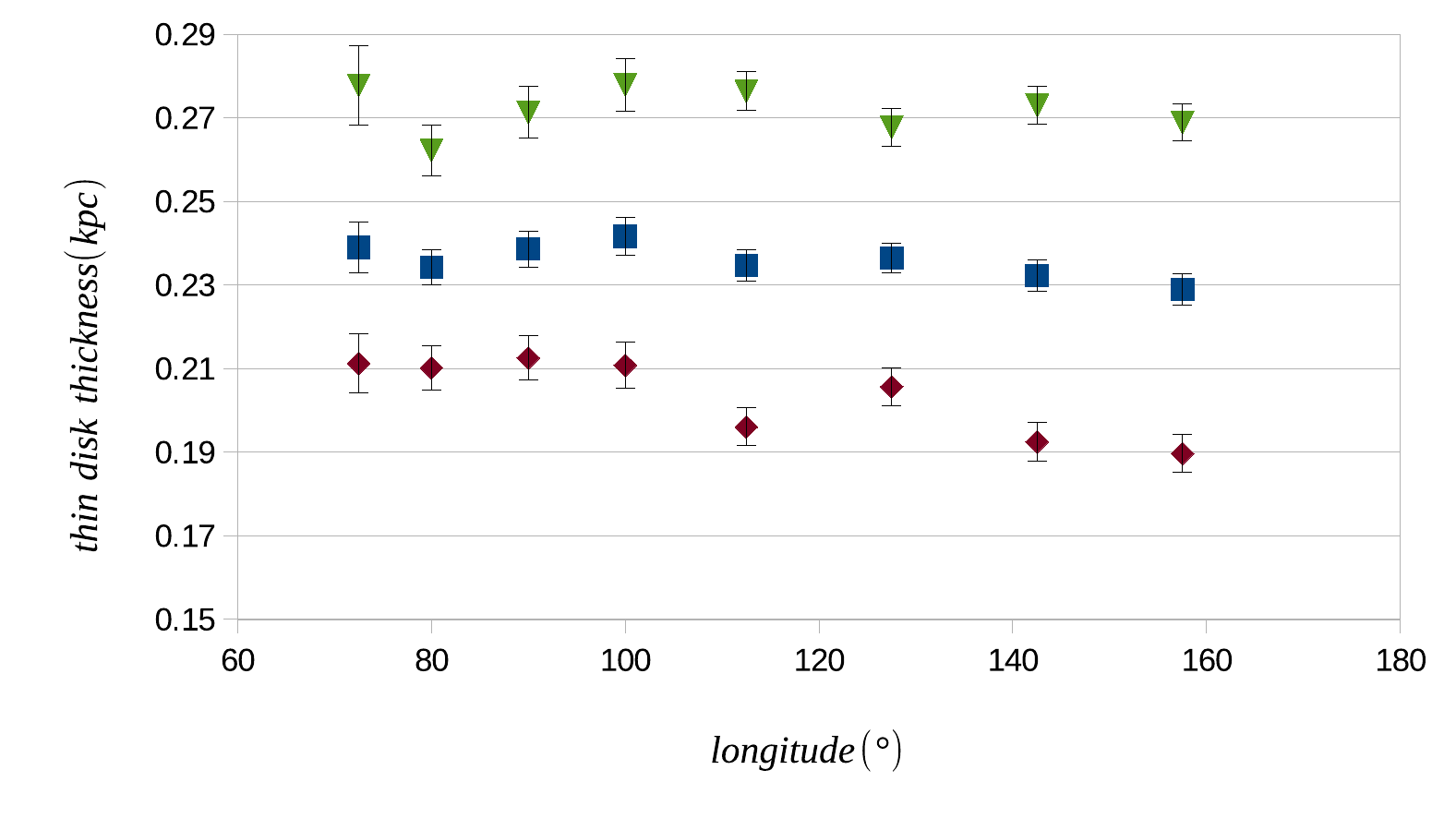}}
\subfloat[]{\includegraphics[scale=0.5]{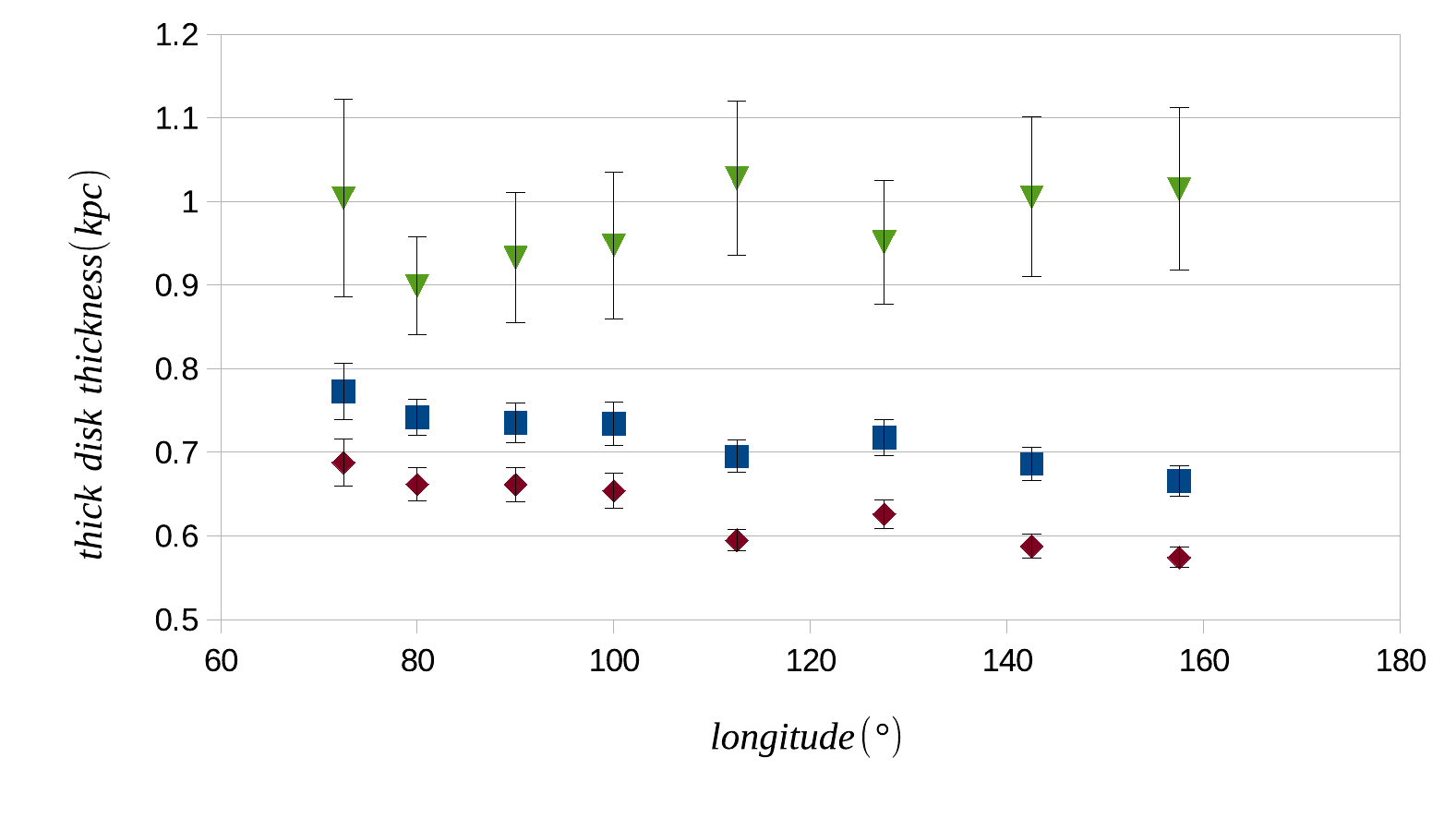}}
\caption[
Plots of the thickness of the thin (a) and thick (b) disks 
%Thin and 
%thick disk thicknesses 
as functions of longitude comparing the north and the south]{
Plots of the thickness of the thin (a) and thick (b) disks 
%Thin (a) and thick (b) disk thickness plots 
as functions of longitude obtained from best-fit 
models on the stellar density histograms using $(g-i)_0$ color 
for the uniform latitude regions shown in Figure \ref{fig:footprintmatched}(a). We compare 
regions in the north (green triangles), the south (maroon diamonds), and the north and 
south combined (blue squares). 
}
\label{fig:thinthicknorthsouth}
\end{center}
\end{figure}

%Figure 11
\begin{figure}[hp!]
\begin{center}
\subfloat[]{\includegraphics[scale=0.5]{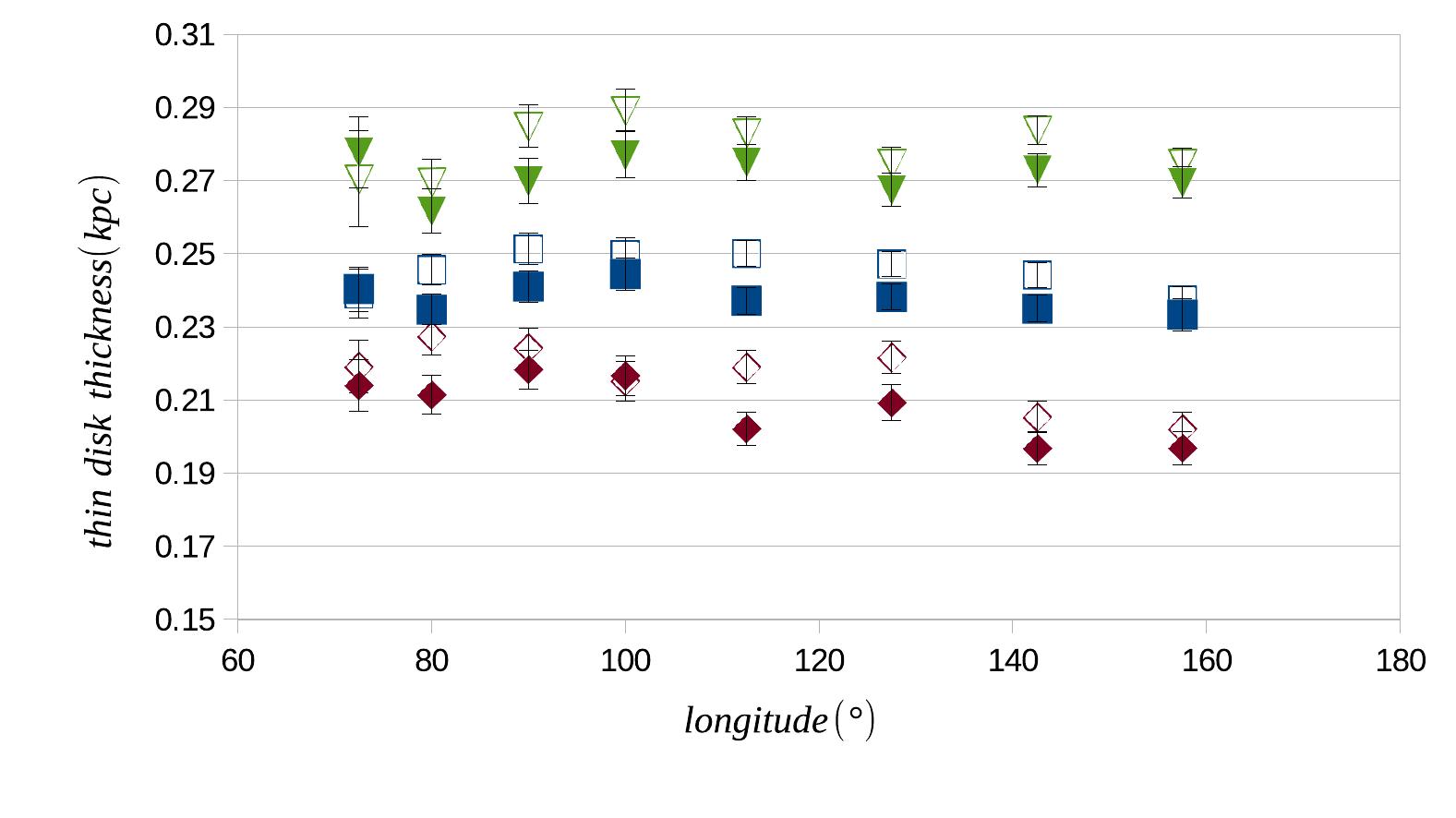}}
\subfloat[]{\includegraphics[scale=0.5]{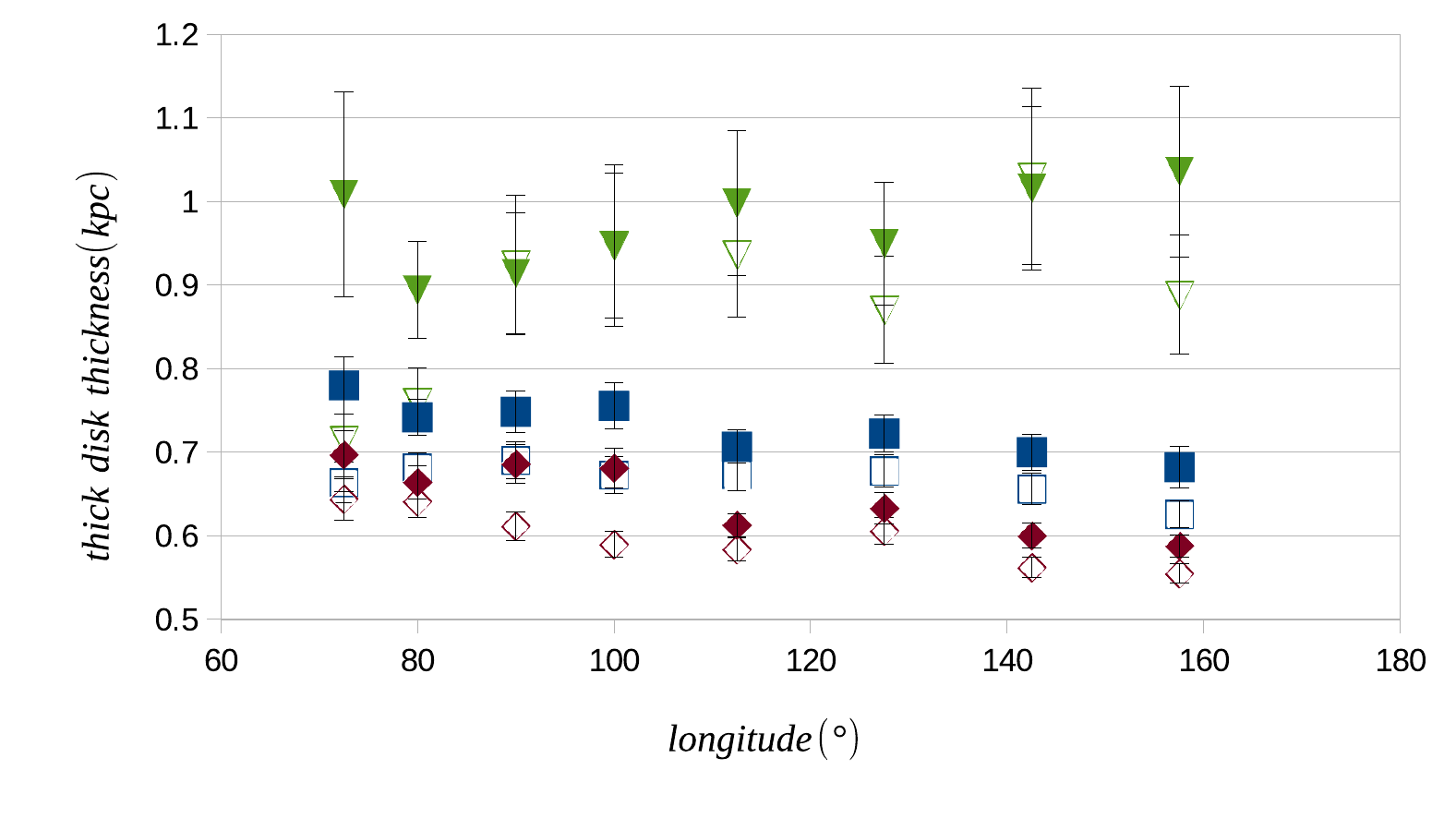}}
\caption[Color calibration comparing the north and the south]{
Thicknesses of the thin (a) and thick (b) disks 
%Thin (a) and thick (b) disk thicknesses 
as functions of longitude, repeating the analysis of 
Figure \ref{fig:thinthicknorthsouth} and adopting its notation, but now testing for sensitivity 
to color effects. 
The filled symbols represent the analysis done with $(g-i)_{0}$ color, implementing 
an overall shift of -16 mmag shift in 
the $(g-i)_{0}$ color in the south, 
whereas the open symbols represent the analysis performed using $(r-i)_{0}$ color.
}
\label{fig:colorcalibration}
\end{center}
\end{figure}

\begin{figure}[hp!]
\begin{center}
{\subfloat[]{\includegraphics[scale=0.27]{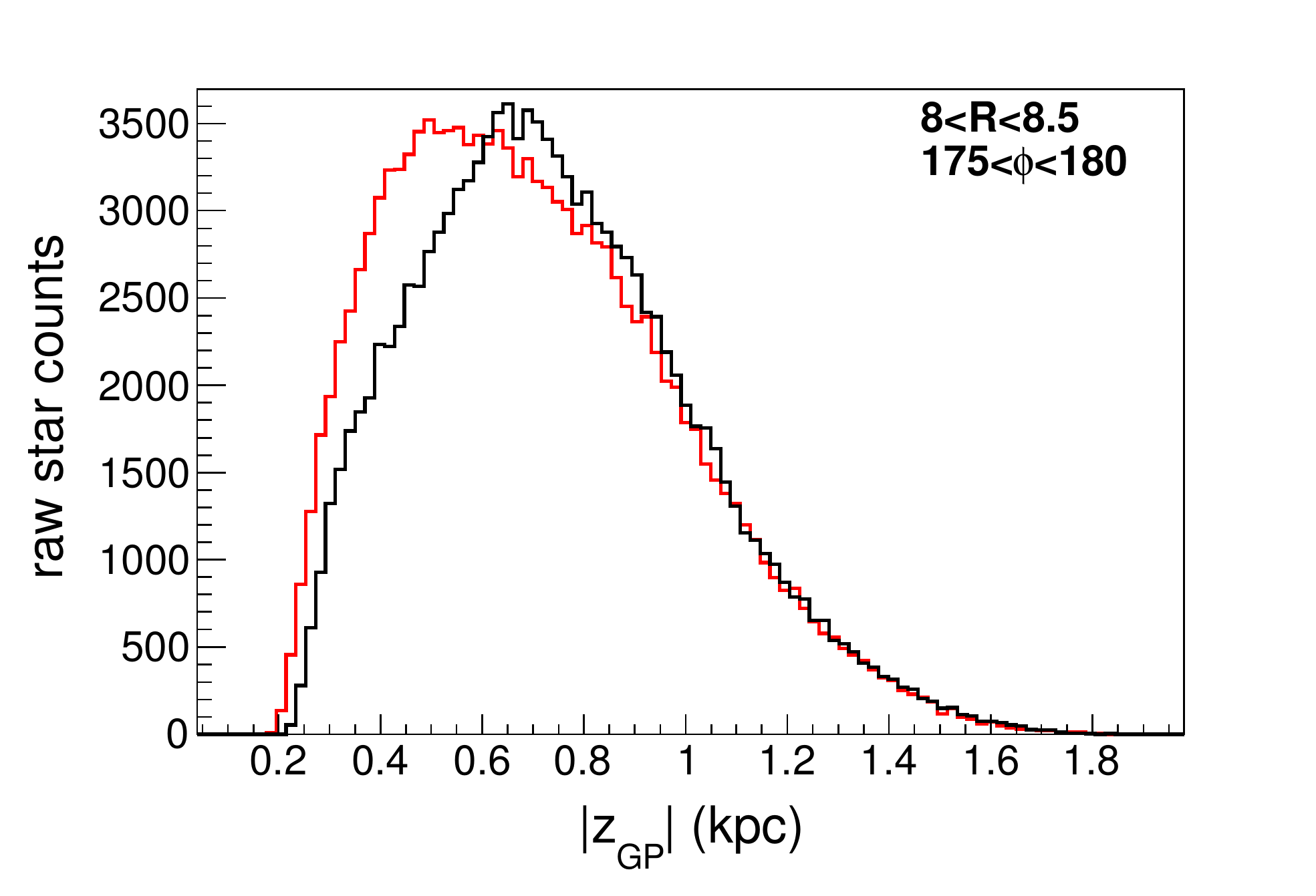}}
\subfloat[]{\includegraphics[scale=0.27]{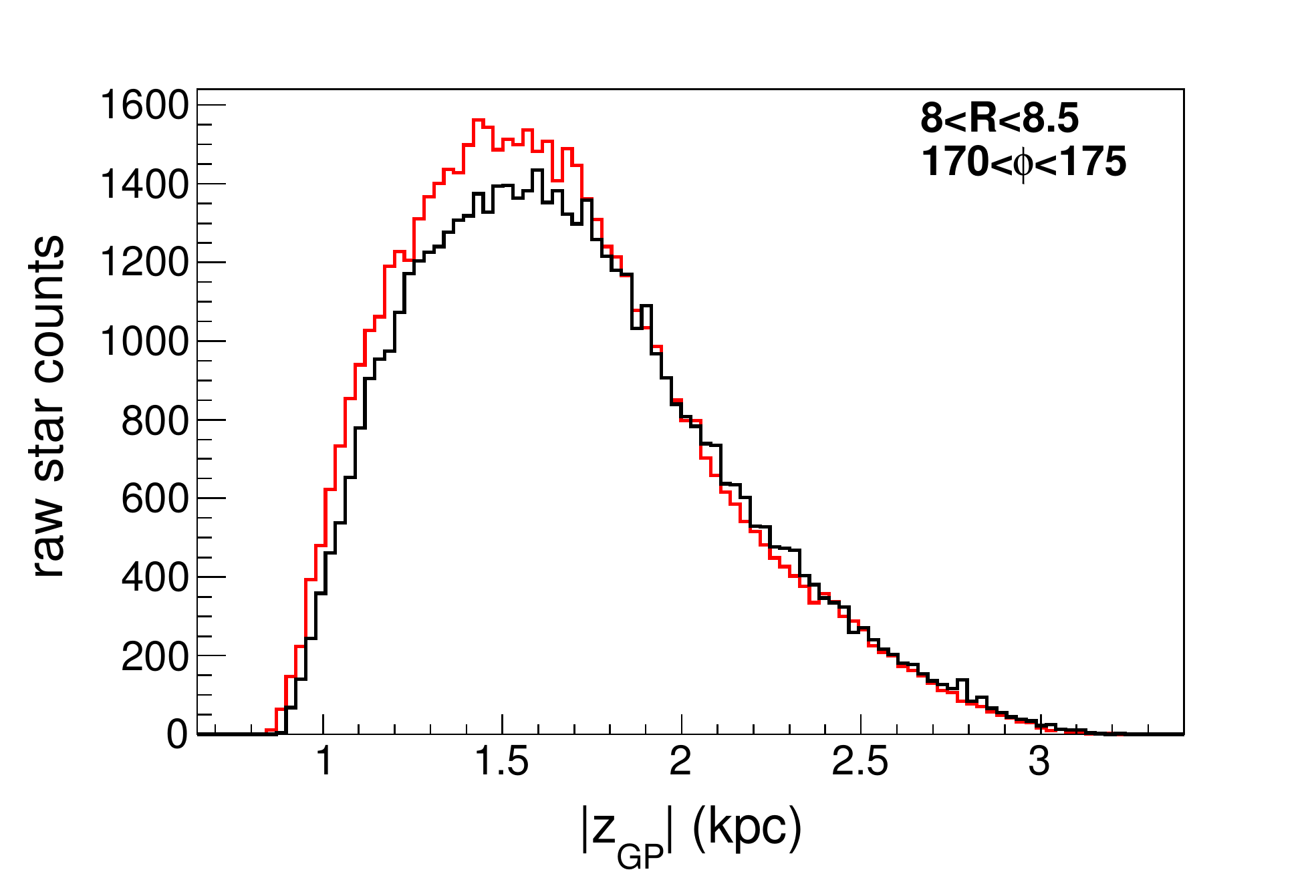}}}
{\subfloat[]{\includegraphics[scale=0.27]{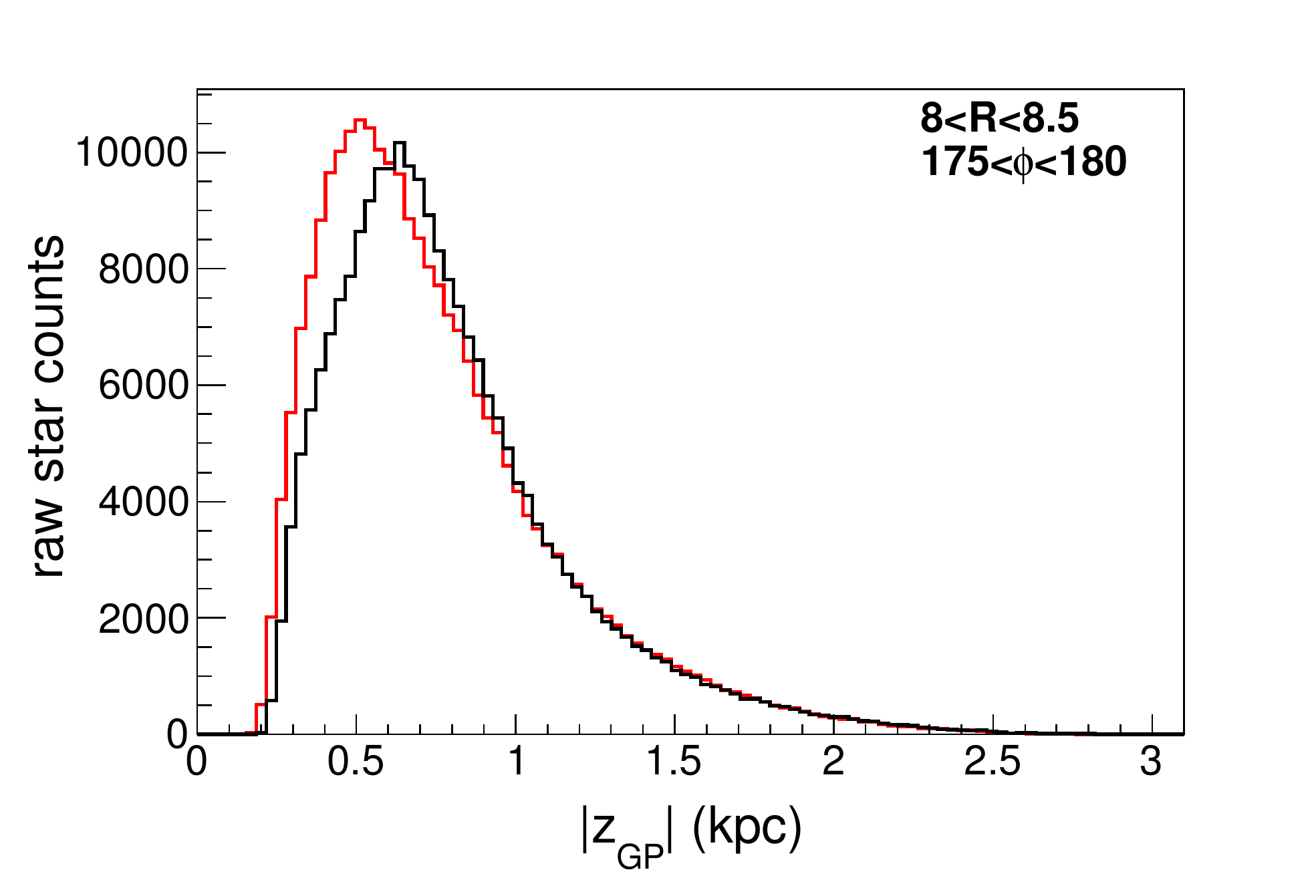}}
\subfloat[]{\includegraphics[scale=0.27]{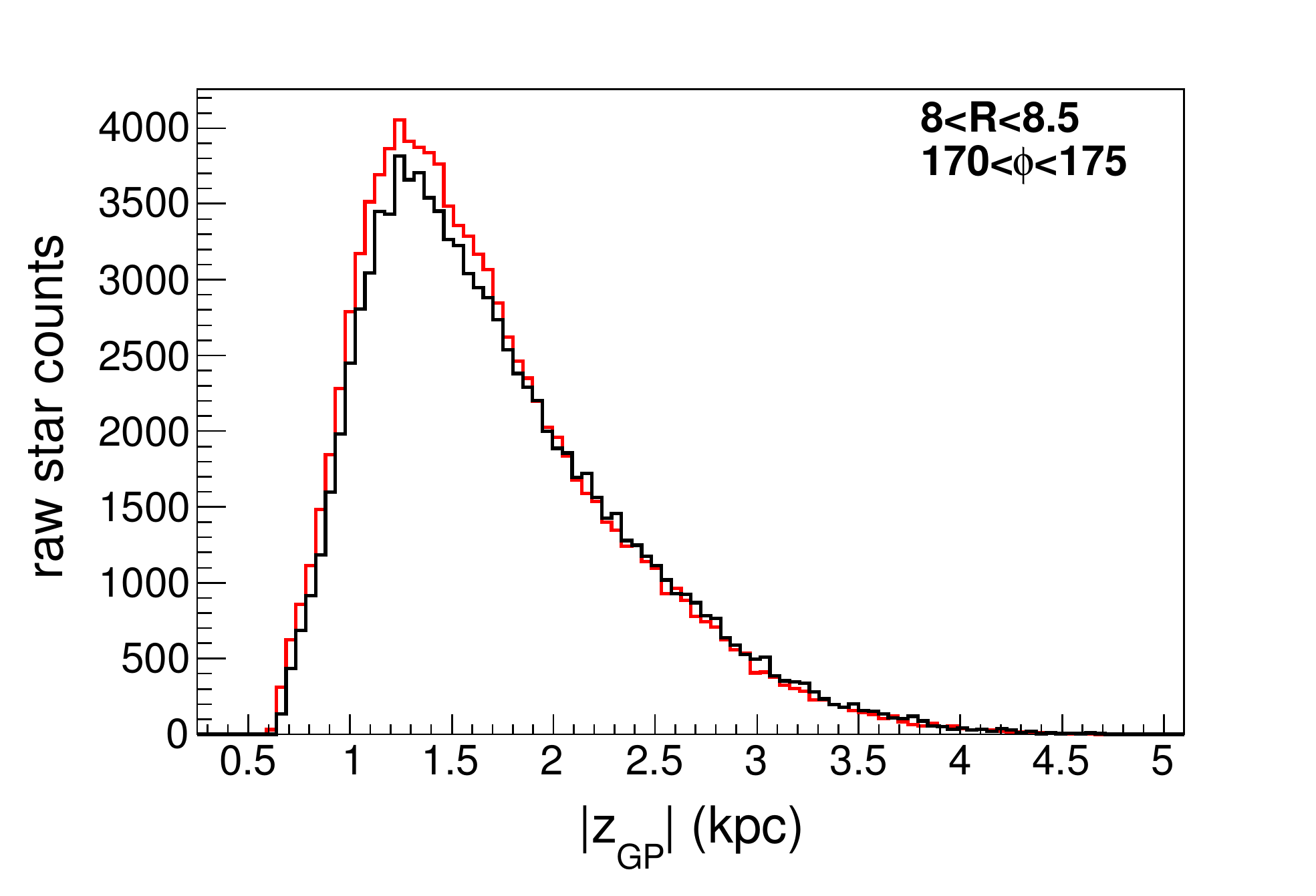}}
\subfloat[]{\includegraphics[scale=0.27]{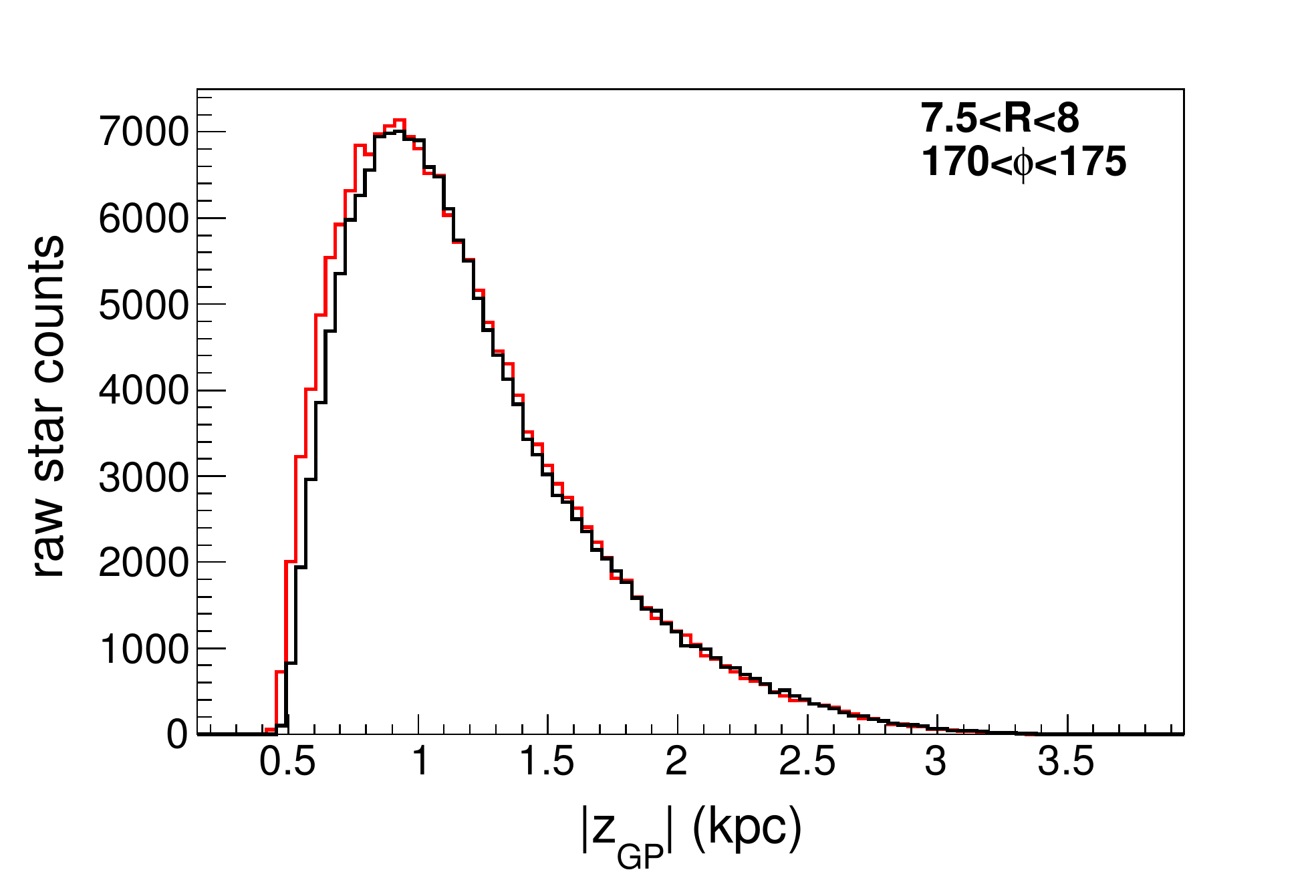}}}
\caption[R-phi distribution]{
 We illustrate the origin of the N/S difference in the Galactic disk 
parameters by plotting the raw number counts, in the north (black) and 
south (red), using the samples of Figures \ref{fig:footprintmatched}(a) and (b), 
restricted to those stars that satisfy $1.8 \le (g-i)_0 \le 2.4$. 
We plot particular ranges in the Galactocentric coordinates
$R$ and $\phi$, as a function of the vertical distance determined 
from the center of the Galactic plane. We use the 
average value of $z_\odot = 14.9\, {\rm pc}$ that emerged from our (uniform) 
analysis of Figure \ref{fig:footprintmatched}(a) for this purpose. 
We have made the following selections: 
in (a) $R\in [8.0, 8.5]\,{\rm kpc}$ and $\phi \in [175^\circ, 180^\circ]$, 
in (b) $R\in [8.0, 8.5]\,{\rm kpc}$ and $\phi \in [170^\circ, 175^\circ]$, both from 
Figure \ref{fig:footprintmatched}(a), 
and
in (c) $R\in [8.0, 8.5]\,{\rm kpc}$ and $\phi \in [175^\circ, 180^\circ]$, 
in (d) $R\in [8.0, 8.5]\,{\rm kpc}$ and $\phi \in [170^\circ, 175^\circ]$, 
in (e) $R\in [7.5, 8.0]\,{\rm kpc}$ and $\phi \in [170^\circ, 175^\circ]$, 
all from Figure \ref{fig:footprintmatched}(b). 
}
\label{fig:reftable}
\end{center}
\end{figure}

\begin{figure}[hp!]
\begin{center}
\includegraphics[scale=0.60,angle=-90]{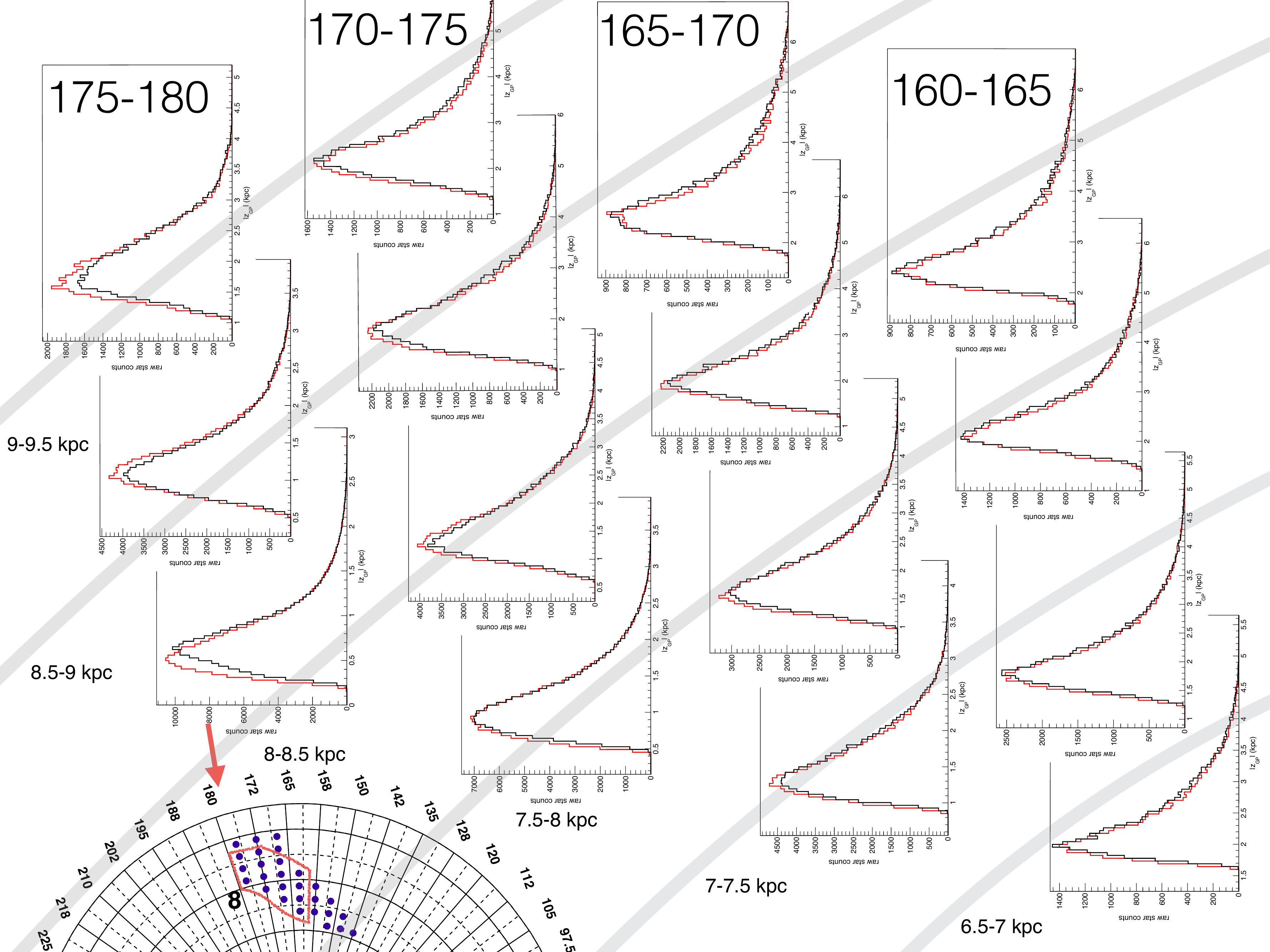}
\caption[R-phi distribution in footprint]{
 We illustrate the N/S variations in the raw number counts with $R$ and $\phi$ 
using the 
samples of Figure \ref{fig:footprintmatched}(b) in the red box and the notation and choices of 
Figure \ref{fig:reftable}.}
\label{fig:Phibreaking}
\end{center}
\end{figure}

\tabletypesize{\footnotesize}
\begin{deluxetable}{l|l|lllll}
\tablecaption{Best-fit Results to the North only (n), South only (s), and
North-South Combined (c) Selections of the Uniform Latitude Sample.}
\tablehead{
\multicolumn{1}{c} {$l$ range (deg)} &
\multicolumn{1}{c} {} &
\multicolumn{1}{c} {$N_{0}$ $(\times 10^6)$} &
\multicolumn{1}{c} {$H_1$ (kpc)} &
\multicolumn{1}{c} {$H_2$ (kpc)} &
\multicolumn{1}{c} {$f$} &
\multicolumn{1}{c} {$\chi^2 /{\rm dof}$} \\
}
\startdata
70 $< l <$ 75&      s&        6.11 $\pm$ 0.33&	0.211 $\pm$ 0.007&	0.688 $\pm$ 0.028&	0.112 $\pm$ 0.008&	1.18\\
&      c&         4.83 $\pm$ 0.17&	0.239 $\pm$ 0.006&	0.773 $\pm$ 0.034&	0.108 $\pm$ 0.008&	1.19\\
&      n&         3.86 $\pm$ 0.16&	0.278 $\pm$ 0.009&	1.004 $\pm$ 0.118&	0.083 $\pm$ 0.013&	0.96\\\hline
75 $< l <$ 85&      s&        6.01 $\pm$ 0.24&	0.210 $\pm$ 0.005&	0.662 $\pm$ 0.019&	0.116 $\pm$ .006&	1.07\\
&      c&         5.07 $\pm$ 0.13&	0.234 $\pm$ 0.004&	0.742 $\pm$ 0.022&	0.104 $\pm$ 0.006&	1.12\\
&      n&         4.35 $\pm$ 0.13&	0.262 $\pm$ 0.006&	0.899 $\pm$ 0.059&	0.082 $\pm$ 0.008&	0.95\\\hline
85$ < l <$ 95&      s&        5.71 $\pm$ 0.22&	0.213 $\pm$ 0.005&	0.662 $\pm$ 0.020&	0.113 $\pm$ 0.007&	1.23\\
&      c&         4.84 $\pm$ 0.12&	0.239 $\pm$ 0.004&	0.735 $\pm$ 0.024&	0.100 $\pm$ 0.006&	1.32\\
&      n&         4.19 $\pm$ 0.12&	0.271 $\pm$ 0.006&	0.933 $\pm$ 0.077&	0.069 $\pm$ 0.009&	1.14\\\hline
95$ < l <$ 105&     s&        5.76 $\pm$ 0.23&	0.211 $\pm$ 0.005&	0.654 $\pm$ 0.021&	0.108 $\pm$ 0.007&	0.83\\
&      c&         4.60 $\pm$ 0.11&	0.242 $\pm$ 0.004&	0.734 $\pm$ 0.026&	0.098 $\pm$ 0.007&	1.20\\
&      n&         3.86 $\pm$ 0.10&	0.278 $\pm$ 0.006&	0.948 $\pm$ 0.088&	0.066 $\pm$ 0.009&	1.18\\\hline
105$ < l <$ 120&    s&        6.17 $\pm$ 0.24&	0.196 $\pm$ 0.005&	0.595 $\pm$ 0.013&	0.119 $\pm$ 0.005&	1.10\\
&      c&         4.74 $\pm$ 0.10&	0.235 $\pm$ 0.004&	0.695 $\pm$ 0.019&	0.099 $\pm$ 0.006&	1.48\\
&      n&         4.01 $\pm$ 0.09&	0.276 $\pm$ 0.005&	1.028 $\pm$ 0.092&	0.052 $\pm$ 0.006&	1.13\\\hline
120$ < l <$ 135&    s&        5.69 $\pm$ 0.20&	0.206 $\pm$ 0.005&	0.626 $\pm$ 0.017&	0.105 $\pm$ 0.006&	1.09\\
&      c&         4.72 $\pm$ 0.10&	0.236 $\pm$ 0.004&	0.718 $\pm$ 0.022&	0.085 $\pm$ 0.005&	1.38\\
&      n&         4.18 $\pm$ 0.10&	0.268 $\pm$ 0.005&	0.952 $\pm$ 0.074&	0.051 $\pm$ 0.006&	1.10\\\hline
135$ < l <$ 150*&   s*&       0.67 $\pm$ 0.04&	0.588 $\pm$ 0.015&	0.193 $\pm$ 0.005&	9.580 $\pm$ 0.471&	1.20\\
&      c&         4.76 $\pm$ 0.10&	0.232 $\pm$ 0.004&	0.686 $\pm$ 0.020&	0.088 $\pm$ 0.005&	1.63\\
&      n&         4.00 $\pm$ 0.09&	0.273 $\pm$ 0.005&	1.006 $\pm$ 0.095&	0.045 $\pm$ 0.006&	1.21\\\hline
150$ < l <$ 165&    s&        6.20 $\pm$ 0.27&	0.190 $\pm$ 0.005&	0.574 $\pm$ 0.012&	0.115 $\pm$ 0.005&	1.12\\
&      c&         4.78 $\pm$ 0.11&	0.229 $\pm$ 0.004&	0.666 $\pm$ 0.019&	0.092 $\pm$ 0.006&	1.65\\
&      n&         4.14 $\pm$ 0.10&	0.269 $\pm$ 0.004&	1.015 $\pm$ 0.097&	0.041 $\pm$ 0.005&	1.15\\\hline
\enddata
%\vspace{-3cm}
\tablecomments{ \label{table:f} 
The asterisk denotes a case in which the fit switched 
the thin and thick disks.}
\end{deluxetable}

%\subsection{North and South Comparison Analysis}
Given the visual differences between the north and the south shown 
in Figure \ref{fig:residualoverlays}, a quantitative comparison has been made by 
performing fits on the north and south independently. 
We employ the same thin- and thick-disk 
model but for one modification. 
The original model had difficulties fitting stars that are only in the north or only in the south; 
this is solved by fixing the value of $z_{\odot}$ 
--- the other parameter values are determined by fitting. 
We use a $z_{\odot}$ of $14.3\,{\rm  pc}$, 
the average value obtained by \citet{yanny13}, though our results are not 
sensitive to that particular choice. 
The results of these fits are shown in Figure \ref{fig:thinthicknorthsouth}, 
which shows the thicknesses of the thin and thick disks as functions of longitude. 
In these fits we found it pertinent to employ the uniform sample of 
Figure \ref{fig:footprintmatched}(a) in order to focus on the possibility of longitudinal 
variations in a crisp way. 

 Separate fits to the northern and southern samples reveal a much greater
      thickness in the north 
for both the thin and thick disks, though it is 
also the case that the fraction $f$ of the thick disk relative to the thin disk 
is also smaller in the north. 
We show an explicit illustration of this correlation in Table~\ref{table:f}, 
which lists the fit parameters for the two-disk component fits (note Eq.~(\ref{fit})) 
        to the matching north and south data samples in bins of $5^\circ-15^\circ$ 
        bins in longitude (see Figure \ref{fig:footprintmatched}(a)), including 
the normalization $N_0$, the thin-
        and thick-disk scale heights $H_1$ and $H_2$, respectively, 
        the relative fraction $f$ of the thick- and thin-disk 
        populations, and the $\chi^2/{\rm dof}$ 
for each fit. We show the fit results 
        for the sample in the north fitted by itself, the sample in the south fitted
by itself, and for the combined north-south sample. 
        In the combined north-south fits, $f$ is approximately 10\%, 
        and the thin- and thick-disk scale heights
        are 0.24 and 0.7 kpc, respectively, with a relatively good $\chi^2/{\rm dof}$. 
        When the fits are done separately for the north and south, however,
        the scale heights for both the thin and thick disks are always
        larger in the north, by about 20\%-30\%, with a corresponding decrease
        in $f$, by about 10\%-50\%.  This strong correlation
        between $f$ and the scale heights has been noted several times previously
        in the literature, and it has been speculated that
        it is a consequence of a fitting degeneracy, rather than an indication of a 
        variation in the scale heights and $f$ with position or hemisphere. 
        We refer to Figure 9 of \citet{chen01}, as well as Figure 21 
of \citet{juric08} for concrete 
        examples. 
        The combined values of the scale heights and thick-disk fraction are in good
        agreement with \citet{juric08}, which analyzed observations 
        from both the northern and southern hemispheres, finding $H_1=0.24$ kpc and $H_2=0.8$ 
        kpc with $f$ around 0.1; they, too, noted a 
        strong correlation between $f$ and the  disk scale heights. 
        Given the good $\chi^2/{\rm dof}$ of the separate
        north and south fits shown in Table \ref{table:f} 
        (to our knowledge such separate fits have not been done before), 
        and the poorer combined fits, we suggest, rather, that the difference in
        scale heights and fractions with hemisphere is a real effect. 
        Its existence supports that of a large north-south 
        asymmetry in the vertical stellar distribution in
        the vicinity of the Sun. Figure 10 shows this is not due to calibration errors. 

The N/S differences shown in Figures \ref{fig:residualoverlays}--\ref{fig:colorcalibration}
are depicted in an $(l,b,z)$, Sun-centered coordinate system.  One may also present the 
        same star-count data in an $(R,\phi,z)$ coordinate
        system based on the Galactic center, and this is shown in
        Figures \ref{fig:reftable} and \ref{fig:Phibreaking}. 
        Panels (a) and (b) of Figure \ref{fig:reftable} shows two azimuthal bins of star-count
        data from the footprint of Figure 3(a) at radial distances up
        to 0.5 kpc from the sun, and up to 3 kpc from the Galactic plane.
        The pattern of asymmetries seen in several
        panels of Figure \ref{fig:residualoverlays} is clearly present in panel (a), which combines
        several $(l,b)$ bins into one nearby $(R,\phi)$ bin.
        Panels (c)-(e) offer $(R,\phi)$ presentations of the
        asymmetries of the data in the expanded footprint of Figure \ref{fig:footprintmatched}(b). 
        While the selection function of the data points with $|z|$  has not been applied, though 
it has been in Figure \ref{fig:residualoverlays}, the data
        samples are in all cases taken from matching north and south
        regions, plotting $|z_{\rm GP}| = |z + z_\odot|$ with $z_\odot$ = 14.9 pc.
Note that the N/S thickness difference emerges as a gross feature close
to the Galactic plane in both the uniform and expanded latitude samples.

        Figure \ref{fig:Phibreaking} offers the broadest look at the data 
        arranged in Galactic $(R,\phi)$ coordinates.
        Here raw star counts, uncorrected by the selection function, are compared 
        in N/S matched samples. 
        In this figure we relax the $|z| < $ 2 kpc constraint for better
        illustration.  The circular inset at left
        portrays the plane of the Galaxy marked with the azimuthal coordinate $\phi$ and
        radial coordinate $R$, where
        ($R_\odot,\phi_\odot$) = (8 kpc, 180$^\circ$).  We have broken our 
coverage in the ($R$, $\phi$) plane into smaller regions indicated 
by the small solid dots, and we have picked out
the dots (regions) with the most star counts, as indicated by the red box, 
for explicit illustration. 
The faint large gray arcs
        indicate curves of constant $R$ in steps of 0.5 kpc. Each row of
        overlapping inset panels selects a different $\phi$ range, as marked.
        Each panel shows in black (red) the northern (southern) Galactic
        hemisphere counts as a function of $|z_{\rm GP}|$ from 0-6 kpc,
        where $|z_{\rm GP}|$ has been adjusted to account for a universal offset
        of the Sun above the plane of $z_\odot$ = 14.9 pc.  The fact that
        the north and south counts converge at large $|z_{\rm GP}|$ suggests that
        this offset is appropriate.  While the fine asymmetrical structure
        in $(l,b)$ is not visible, one still may clearly see, especially in
        the panel for $8 < R < 8.5$ kpc, $175^\circ < \phi < 180^\circ$, that there are 
        vertical oscillations in the stellar density near the Sun.
        Several other panels show an excess of counts in the south at
        vertical distances up to 1 kpc from the plane.
        We see explicitly that (i) there are more stars in the south
        over most of the footprint and (ii) the feature of
        Figure \ref{fig:reftable}(c), which drives the wave-like N/S asymmetry we have
        found, as well as the N/S variation in Galactic parameters, also
        breaks axial symmetry. It also has no analogue at slightly larger $R$. 
We encourage further studies of the 
        combined stellar and dark matter mass distributions in the solar neighborhood
        to delve into its origin. 

It should be noted that a partial explanation of these differences could come from 
a global calibration difference. 
The photometric calibration of 
the SDSS survey is done by studying adjacent regions across the footprint and 
calibrating overlapping regions together~\citep{padmanabhan08}. 
Since the entire footprint in the north is connected, it can be calibrated in a uniform way. 
However, the northern and southern regions of the footprint do not overlap, 
so that they cannot be calibrated in this manner. 
Studies of the blue-tip stars suggest that the stars are systematically redder in the south~\citep{schlafly10,yanny13}. 
To study the implications of this, a downward shift of -16 mmag for $(g-i)_{0}$ color 
was implemented in the south, 
which would redress the typical differences in reddening found by \citet{yanny13}. 
By repeating the fits with 
$(g-i)_{0}$ color 
as well as repeating the fits using distances determined with $(r-i)_{0}$ color, 
Figure \ref{fig:colorcalibration} is obtained. 
The shift in color calibration 
has a negligible effect on the thicknesses, as we would expect from our discussion in 
Section \ref{photod}. 
Although using the $(r-i)_{0}$ color relation
reveals slightly different thicknesses, 
the north-south offsets still exist, encouraging the stance 
that there is a real difference in the stellar densities between north and south. 

%figure 12
\begin{figure}[hp!]
\begin{center}
\subfloat[]{\includegraphics[scale=0.33]{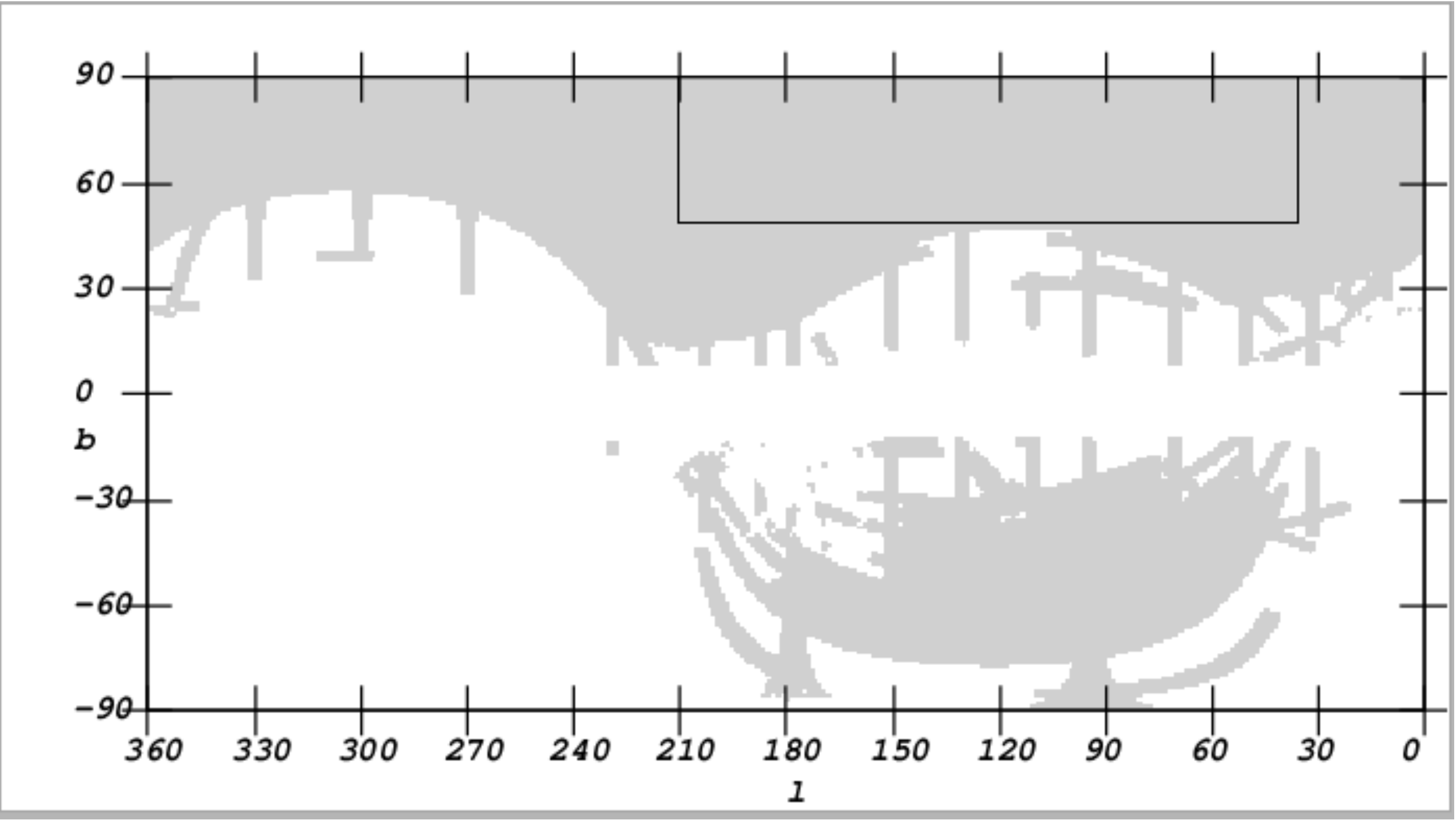}} 
\subfloat[]{\includegraphics[scale=0.33]{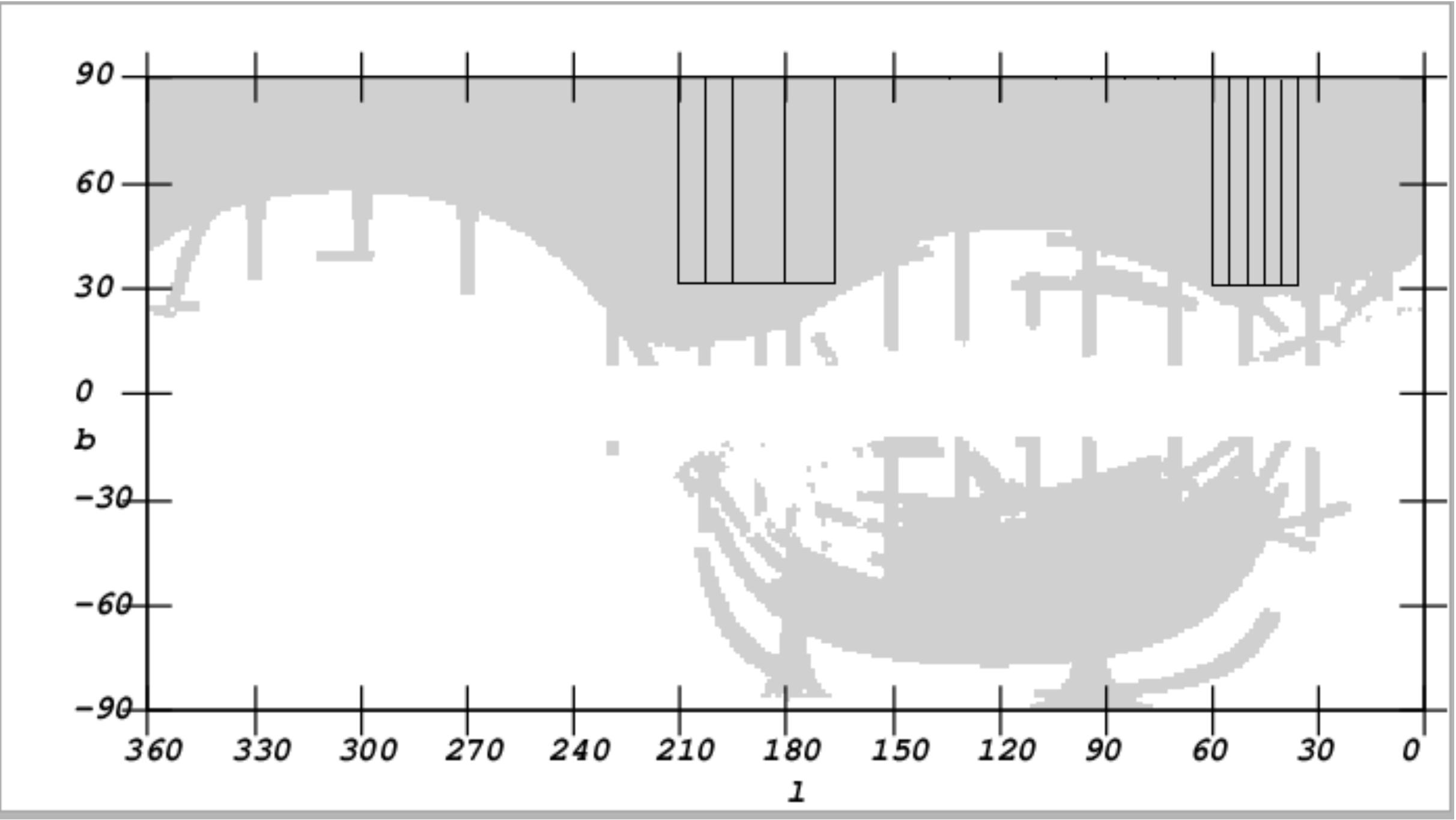}}
\caption[Sloan Digital Sky Survey footprint for north only analysis]{
The SDSS footprint~\citep{aihara11} as a map of $b$ vs. $l$. 
The blocked-in regions represent the sections this analysis studies. 
%Fig.~\ref{fig:footprintnorth}a shows 
(a) A longitude range of $35^\circ<l<210^\circ$ with latitude of $50^\circ< b < 90^\circ$. 
(b) Two regions of longitude $35^\circ <l< 60^\circ$ and $165^\circ< l< 210^\circ$ with latitude $30^\circ< b< 90^\circ$.
}
\label{fig:footprintnorth}
\end{center}
\end{figure}

\subsection{North-only Analysis: Scale Height Changes Across the Footprint}
\label{Nonlyscale}
Since more stars have been observed in the north 
we use these observations to 
study variations in Galactic parameters across the footprint. 
Once again, we have made two different types of selections. 
Both divide the footprint into slices of 
$5^\circ$, $10^\circ$, and $15^\circ$ slices in longitude, 
but each uses different latitudes. 
The first analysis uses a fixed latitude range of $50^\circ < b <90^\circ$ 
and longitude range $35^\circ < l < 210^\circ$ (Figure \ref{fig:footprintnorth}(a)), in which 
we have 1.67 million stars in total. 
The second analysis uses a fixed latitude range of $30^\circ < b <90^\circ$ but only 
looks at the longitudes $35^\circ < l < 60^\circ$ and $165^\circ < l < 210^\circ$ (Figure \ref{fig:footprintnorth}(b)). 
These regions are selected in order to have the largest latitude coverage possible, and 
we analyze 2.22 million stars in total.  
Since we use $ b >30^\circ$ in order to avoid dust effects, 
this is the maximum range of latitude allowed by our analysis. 

%figure 13
\begin{figure}[hp!]
\begin{center}
\subfloat[]{\includegraphics[scale=0.5]{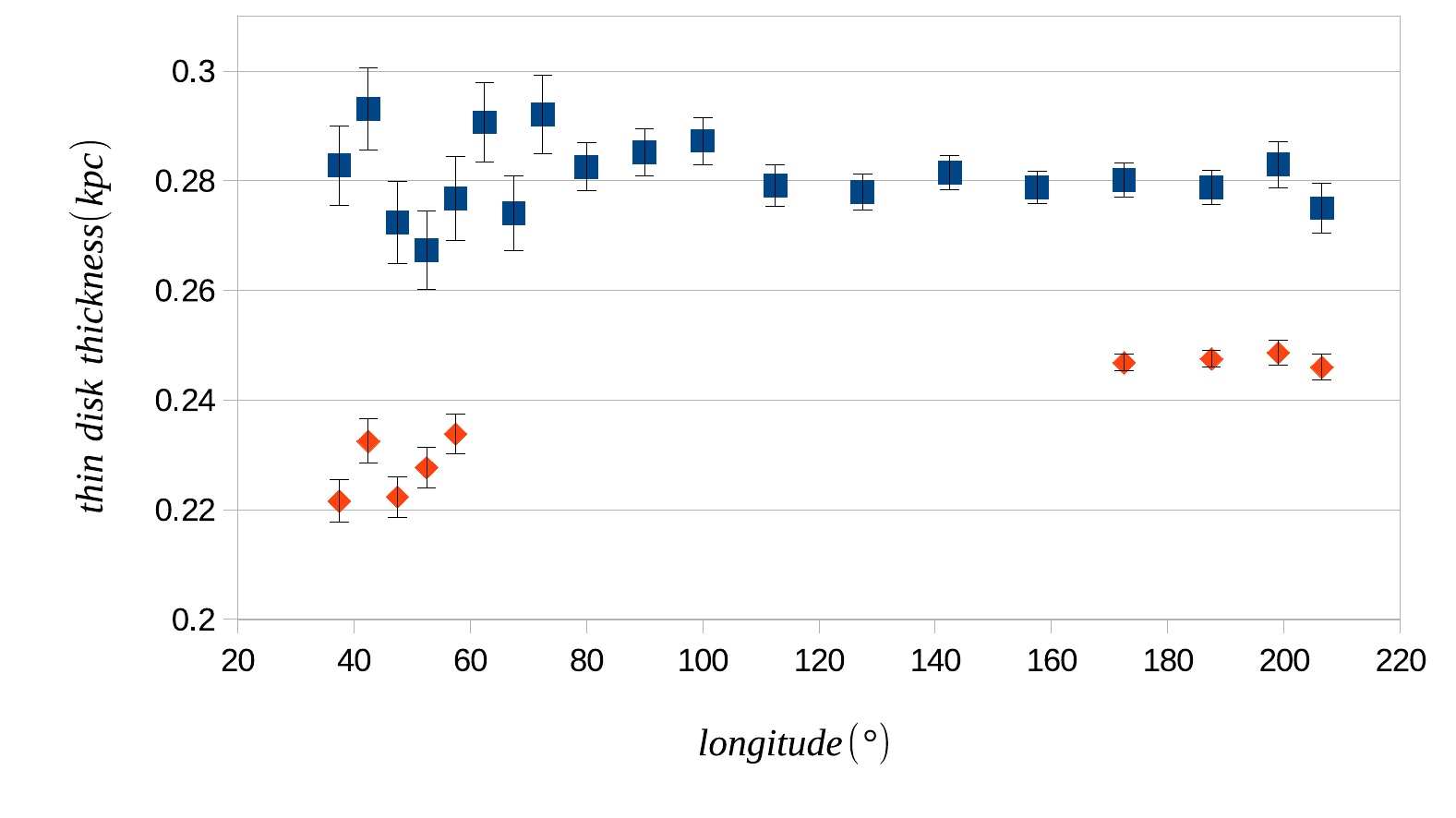}} 
\subfloat[]{\includegraphics[scale=0.5]{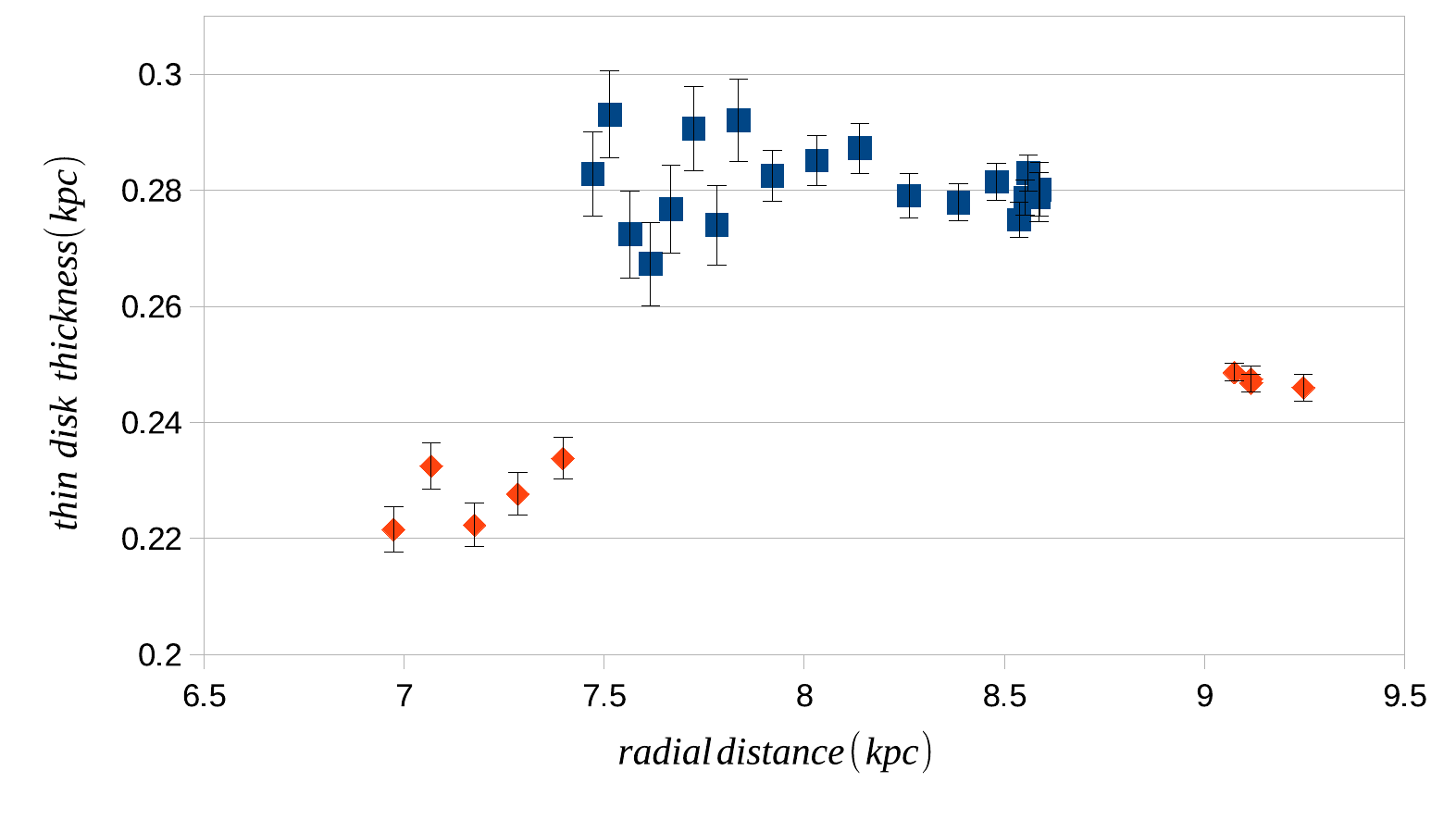}}
\caption[Thin disk thickness as a function of longitude and radial distance for the north only analysis]{
Thin-disk thickness as a function of longitude (a) and radial distance (b). The graphs overlay the values obtained from Figure \ref{fig:footprintnorth}(a) (blue squares) and the values obtained from Figure \ref{fig:footprintnorth}(b) (red diamonds). 
In (b) horizontal errors of ${\cal O}(0.5-0.7 {\rm kpc})$ in $R$ have not been included. 
}
\label{fig:thinnorth}
\end{center}
\end{figure}

%figure 14
\begin{figure}[hp!]
\begin{center}
\subfloat[]{\includegraphics[scale=0.5]{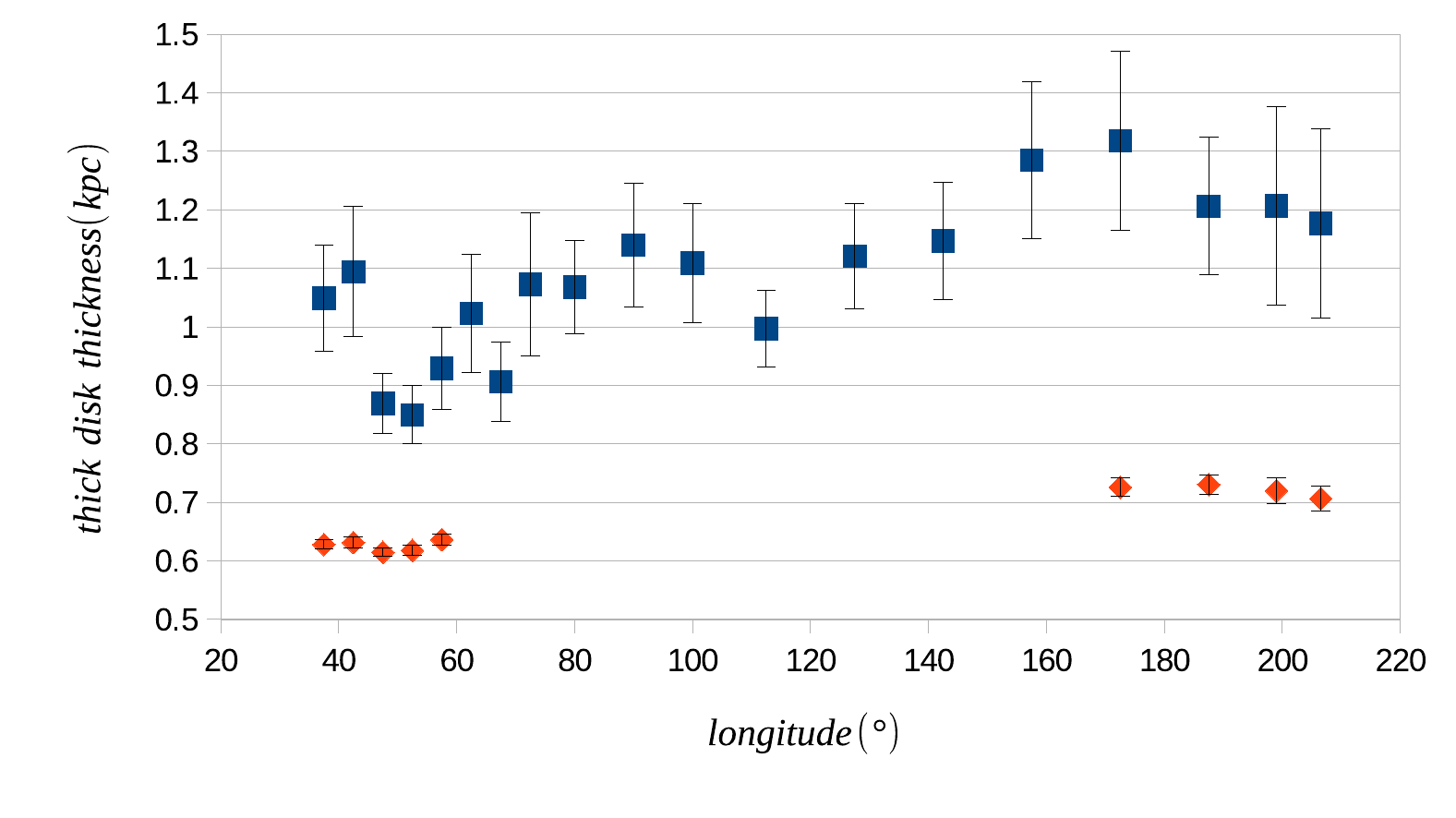}} 
\subfloat[]{\includegraphics[scale=0.5]{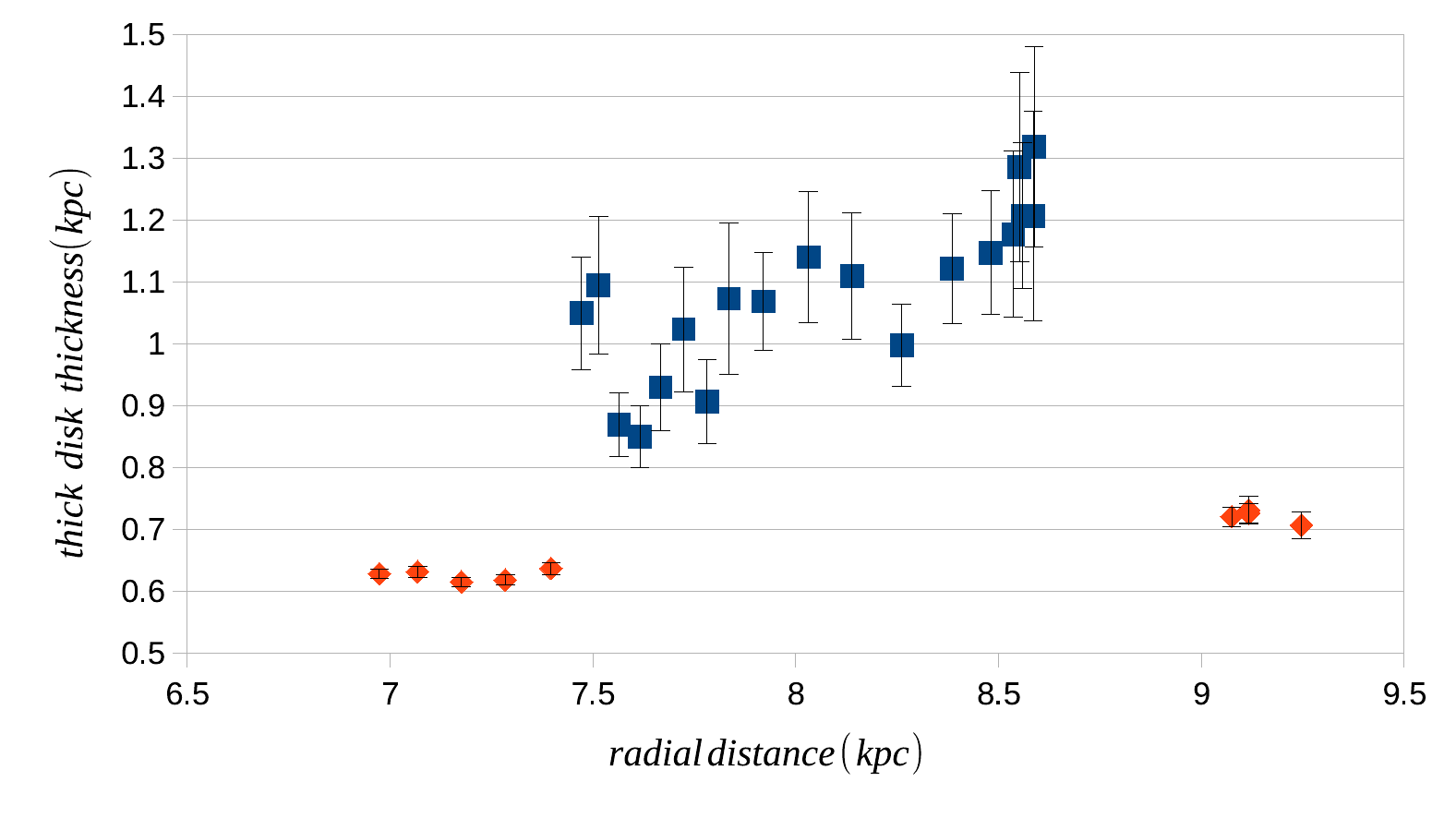}}
\caption[Thick disk thickness as a function of longitude and radial distance for the north only analysis]{
Thick-disk thicknesses as a function of longitude (a) and radial distance (b). 
The graphs overlay the values obtained from Figure \ref{fig:footprintnorth}(a) (blue squares) and the values obtained from Figure \ref{fig:footprintnorth}(b) (red diamonds). 
In (b) horizontal errors of ${\cal O}(0.5-0.7 {\rm kpc})$ in $R$ have not been included. 
}
\label{fig:thicknorth}
\end{center}
\end{figure}

%Figure 15
\begin{figure}[hp!]
\begin{center}
\subfloat[]{\includegraphics[scale=0.5]{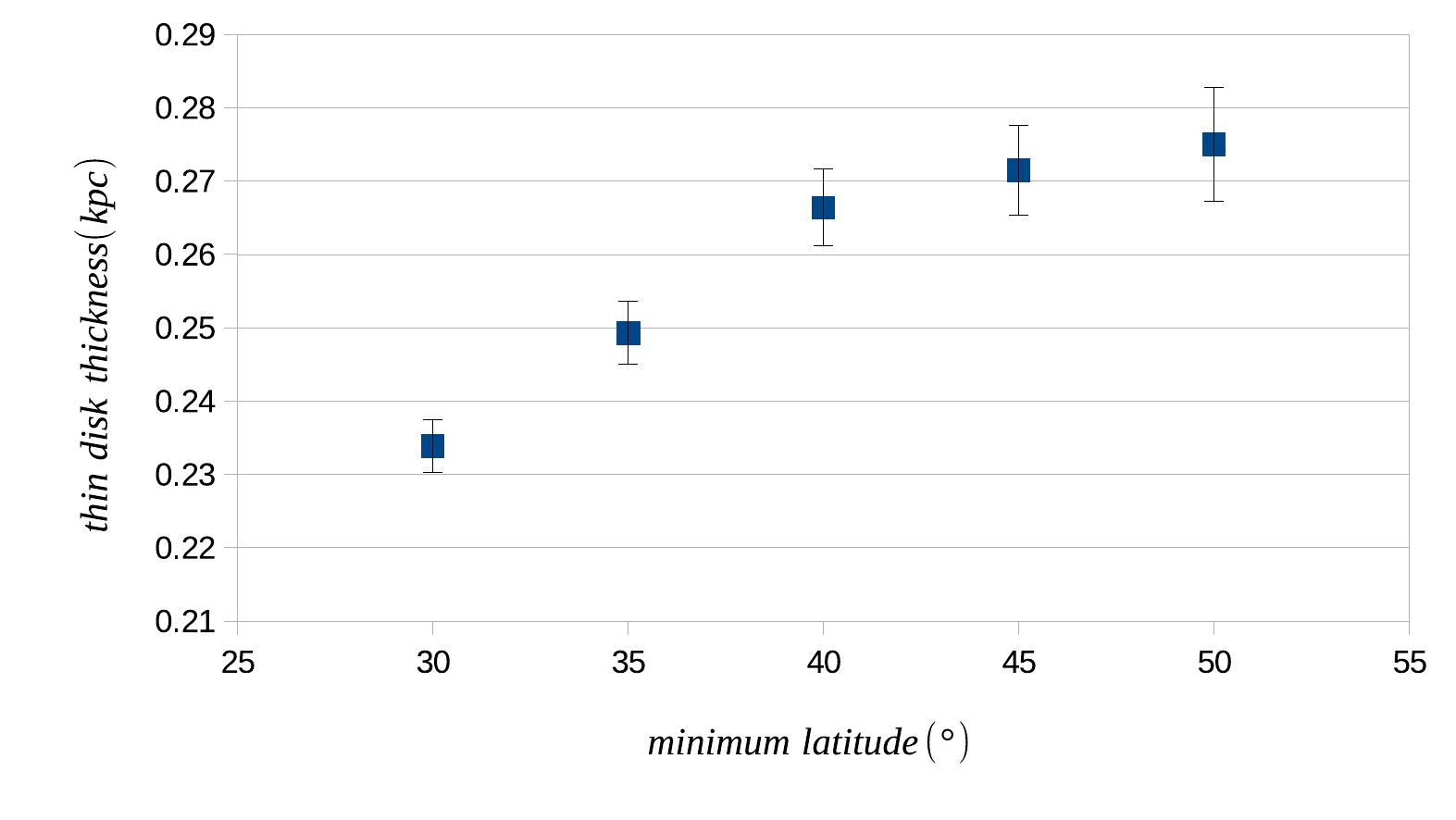}} 
\subfloat[]{\includegraphics[scale=0.5]{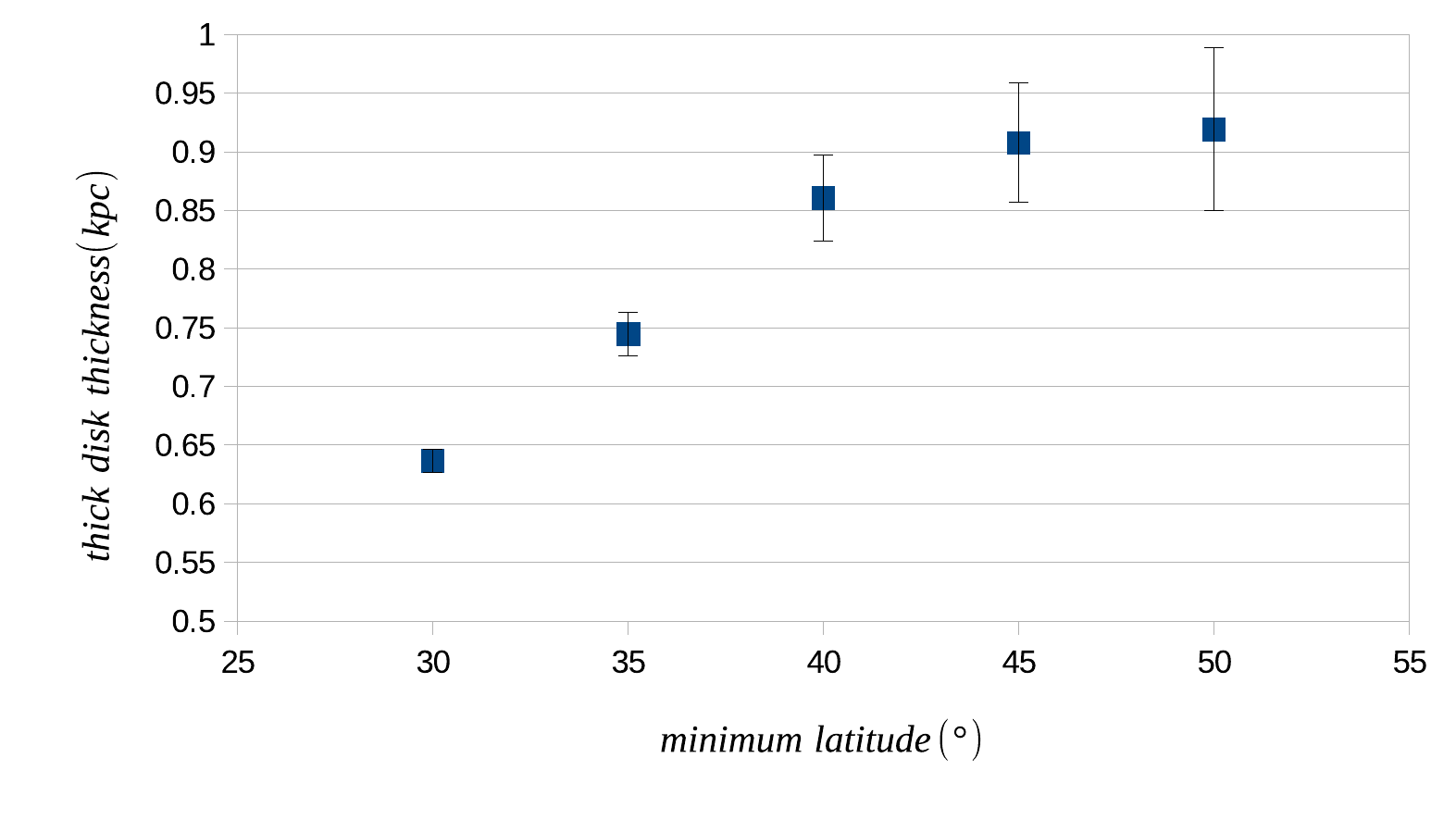}}
\caption[Thin and thick disk thicknesses as functions of minimum latitude]{
Thicknesses of the thin (a) and thick (b) disks 
as functions of minimum latitude for an analysis of the region $55^\circ<l< 60^\circ$ 
in the north with a latitude up to $90^\circ$. 
}
\label{fig:latitudechange}
\end{center}
\end{figure}

Figures \ref{fig:thinnorth} 
and \ref{fig:thicknorth} show the thicknesses of the thin and thick disks as functions of 
longitude and the in-plane radial distance $R$ from the Galactic center
for both of the regions depicted in Figure \ref{fig:footprintnorth}. 
The difference in thickness for regions 
with the same longitudes but different latitudes reveals that the 
scale heights are sensitive to the value of the lower latitude cut. We have checked
that this result persists once we repeat our fits to a data sample in which the $R$ dependence 
has been scaled out, as per the modified selection function of Eq.~(\ref{selectionR}).
It can also be observed, 
particularly for the analysis of Figure \ref{fig:footprintnorth}(b) shown in Figures \ref{fig:thinnorth} and \ref{fig:thicknorth}, 
that there is an increase in thickness, for both disks, with an increase in 
$R$. 
The value of $R$ used is the average value of $\sqrt{x^2+y^2}$ 
for the stars in that wedge that are included in the fit. 
The red points, corresponding to the inclusion of stars with latitudes down 
to $30^\circ$, sample distances further from the Galactic center at large $l$ and 
those closer to the Galactic center at small $l$. 
In a single-disk model,
% the increase in 
%thickness with increasing $R$ 
%corresponds to a tapering in the surface (mass) density of the stellar disk with $R$. 
 in particular, if $\rho(z) = \rho_0 {\rm sech}^2 (z/(2 z_0))$, then the surface mass 
density $\Sigma$ has the form $\Sigma = 4\rho_0 z_0$~\citep{spitzer42,binney08}. 
Figure \ref{fig:thinnorth} shows an increase in scale height of about 10\% over a change in 
$R$ from 7 to 9 kpc, but from our fits we determine the parameter $z_0$ to be more 
nearly constant and, moreover, 
we infer that $\Sigma$ is decreasing. 
%as well. The ratio of
%thick disk scale heights is slightly smaller than that of the thin disk scale heights;
%this, in turn, suggests that radial scale length could be shorter for the thick disk 
%population, as discussed in \citet{bovy13}. (We emphasize, though, that our assay is for
%the stars alone.)
A change in the vertical scale height with $R$ has been noticed previously by \citet{kent91}
(and in galaxies other than the Milky Way by \citet{degrijs97}), 
with an estimated scale height of $\sim 247$ pc at 
$R=8$ kpc, in good agreement with our result for the thin-disk scale height. 
Its origin has been discussed by \citet{narayan02}. 
It is important to note that this analysis 
covers the entire range of latitude from $30^\circ < b <90^\circ$; this provides us with 
more statistics than our analyses limited to higher latitudes.

In all of our results, the thin- and thick-disk thicknesses appear to depend 
on the lower limit of the latitudes included in the analysis. 
To show this effect explicitly, Figure \ref{fig:latitudechange} shows how the thickness
depends on te minimum latitude chosen, 
for a region with $55^\circ < l < 60^\circ$ and a maximum latitude of $90^\circ$. 
%how the thickness depends on the minimum latitude chosen. 
We thus see that a higher minimum latitude cut 
leads to a larger thickness, for both the thin and thick disks; the change with minimum latitude  can be as large as 20\%. 
Since we employ a selection function to eliminate geometric effects and the 
sizes of the systematic errors that could account
for such a change are ruled out (see Section 2), 
%that trivially small by comparison, 
we expect that 
this apparent change arises from one or more physical effects. 

\begin{figure}[hp!]
\begin{center}
\includegraphics[scale=0.27]{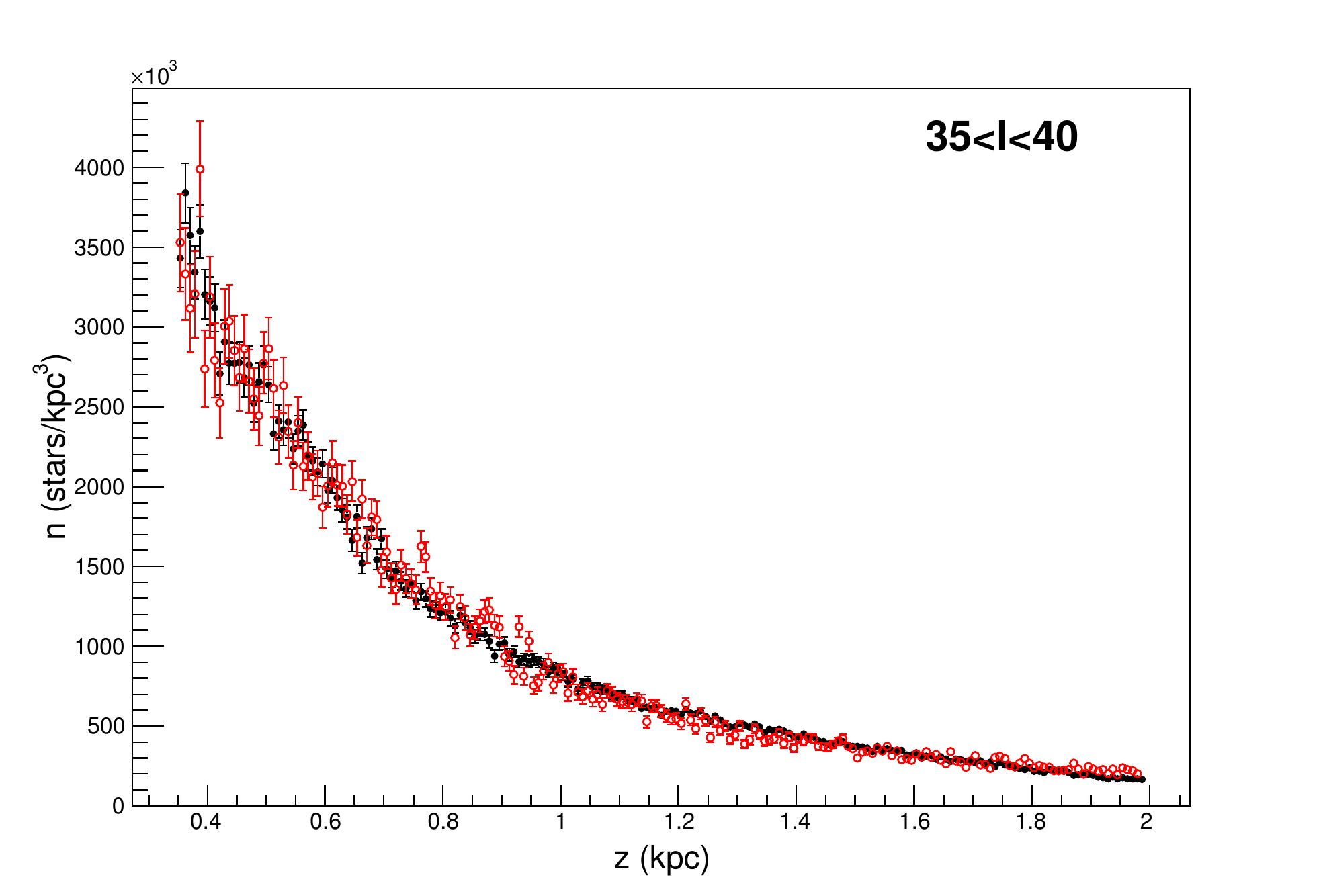}
\includegraphics[scale=0.27]{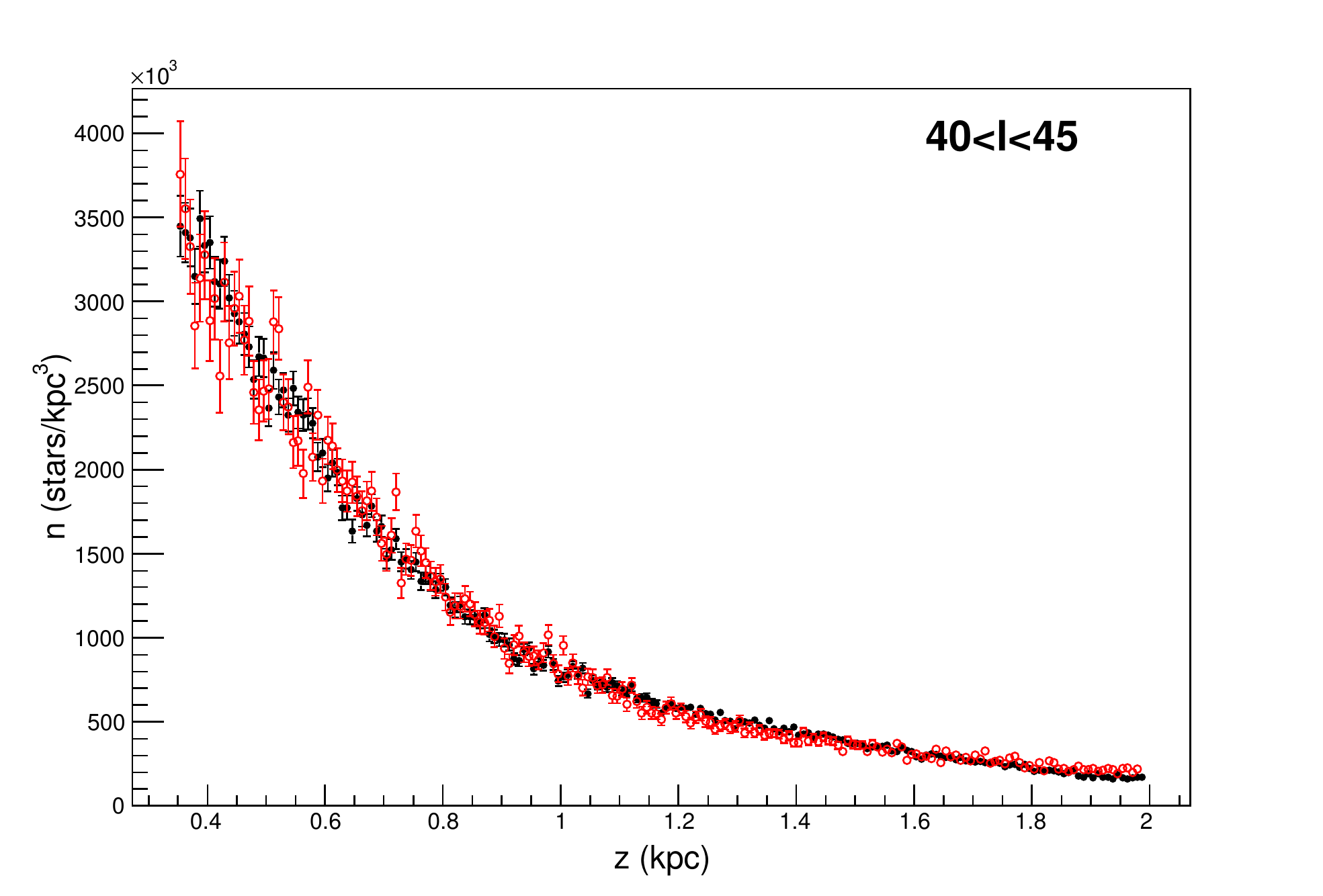}
\includegraphics[scale=0.27]{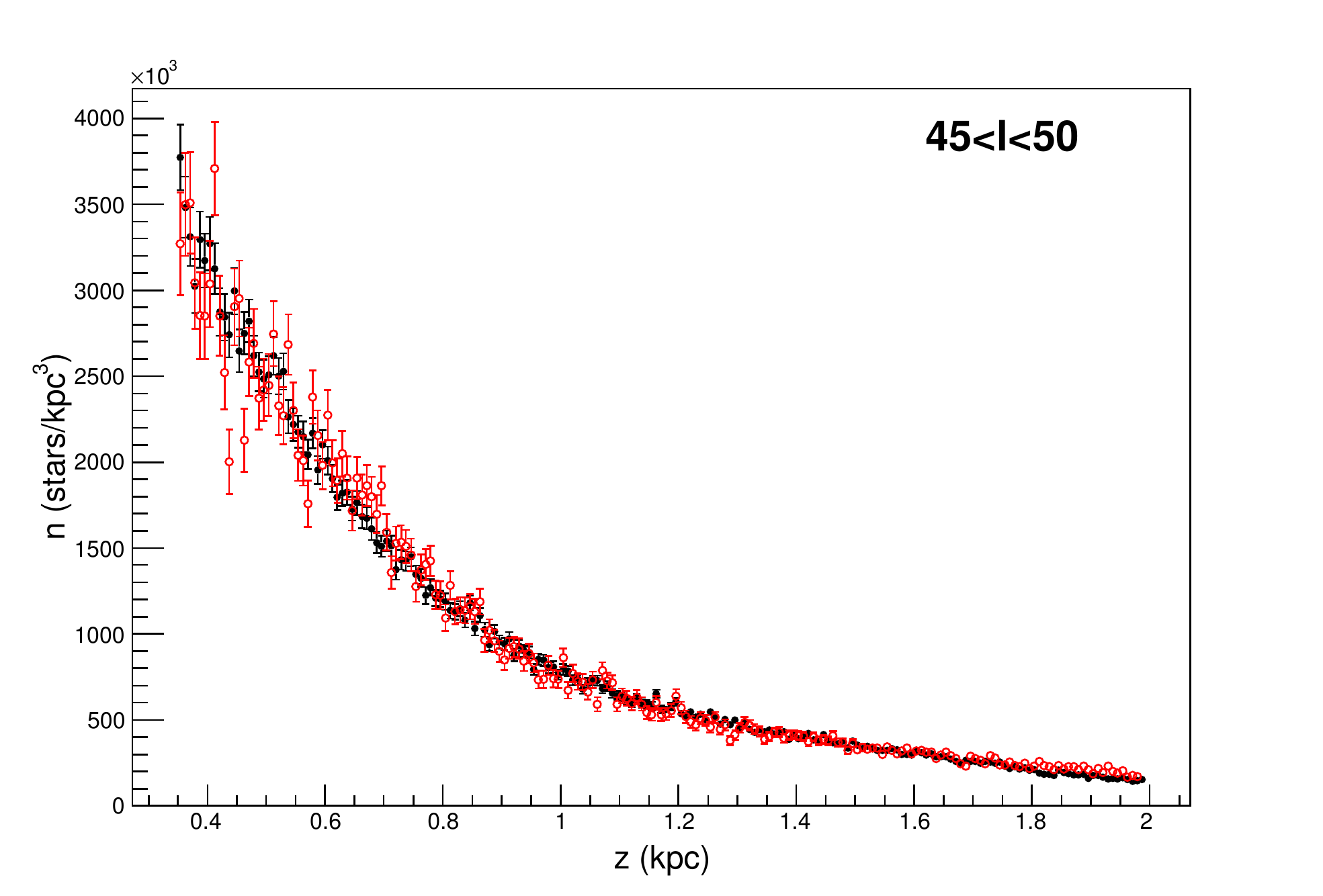}
\includegraphics[scale=0.27]{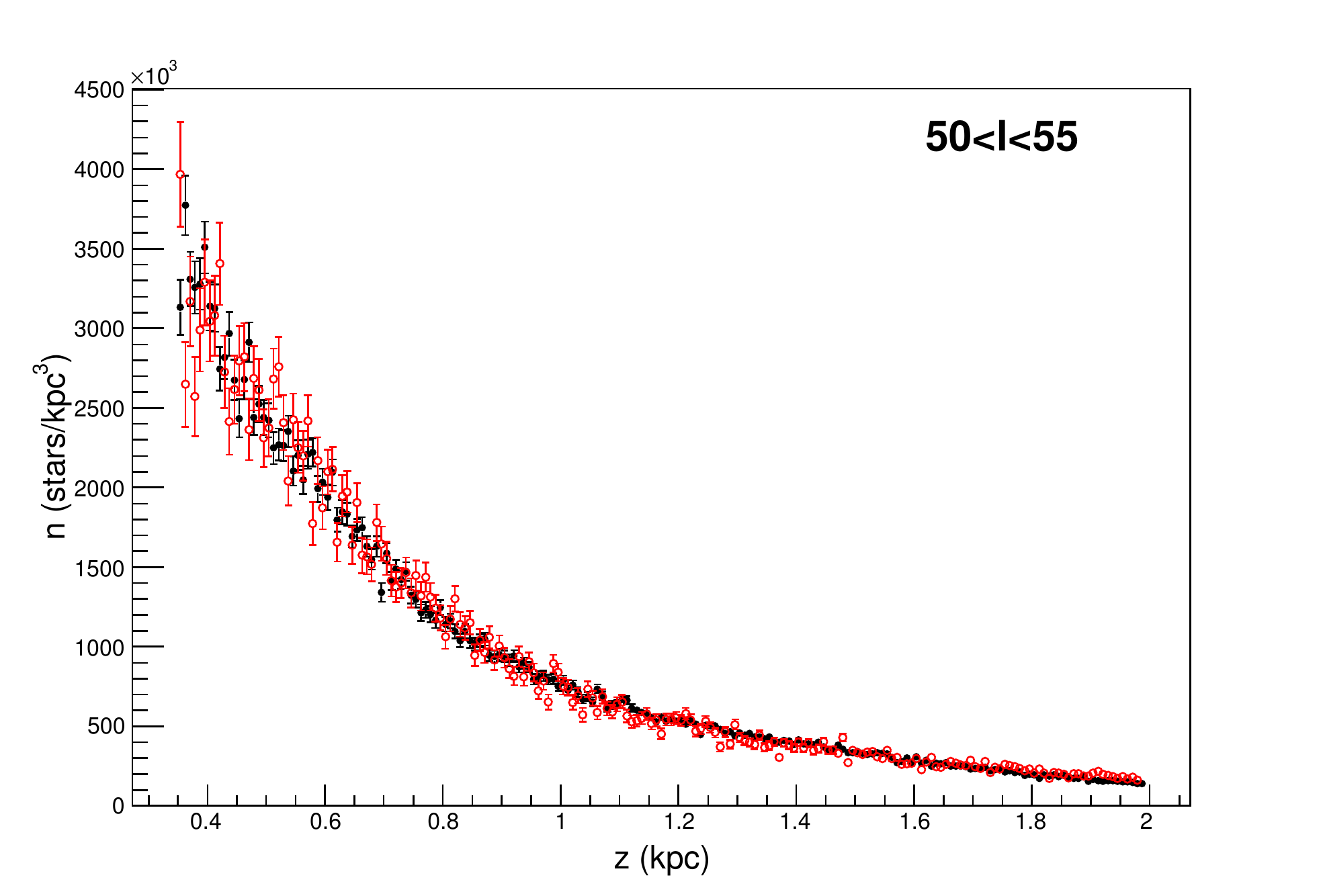}
\includegraphics[scale=0.27]{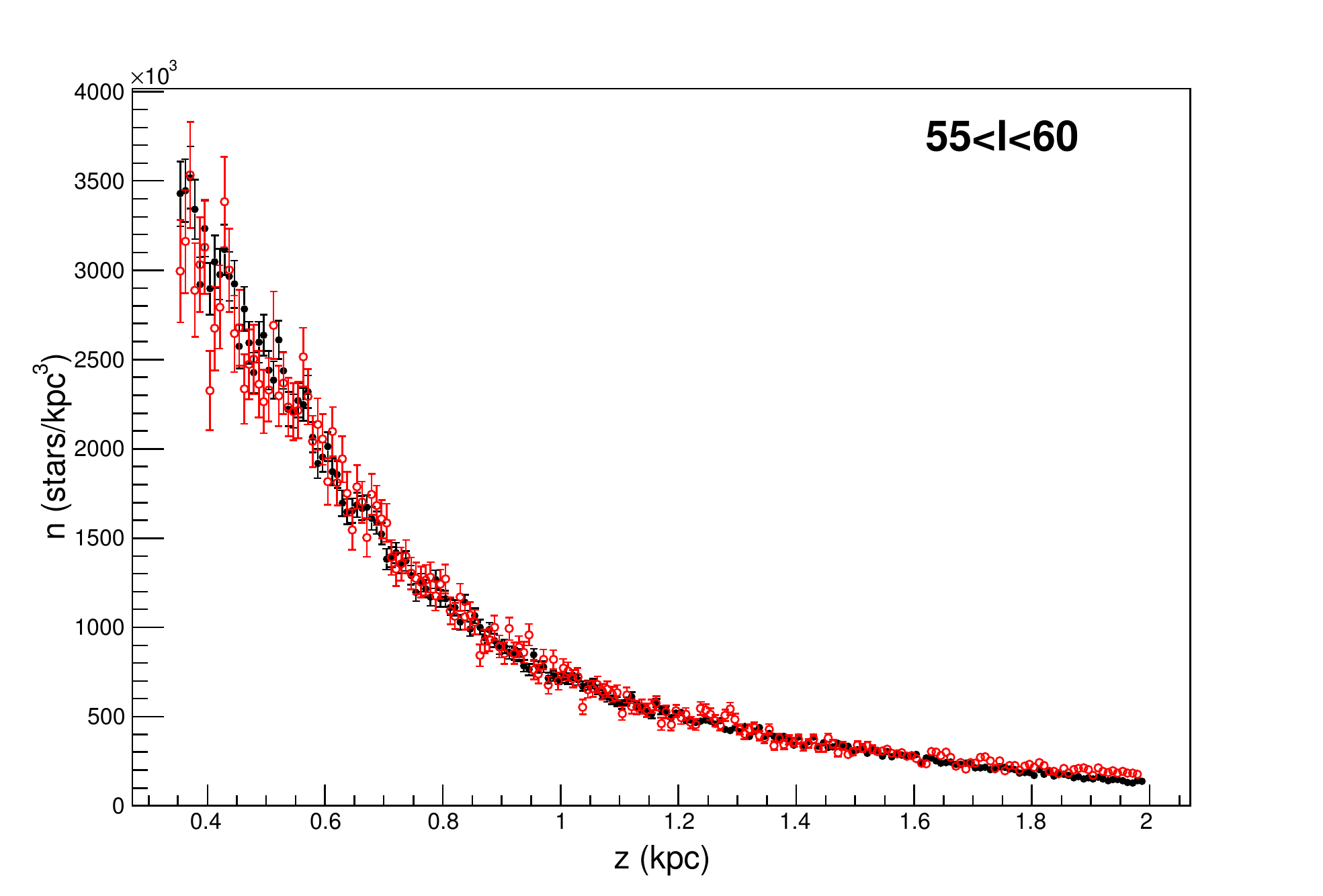}
\includegraphics[scale=0.27]{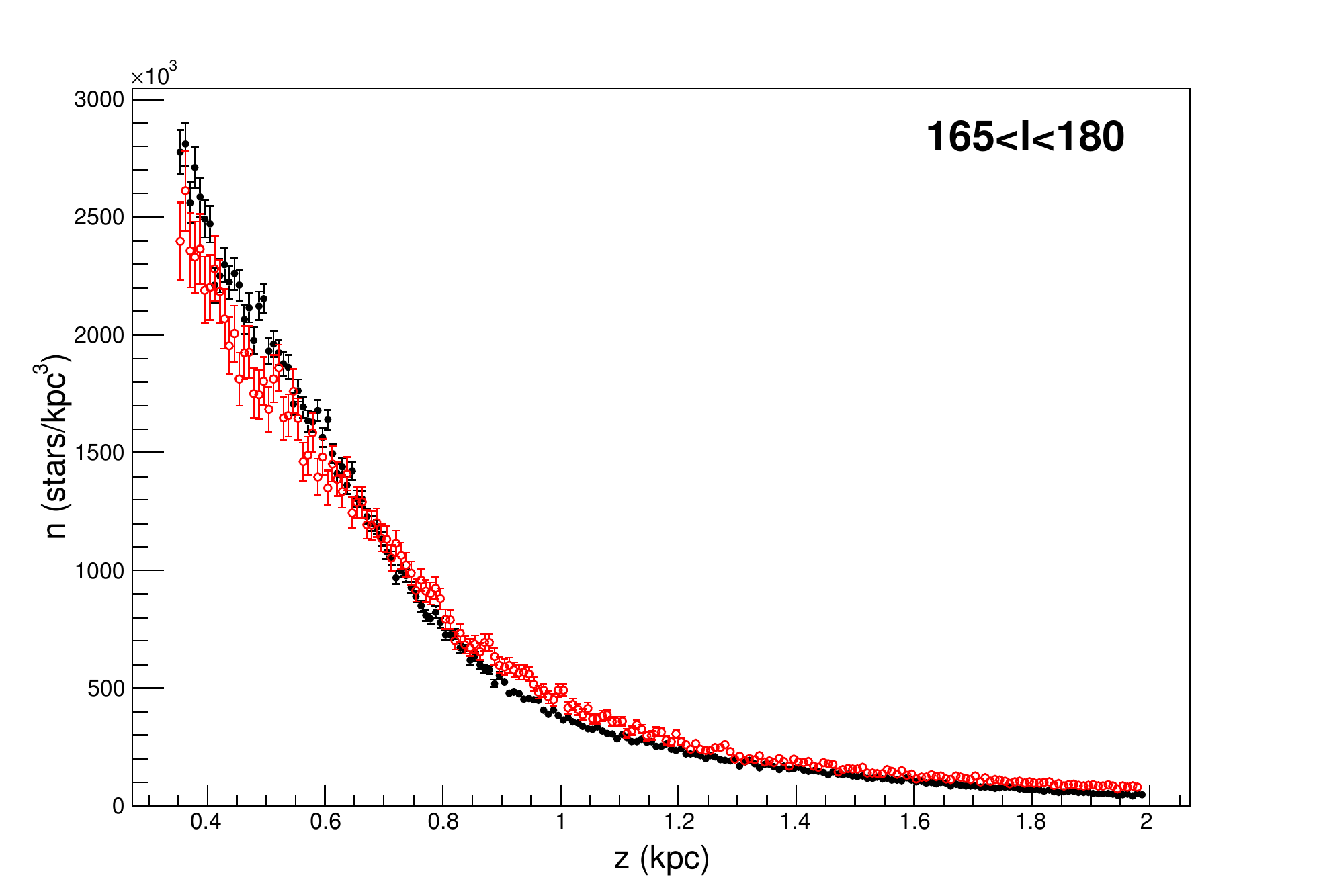}
\includegraphics[scale=0.27]{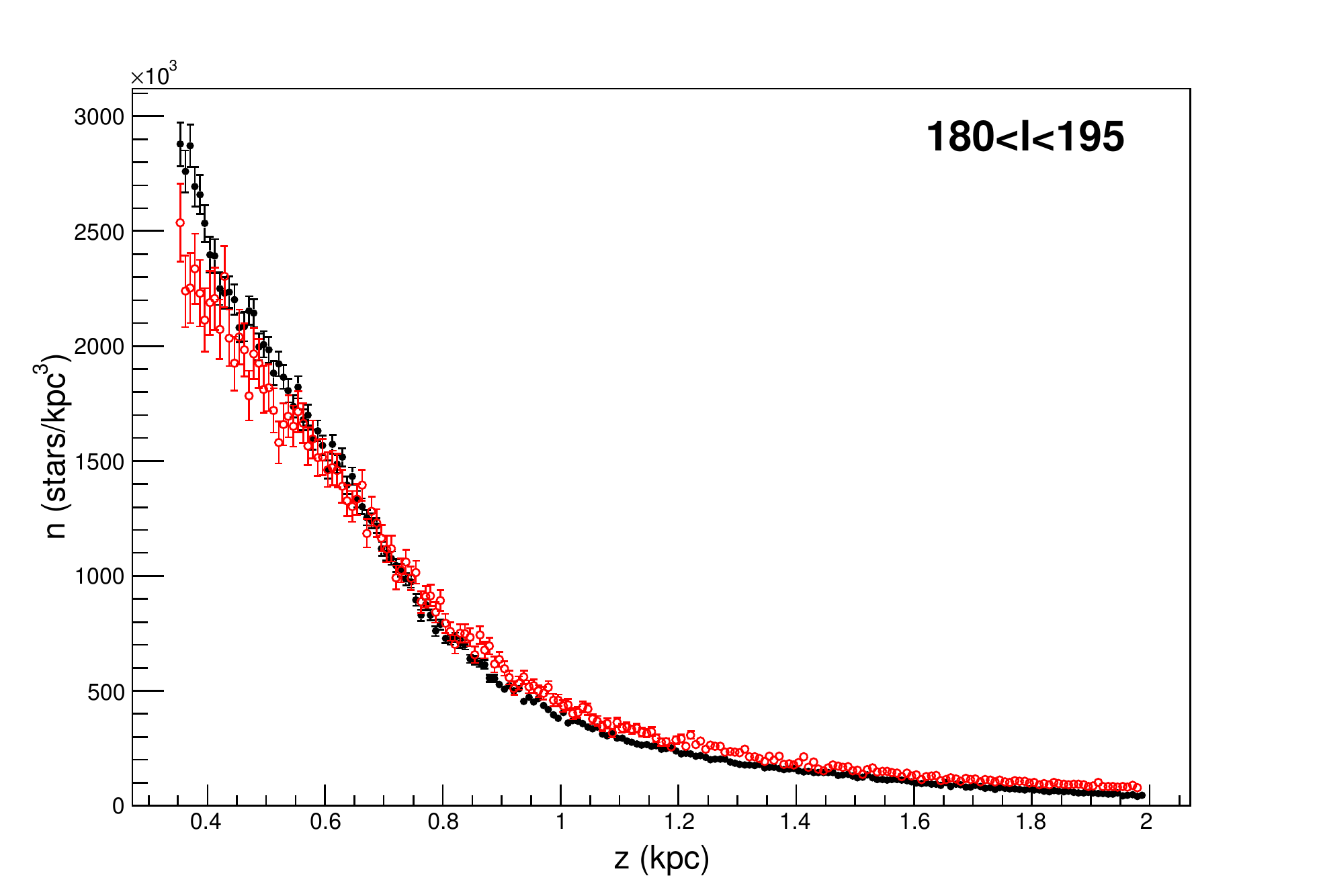}
\includegraphics[scale=0.27]{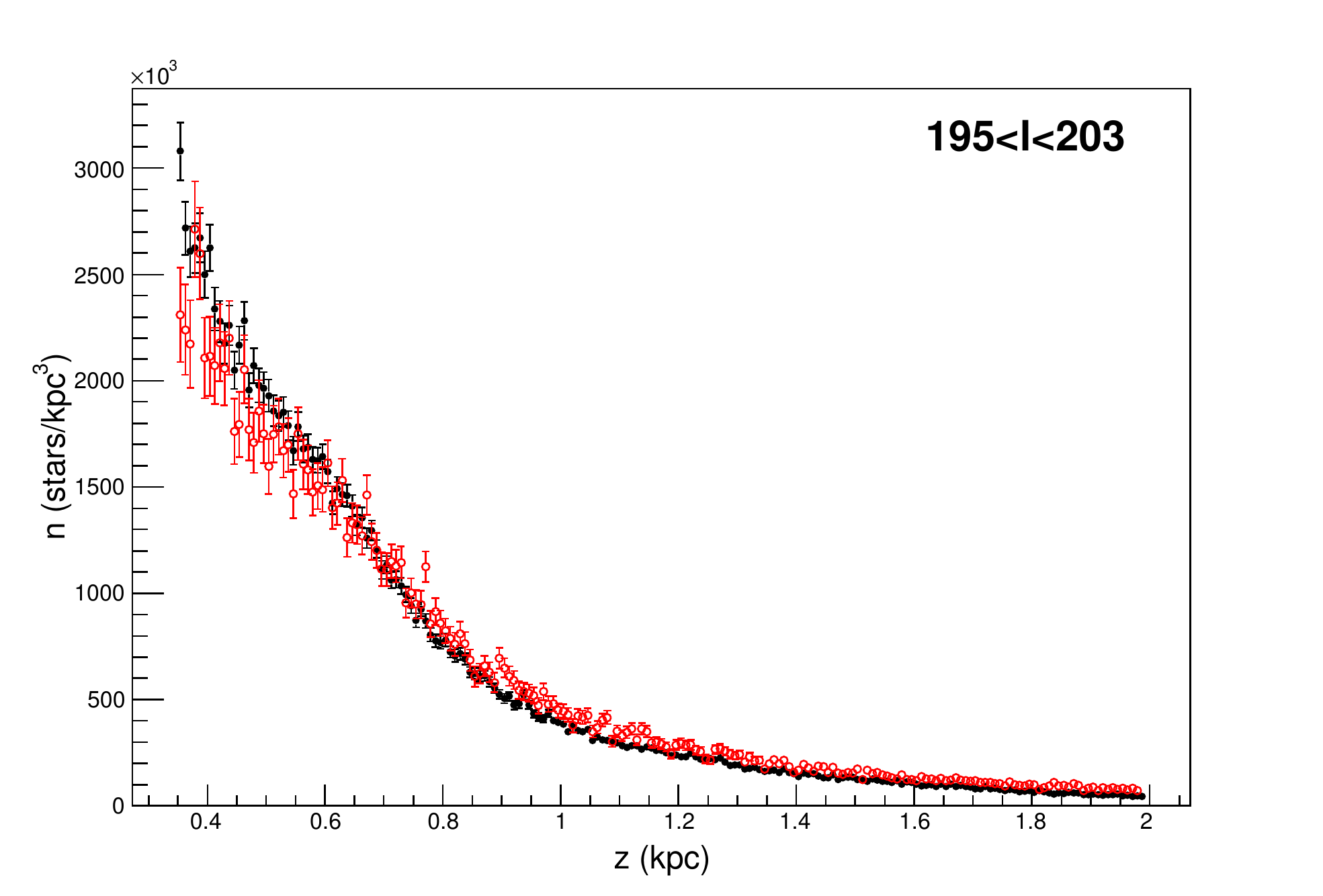}
\includegraphics[scale=0.27]{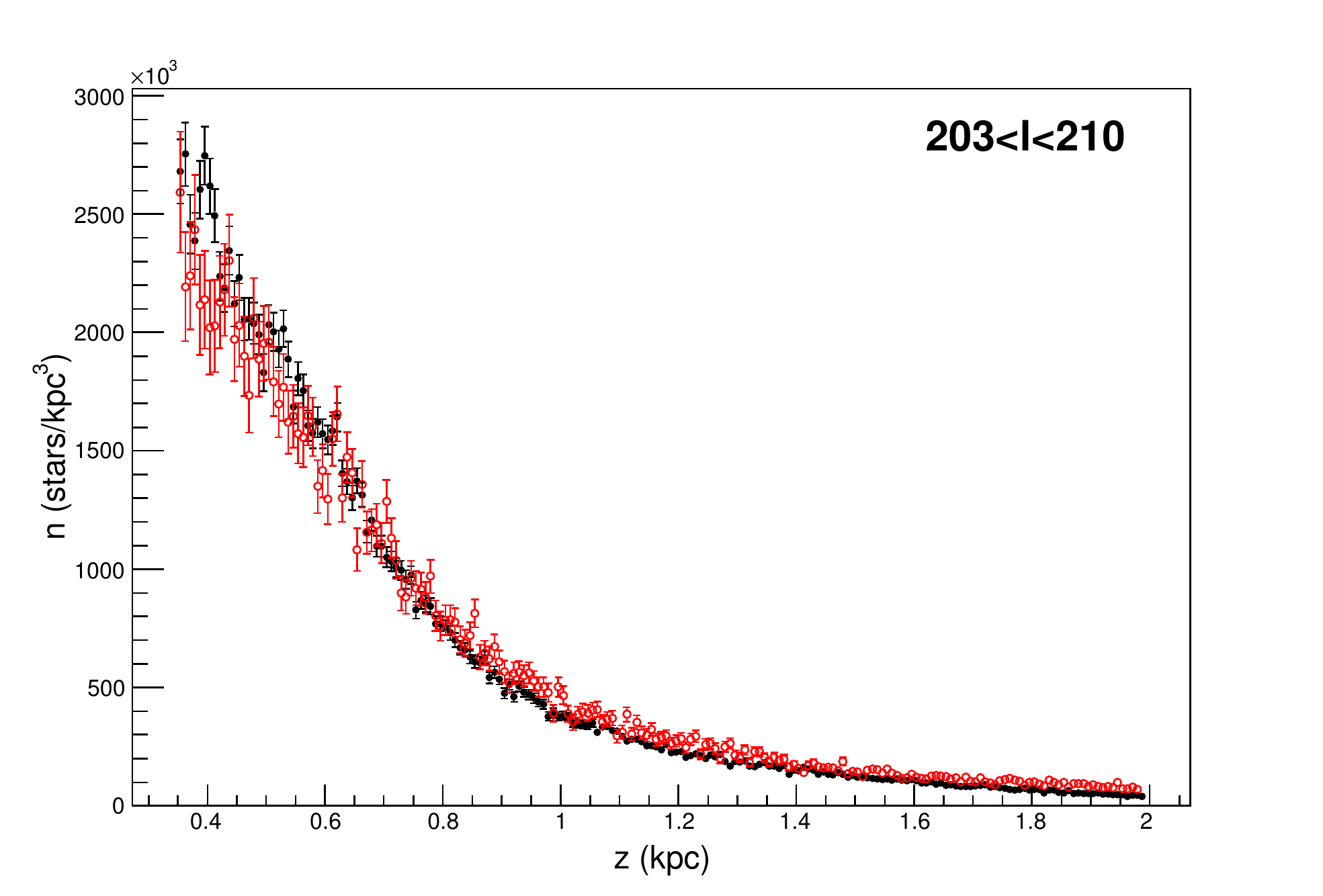}
\caption[Stellar density histogram overlay of high and low latitude]{
Stellar density histograms in the north for the regions in Figure \ref{fig:footprintnorth}(b) overlaying a region with latitude $30^\circ< b<50^\circ$ (filled, black) and a region with latitude $50^\circ< b<90^\circ$ (open, red).
}
\label{fig:latitudeoverlay}
\end{center}
\end{figure}

In order to understand this apparent change in thickness with latitude cut, 
regions of a low minimum latitude cut ($30^\circ < b <50^\circ$) and a high minimum latitude cut ($50^\circ < b <90^\circ$) 
are compared in Figure \ref{fig:latitudeoverlay}. This has been repeated 
for all the regions shown in Figure \ref{fig:footprintnorth}(b). Although the shapes do not change grossly, 
it can nevertheless be seen at high $z$ that the higher-latitude data points 
have a greater stellar density than the lower-latitude data points. 
This 
illustrates how the measured thickness could increase with increasing minimum latitude cut. 
The apparent increase in thickness could be the result of 
vertically changing stellar populations or of a change in in-plane structure. 
The manner in which a change in the selected latitude window changes 
the sampling of the Galactic plane is 
illustrated in Figure \ref{fig:selectionfunction}.
The different latitude selections also sample the stellar populations
at different average heights above the Galactic plane as shown in Table \ref{table:avgz}. Note that $\langle z_{\rm raw} \rangle$ 
determines the average in $z$ of the raw stellar number counts --- no selection function has been applied. 
We observe that the approximate difference in $\langle z_{\rm raw} \rangle$ in the two latitude samples is about
$0.5$ kpc, with the difference increasing slightly, to about $0.6$ kpc, at large $l$. Since we expect the 
metallicity of a selected star to be smaller as its value of $z$ grows larger, noting that this has been 
established 
in a recent spectroscopic study of red giants~\citep{hayden14}, we expect that the vertical changes in the metallicity of
the stellar populations should play a role to some extent. 
%Figure 16

\tabletypesize{\footnotesize}
\begin{deluxetable}{c|cc|ccc}
\tablecaption{Quantifying the Effects of Latitude Selection in the North-only Analysis
\label{table:avgz}}
\tablehead{
\multicolumn{1}{c} {$l$ range (deg)} &
\multicolumn{1}{c} {$b$ range (deg)} &
\multicolumn{1}{c} { $\langle z_{\rm raw} \rangle$ (kpc)} &
\multicolumn{1}{c} { $z$ range (kpc) } &
\multicolumn{1}{c} { $\Delta$[Fe/H] (dex)} &
\multicolumn{1}{c} { $\Delta M_r$ (mmag)}
\\
}
\startdata
\hline 
35 $< l <$ 40&      30 $< b <$ 50&        1.57 &  0.35 $< z <$ 0.5&        0.01 & -10\\
&      50 $< b <$ 90&         2.09 &  0.8 $< z <$ 1.3&        -0.08 & 80 \\
& &  &  1.7 $< z <$ 2.0&         -0.49 & 450 \\
\hline
40 $< l <$ 45&      30 $< b <$ 50&         1.55 &      0.35 $< z <$ 0.5&        -0.04 & 40 \\
&      50 $< b <$ 90&        2.10  &  0.8 $< z <$ 1.3&        -0.1 & 100 \\
& & &  1.7 $< z <$ 2.0&         -0.6 & 500 \\
\hline
45$ < l <$ 50&      30 $< b <$ 50&         1.54 &    0.35 $< z <$ 0.5&        -0.03 & 30\\
&      50 $< b <$ 90&       2.08 &      0.8 $< z <$ 1.3&        -0.05 & 50 \\
& & &      1.7 $< z <$ 2.0&        -0.7 & 600 \\
\hline
50$ < l <$ 55&      30 $< b <$ 50&        1.53 &      0.35 $< z <$ 0.5&       0.05 & -50 \\
&      50 $< b <$ 90&       2.07 &      0.8 $< z <$ 1.3&        -0.06 & 60 \\
& & &      1.7 $< z <$ 2.0&        -0.53 & 480 \\
\hline
55$ < l <$ 60&      30 $< b <$ 50&        1.51   &  0.35 $< z <$ 0.5&       0.05 & -50 \\
&      50 $< b <$ 90&       2.07 &      0.8 $< z <$ 1.3&        -0.06 & 60 \\
& & &      1.7 $< z <$ 2.0&        -0.62 & 550 \\
\hline
165$ < l <$ 180&    30 $< b <$ 50&        1.29 &    0.35 $< z <$ 0.5&       -0.3 & 280 \\
&      50 $< b <$ 90&      1.95 &    0.8 $< z <$ 1.3&        -0.48 & 440 \\
& & &    1.7 $< z <$ 2.0&        -1.32 & 1010 \\
\hline
180$ < l <$ 195&    30 $< b <$ 50&         1.29 &    0.35 $< z <$ 0.5&        -0.44 & 410 \\
&      50 $< b <$ 90&       1.95 &    0.8 $< z <$ 1.3&        -0.46 & 420 \\
& & &    1.7 $< z <$ 2.0&        -1.4  & 1100 \\
\hline
195$ < l <$ 203&    30 $< b <$ 50&        1.30 &    0.35 $< z <$ 0.5&       -0.5 & 500 \\
&      50 $< b <$ 90&       1.95 &    0.8 $< z <$ 1.3&        -0.34 & 320 \\
& & &    1.7 $< z <$ 2.0&        -1.26 & 977 \\
\hline
203$ < l <$ 210&    30 $< b <$ 50&        1.28 &    0.35 $< z <$ 0.5&       -0.29 & 280 \\
&      50 $< b <$ 90&        1.92 &    0.8 $< z <$ 1.3&        -0.38 & 360 \\
& & &    1.7 $< z <$ 2.0&        -1.3 & 1000 \\
\hline
\enddata
\end{deluxetable}

We now quantify the latitude-dependent difference in shape in 
each panel of Figure \ref{fig:latitudeoverlay}. 
To do this, we have determined the metallicity shift required to minimize the shape differences 
between the high- and low-latitude 
samples. Applying the selection function to each latitude sample, we 
multiplied the number of counts in each  bin in the 
high-latitude sample by an overall scaling factor such that the total counts for 
each latitude sample are the same. This has been 
done to permit a comparison of the vertical distribution of the stars 
regardless of their net number. 
We then examined the three different regions in $z$, in which we have seen 
shape differences, separately. 
In each one we executed a $\chi^2$ test to determine 
whether the shapes of the 
two latitude samples are the same, computing, namely, 
\begin{equation}
\chi_{b}^2 = \sum_{i=1}^{N_{\rm bin}} \frac{\left(n_i^{>} - n_i^{<}\right)^2}{n_i^{<}} \,,
\end{equation}
where $n_i^>$ and $n_i^<$ are the numbers of stars in the (rescaled) high-latitude and
low-latitude samples in bin $i$ of $N_{\rm bin}$ equal-width bins. We then modified 
the metallicity of the higher-latitude sample by steps of -0.01 
until $\chi_b^2$ is  minimized. 
We repeated this for each region in $z$ and then repeated the entire process 
for each slice of longitude. 
The results of this analysis 
are shown in Table \ref{table:avgz} as a metallicity shift $\Delta$[Fe/H] in dex and as a shift
in the intrinsic magnitude $\Delta M_r$ in mmag as per Eq.~(\ref{ppgmi0}). 
We can see that the shifts in 
$\Delta M_r$ are typically much larger, particularly at large $z$, than 
the photometric calibration 
shifts of less than 10 mmag discussed by \citet{finkbeiner16} and thus speak to a physical effect. 
(We have also checked 
that similar features emerge when we repeat our analysis in $(r-i)_0$ color.) 
These shifts are largely, but not completely,  
explained by the common paradigm of two Galactic disks ---
a thin disk with a [Fe/H] of -0.3 and a thick 
disk with a [Fe/H] of -0.8 with different scale heights. 
  Different $(l,b)$ bins sample different ratios of thin- and thick-disk stars, and thus our 
  approximation that all stars have a [Fe/H] of -0.3 is more incorrect as $|z|$ moves further
  above 0.5 kpc. 
Interestingly, too, the manner in which the metallicity changes, that it 
decreases substantially at large $z$, 
is also consistent with the trend found in spectroscopic studies~\citep{hayden14}, 
though the change is also larger in 
the large $l$ samples at large $z$ than the typical 
shifts observed in those studies~\citep{hayden14}. 
It is intriguing that a simple thick- and thin-disk paradigm
        is not enough to explain the very large shifts seen for $1.7 < |z| < 2.0$ kpc, toward
        the anticenter ($165^\circ < l < 210^\circ$), 
and here a significantly different stellar population,
        with lower metallicity than the thick disk may be present in significant amounts.  This
        is worth exploring further. It is also possible that the vertical structure changes 
as one moves across the Galactic plane. 
We anticipate that the rich spectroscopic data sets 
to emerge from the {\it Gaia} mission~\citep{gaia16a,gaia16b} 
will help to resolve the origin of the
effect we have found regarding the 
change in the vertical distributions of stars with latitude.
We nevertheless view photometric studies of the sort we have pioneered here 
as being of continuing utility, because they serve as an 
efficient way of 
discerning what the most interesting 
regions of the sky might be for detailed spectroscopic studies.

\section{Summary and Future Prospects} 

We have used a photometric sample of up to 3.6 million K and M dwarf stars from the SDSS to 
study the structure of the Galactic disk in the vicinity of the Sun. 
By selecting different regions of Galactic latitude and longitude, we have been able 
to study changes in structure as a function of the stars' location within 
and above the Galactic plane. As shown in Figure \ref{fig:histogramsmatchedset}, 
the stellar number count distributions are remarkably smooth across the footprint, though 
we have discovered significant changes in their distribution nonetheless. 
Our sample of red, main-sequence stars are the longest-lived stars, so that we would expect
its one-body distribution function to be the solution of a collisionless Boltzmann equation, 
which describes the interaction of a star with a mean-field mass distribution. 
We have presumed this solution to be both axially and mirror (north-south) symmetric in 
adopting a ${\rm sech}^2 (z/2H_1)$ form for its spatial projection in $z$, $n(z)$, as
we have used in Eq.~(\ref{fit}). With this expectation, 
we have analyzed the stars and their distribution through matched observations 
in the north and the south, 
as well as through observations made in the north only. The departures we have 
found break these supposed symmetries, though we cannot definitely say whether this is the
result of recent dynamical interactions of the disk stars with external agents or 
because of other internal effects, such as the presence of the spiral arms or the Galactic bar, 
so that they do not hold precisely. 

Our combined north-south analysis 
reveals a difference in the stellar densities north and south, providing further evidence for 
a vertical wave in number counts found by \citet{widrow12} and \citet{yanny13}, 
as well as circumstantial 
evidence for dynamical symmetry breaking of the presumed integrals of motion.  
       We have presented, in Figure \ref{fig:residualoverlays} 
 and Table \ref{table:f}, quantitative measures of the
        differences in star counts in matched areas above and below the plane of the Milky Way
        as a function of distance in  $(l,b)$ bins of ${\cal O}(10^\circ)$.  
        We find significant offsets
        from north-south symmetry 
        in individual bins of up to 20\% in star counts, even after correcting
        for the location of the Sun at some 14.9 pc above the plane.  These differences cannot
        be explained solely by misestimations of the metallicities of the stars, because 
the differences
        are too large and oscillate in $|z|$ with a vertical period of about $400$ pc.

We have also studied the evolution of the north-south 
differences with longitude and find that they 
become more pronounced as $l$ approaches $l \approx 180^\circ$. 
 Panels (c)-(d) in Figure \ref{fig:reftable} (and Figure \ref{fig:Phibreaking}) indicate 
variations 
%part                 of a wave-like oscillation 
in the azimuthal, 
                in addition to the vertical, direction.  
Although we have observed vertical wave-like features (see Figure 8), 
such are not apparent in the breaking of axial symmetry we observe. 
A close look at Figure 12 reveals, rather, an impulse localized within 
5 degrees 
of the Galactic anticenter. 
This is intriguingly reminiscent of the numerical study of the impact of the 
Sagittarius dSph 
galaxy with the Galactic disk by \citet{purcell11}. 
%Purcell et al. 2011. 
%Though this
%                azimuthal oscillation cannot be traced
%                for a whole period (only for about 20 degrees or 1/3
%                radian), its amplitude may thus be estimated at roughly 
%                 1/3 of the 0.5-1.5 vertical amplitude, or 0.2-0.5
%                kpc.
A comparison of the disk thicknesses, north and south, also reveals  
that the thicknesses are larger in the north, though this is also correlated with a 
smaller thick-disk fraction in the north. 
These results remain after correcting for possible color calibration offsets and 
repeating the analysis with $(r-i)_{0}$ color. 

In addition, we have used our comparative analysis 
to determine the inferred location of the Sun above the Galactic plane 
across the footpoint: we can say %definitively 
that 
this distance, $z_\odot$, changes across the footprint, speaking 
possibly to ripples in the stellar density across the plane. 
     No significant correlation is seen between sightlines toward spiral arms where the
     stellar spiral arms of the Galaxy appear out of the plane by up to several hundred parsecs
     and the wave-like overdensities we have seen at distances of $> 500$ pc --- but the data
     are very limited: such a correlation cannot be ruled out.

Studying the observations in the north exclusively admits the possibility of studying the effects of changing
latitude, allowing the analysis of observations up to $b=90^\circ$ with improved statistics. We have found that the
determined thicknesses, for both thin and thick disks, 
do depend on the value of the minimum latitude analyzed. 
%We have also shown that the thin disk thickness grows larger away from the Galactic center, which 
%is consistent with a decrease in surface density of the disk as one moves away from the Galactic center, as noted
%elsewhere~\citep{kent91,narayan02}. 
An increase in thickness was also found with increasing lower latitude cut, which is
likely representative of vertically changing stellar 
populations or of changes in the in-plane structure. 
Finally we have compared the shape in the stellar distributions with vertical height for different 
latitude windows, finding definite changes in shape that we think are reflective of either or 
both stellar population changes and/or changes in the in-plane structure.

A variety of systematic effects have been considered in this study. 
The effects of dust are minimized by employing dust corrections and restricting our analysis to 
out-of-plane sightlines with $|b|>30^\circ$. We have also performed a photometric test
for giants and have found that their infiltration into our analysis sample is 
completely negligible. The possibility of a color 
calibration effect that could be different in the north and south has also been 
studied explicitly and determined to be insignificant. 

Our error in converting from stellar colors to distance is a combination of that in the 
photometric
        calibration in magnitude and color and of metallicity misestimation, noting that
we have used [Fe/H] of the thin disk 
for a mixed population of thin- and thick-disk stars in the absolute
        magnitude calculation.  
The distance errors from calibration and color error have been
        estimated to be approximately 
10\% rms, or 0.2 in absolute magnitude~\citep{juric08}, and we have refined our
distance assessment using red globular clusters. As for the metallicity error, 
we reiterate that 
we fix [Fe/H] to be -0.3, i.e., we assume a 100\% thin-disk population 
for our photometric metallicity correction. An additional error would 
appear if we were, rather, sampling a 100\% thick-disk population with [Fe/H] of -0.8.
       The empirical effects of sampling different ratios of thin- and thick-disk populations
        stars in different $(l,b)$ bins are tabulated in Table \ref{table:avgz}. 
We leave the exercise of 
``inverting''
        the results of this table 
to deduce the thin-disk and thick-disk fractions in each bin
        to a future work. 
We note that the results are broadly consistent with a switchover from a thin-disk 
to a thick-disk population as one moves from heights of $< 0.5$ kpc 
to over 1.5 kpc above 
        or below the plane, with a few places where halo or other low-metallicity
        stars, with [Fe/H] of -1.6, 
may represent a significant fraction of the mix toward the Galactic
        anticenter.

      We ask the question of whether the larger differences of up to 20\% 
in stellar density between
      the north and south in various longitude bins seen in Figure \ref{fig:residualoverlays}, 
could be due to a significant
        difference in the metallicity of the population of stars between the north and the 
        south, rather than a difference in overall star density at a given distance.
        For instance, following the lower right panel in Figure \ref{fig:residualoverlays}, 
which has $165^\circ < l < 180^\circ$, 
$46.6^\circ < |b| < 64.3^\circ$, 
there is a 20\% excess in density of star counts at $|z|=0.7$ kpc in the
     north versus the south. 
To explain this difference as a difference in population metallicity 
    and not as an overall stellar density wave, one would need, for instance, a very large
        difference in the metallicity 
as a function of $|z|$ in the ratio of thin- to thick-disk stars between 
       north and south, and that large difference in metals is not compatible
        with the relatively smooth change in metals content of stars seen around the 
Galaxy~\citep{ivezic08,hayden14}. 

   Plotting the matched N/S stellar count data in Galactic
        $(R,\phi,z)$ coordinates shows a 20\% excess of counts in the south
        just beyond (0-0.5 kpc) the solar radius toward the anticenter at
        a height of 0.4 kpc below the Galactic plane.  This excess becomes
        a 10\% deficit (or excess in the north) once one reaches a height
        of 0.8 kpc above the plane (see Figure \ref{fig:Phibreaking}).  It is interesting
        to note that at lower vertical heights, the Orion spur is 
        in this same part of the sky.  In other
        directions, as in we see in Figure \ref{fig:residualoverlays}, 
at similar (sub-kpc) distances from the Sun, the
        asymmetry mostly manifests itself as an excess of counts in
        the south at about 0.4 kpc below the plane.  In all subsamples
        of the matched north-south star-count data, we report here
        that the scale height of both the thin- and thick-disk stellar
        populations appears to be systematically larger in the
        north than the south, when the scale heights are fit separately.
        This may indicate a slight displacement of the center plane
        of the thick disk above the thin disk near the Sun or other
        configurations suggesting that the disks are not fully in stationary
        equilibrium.

        These quantitative locations and amplitudes of over- and underdensities 
should be useful in serving as a constraint on dynamical models of the Galaxy, 
which would describe its dark matter
        distribution, its satellites, and its stellar disk 
as they interact and evolve over time. Insights
        into the past history and distribution of matter can be inferred by a model that 
        reproduces these disk asymmetries. We note, too, that
comparing the vertical distribution of 
number counts as a function of selected latitude may prove an efficient way of 
locating possible stellar populations, or streams, 
of ultralow metallicity for subsequent spectroscopic study. 
The $\Lambda$CDM model speaks to dark matter with small-scale phase-space structure
and motivates the search for stellar streams. 
Follow-up studies at yet higher resolution with our methods could look for
variations and asymmetries with greater sensitivity, though they 
would likely require improved photometric calibrations and 
reddening corrections. 
These improvements already largely exist~\citep{green15,finkbeiner16}, 
so that we look forward to data from the {\it Gaia} era~\citep{gaia16a,gaia16b} with 
confidence in the ability to further and refine the studies pioneered here.

\acknowledgements{

This manuscript evolved from D.F.'s 2016 May senior honors thesis at the University of Kentucky, 
and she acknowledges the support of 
The University of Kentucky Singletary Scholarship
as well as an Undergraduate Summer Research 
Fellowship from the University of Kentucky during its completion. 

D.F. and S.G. acknowledge partial support from the U.S. Department of Energy under
contract DE-FG02-96ER40989. S.G. thanks Jonathan Feng and his colleagues at the
University of California, Irvine for gracious hospitality 
during the completion of 
this work.
%, as well as for partial support through Simons Investigator (JF) 
%Award \# 376204. 

 We thank the referee for several important suggestions that improved
the presentation of this paper, including presentation of the data in
Galactic coordinates. 

Funding for the Sloan Digital Sky Survey has been provided by the
Alfred P. Sloan Foundation, the U.S. Department of Energy Office of
Science, and the Participating Institutions. SDSS acknowledges
support and resources from the Center for High-Performance Computing at
the University of Utah. The SDSS web site is http://www.sdss.org$\, .$

SDSS is managed by the Astrophysical Research Consortium for the Participating Institutions of the SDSS Collaboration including the Brazilian Participation Group, the Carnegie Institution for Science, Carnegie Mellon University, the Chilean Participation Group, the French Participation Group, Harvard-Smithsonian Center for Astrophysics, Instituto de Astrofísica de Canarias, The Johns Hopkins University, Kavli Institute for the Physics and Mathematics of the Universe (IPMU) / University of Tokyo, Lawrence Berkeley National Laboratory, Leibniz Institut f\"ur Astrophysik Potsdam (AIP), Max-Planck-Institut f\"ur Astronomie (MPIA Heidelberg), Max-Planck-Institut f\"ur Astrophysik (MPA Garching), Max-Planck-Institut f\"ur Extraterrestrische Physik (MPE), National Astronomical Observatories of China, New Mexico State University, New York University, University of Notre Dame, Observatório Nacional / MCTI, The Ohio State University, Pennsylvania State University, Shanghai Astronomical Observatory, United Kingdom Participation Group, Universidad Nacional Autónoma de México, University of Arizona, University of Colorado Boulder, University of Oxford, University of Portsmouth, University of Utah, University of Virginia, University of Washington, University of Wisconsin, Vanderbilt University, and Yale University.
}

{\noindent 
$^*$Present Address: {School of Physics, Georgia Institute of Technology, Atlanta, GA 30332, USA}}

\bibliographystyle{apj}

\end{document}